\documentclass[paperpaper,twocolumn,english,superscriptaddress,aps,prb,floatfix,amsfonts,amssymb]{revtex4}
\usepackage{float}
\usepackage{epsfig}
\usepackage{graphicx}
\usepackage{dcolumn}
\usepackage{bm}
\usepackage{appendix}
\usepackage{subfigure}
\usepackage{color}
\usepackage{natbib}

\begin{document}


\title{One-particle spectral functions of the one-dimensional Fermionic Hubbard model with 
one fermion per site at zero and finite magnetic fields}
\author{Jos\'e M. P. Carmelo}
\affiliation{Center of Physics of University of Minho and University of Porto, P-4169-007 Oporto, Portugal}
\affiliation{Department of Physics, University of Minho, Campus Gualtar, P-4710-057 Braga, Portugal}
\affiliation{Boston University, Department of Physics, 590 Commonwealth Ave, Boston, Massachusetts 02215, USA}
\author{Tilen \v{C}ade\v{z}}
\affiliation{Center for Theoretical Physics of Complex Systems, Institute for Basic Science (IBS), Daejeon 34126, Republic of Korea}
\author{Pedro D. Sacramento}
\affiliation{CeFEMA, Instituto Superior T\'ecnico, Universidade de Lisboa, Av. Rovisco Pais, P-1049-001 Lisboa, Portugal}

\date{28 December 2020}


\begin{abstract}
Although charge-spin separation has important consequences for
the properties of one-dimensional (1D) Mott-Hubbard insulators,
matrix elements in some of its dynamical correlation functions involve
the coupling of spin and charge degrees of freedom.
The corresponding interplay of the 1D Mott-Hubbard insulator's 
charge and spin degrees of freedom is an issue of both fundamental and technological interest,
for instance concerning their dynamics at sub-picosecond timescales.
On the other hand, the up- and down-spin one-particle spectral functions are the
simplest dynamical correlation functions that involve excitation of
both the charge and spin degrees of freedom. They
are thus suitable to extract basic useful physical information on the above interplay 
at finite magnetic fields. Here the line shape of such
functions at and in the $(k,\omega)$-plane's vicinity of their cusp singularities
is studied for the Mott-Hubbard insulator described by
the 1D Hubbard model with one fermion per site at zero and finite magnetic fields. 
At zero field they can be accessed in terms of electrons by photoemission experiments. 
At finite field, such functions and corresponding interplay of correlations and magnetic-field 
effects refer to an involved non-perturbative many-particle problem that is poorly understood.
The Mott-Hubbard gap that separates the addition and removal spectral functions
is calculated for all spin densities and interaction values. 
The qualitative differences in the one-particle properties of the Mott-Hubbard insulator 
and corresponding doped insulator are also investigated. 
The relation of our theoretical results and predictions to both condensed-matter and ultracold 
spin-$1/2$ atom systems is discussed. 
 \end{abstract}


\maketitle

\section{Introduction}
\label{SECI}
Charge-spin separation has important physical consequences for
the properties of one-dimensional (1D) Mott-Hubbard insulators \cite{Mizuno_00}.
Their spin spectra are gapless in spite of the finite charge Mott-Hubbard gap $2\Delta_{MH}$.
However, the matrix elements in some of its dynamical correlation functions involve
the coupling of spin and charge degrees of freedom, which refers both to frequency and time dependencies.
For instance, the coupling of the spin degrees of freedom to the Floquet sector with shifted transfer 
energies $2\Delta_{MH} \pm \omega$ by virtual absorption and emission of photons \cite{Mentink_15} of
energy $\omega$ plays a key role in the dynamics of 1D Mott-Hubbard insulators at sub-picosecond timescales.
This is an issue of both fundamental and technological interest \cite{Mentink_15,Shinjo_18}.
The same applies to the more general problem of the 1D Mott-Hubbard insulators's
interplay of charge and spin degrees of freedom. For instance, after a sudden quench, such insulators 
realize a non-thermal state that is an admixture of spin and charge density wave states \cite{Zawadzki_19}.

The 1D Hubbard model with one fermion per lattice site, onsite repulsion 
$U$, and first neighboring integral $t$ (usually called half-filled 1D Hubbard model)
is exactly solvable by the Bethe ansatz \cite{Lieb_68,Lieb_03,Takahashi_72,Martins_98}.
It corresponds to phase V in the phase diagram shown in Fig. 1 of  Ref. \onlinecite{Patu_20},
which presents a recent study of the 1D Hubbard model in a magnetic field's
quantum critical behavior and thermodynamics. The half-filled 1D Hubbard model
is the simplest and an emblematic example of a 1D Mott-Hubbard insulator. 
However, the interplay between correlation effects associated with the gapped 
charge excitations and its spin degrees of freedom is an involved many-particle problem that 
was little studied, is not well understood, and deserves further investigations.

A very recent study on the spin dynamical properties of that model in a magnetic field
was on the spin longitudinal and transverse dynamical structure factors 
\cite{Carmelo_21}. Beyond previous studies \cite{Carmelo_16}, it accounted for contributions from
spin $n$-strings. It was found that the momentum dependent
exponents that control such factors line shape at and near the cusp singularities depend very little 
on $u=U/4t$ \cite{Carmelo_21,Carmelo_16}. The main effect of correlations was found to be on the
energy bandwidth of the dynamical structure factors's $(k,\omega)$-plane spectra, which increases
upon decreasing $u$, yet preserves the same shape. 

Such a study revealed that {\it both} the isotropic spin-$1/2$ Heisenberg model and 
the half-filled 1D Hubbard model for {\it any} finite $u=U/4t$ value 
and a suitable choice of units for the spectra's energy bandwidths describe the same spin dynamical properties. 
This suggests that the suitable values of the interaction for chain compounds and ultracold spin-$1/2$ atom systems 
described by 1D Mott-Hubbard insulators could be settled by the agreement with experimental results
on the up- and down-spin one-particle spectral functions, at energy scales above the Mott-Hubbard gap.
In addition and importantly, they are the simplest dynamical correlation functions that involve excitation of
{\it both} the charge and spin degrees of freedom, being more sensitive to magnetic effects than the 
charge dynamics \cite{Pereira_12}.

For electrons, the one-particle spectral function describes at zero field 
spectral-weight distributions that can be accessed by angle-resolved photoemission spectroscopy (ARPES) 
\cite{Damascelli_03,Sing_03,Carmelo_06} in low-dimensional Mott-Hubbard 
insulators and corresponding doped insulators \cite{Kim_06}. 
1D Mott-Hubbard insulators can be studied within condensed matter by inelastic neutron scattering in spin chains
whose charge degrees of freedom are gapped, as well as a 
number of quasi-1D organic compounds \cite{Raczkowski_15}. 

On the other hand, in the presence of a magnetic field, 
electrons are deflected and the up- and down-spin one-particle spectral function are not accessible 
via ARPES. Such spectral functions and corresponding interplay of correlations and magnetic-field effects 
refers though to a physically interesting and involved non-perturbative many-particle problem that
is poorly understood. It could in principle be experimentally accessible in systems of ultra-cold spin-$1/2$ atoms on optical lattices
\cite{Batchelor_16,Guan_13,Zinner_16,Dao_2007,Stewart_08,Clement_09,Febbri_12}. 

In the case of the 1D half-filled Hubbard model
at chemical potential $\mu =0$, there is for $u=U/4t>0$ no one-particle spectral weight in the excitation energy 
range within the Mott-Hubbard gap
$2\Delta_{MH}$ \cite{Lieb_68,Ovchinnikov_70}. The spectral-weight distributions of the
removal and addition one-particle spectral functions thus occur in $(k,\omega)$-plane
regions $\omega < -\Delta_{MH}$ and $\omega > \Delta_{MH}$, respectively. 

Most studies of the 1D half-filled Hubbard model's dynamical correlation
functions \cite{Nocera_18,Lante_09,Benthien_07} and related properties 
\cite{Balzer_08,Carmelo_97,Controzzi_02,Baeriswyl_87} refer to zero magnetic field. 
Few of them focused specifically on the one-particle spectral functions. In the case of zero and low temperatures,
the latter considered interaction values in the $u=U/4t\in [1,2]$ range and relied on 
{\it time dependent} density matrix renormalization group (tDMRG) computations \cite{Nocera_18},
combination of Bethe ansatz results, Lanczos diagonalizations, and field theoretical approaches \cite{Lante_09},
and the {\it dynamical} density-matrix renormalization group (DDMRG) method \cite{Benthien_07}.

Other studies of the half-filled 1D Hubbard model at zero magnetic field
concerned, for instance, static and one-particle dynamical quantities \cite{Balzer_08},
the gapped optical conductivity \cite{Carmelo_97}, the dynamical density-density correlation function 
\cite{Controzzi_02}, and the Mott-Hubbard insulator's enhancement of lattice dimerization 
by Coulomb correlations \cite{Baeriswyl_87}. On the other hand, studies on half-filled 1D Hubbard 
model at finite magnetic field refer to the low-energy limit \cite{Frahm_08}.

Our goal is the study of the line shape of the up- and down-spin one-particle 
spectral functions of the 1D Hubbard model with one fermion per site 
both at zero and finite magnetic fields at and in the $(k,\omega)$-plane's vicinity 
of such functions's cusp singularities by means of a suitable dynamical theory. 
The latter refer to peaks in the vicinity of which most one-particle
spectral weight is located. 

The Mott-Hubbard gap that separates the addition and removal spectral functions
is calculated for all spin densities and interaction values. Accounting for the relation
$B_{\sigma,+1} (k,\omega) = B_{\bar{\sigma},-1} (\pi-k,-\omega)$ between the one-particle
addition $B_{\sigma,+1} (k,\omega)$ and removal $B_{\bar{\sigma},-1} (k,\omega)$ spectral functions, 
where here and below $\bar{\sigma}$ is the spin projection opposite to $\sigma$, for simplicity 
our study focuses explicitly on the latter function. 

As shortly reported in Sec. \ref{SECII} and Appendix \ref{A}, the
irreducible representations of the $\eta$-spin $SU(2)$ symmetry group 
are for the 1D Hubbard model naturally described by configurations of rotated 
fermions whose relation to fermions has been uniquely defined
in terms of the corresponding unitary operator's matrix elements between all
Hilbert space's energy eigenstates in the case of electrons \cite{Carmelo_17}. 
Indeed, the symmetries of the Hubbard model on any bipartite lattice include such an $\eta$-spin $SU(2)$
symmetry, the corresponding $\eta$-spins of projection $-1/2$ and $+1/2$ referring 
to $SU(2)$ $\eta$-spin symmetry degrees of freedom of sites doubly occupied and unoccupied, respectively,
by rotated fermions \cite{Carmelo_17,Carmelo_18A,Carmelo_10}.

Before studying the Mott-Hubbard insulator's spectral functions line shape near their cusp singularities,
we account for such a $SU(2)$ symmetry irreducible representations to show that the one-particle addition
upper Hubbard band stems from transitions to excited states of (i) the Mott-Hubbard insulator and
(ii) doped Mott-Hubbard with a small yet finite deviation from fermionic density $1$ that have a
different $\eta$-spin character. Such states refer to creation of one $\eta$-spin of projection $-1/2$ (i) 
associated with an $\eta$-spin doublet and (ii) that emerges within an $\eta$-spin singlet pair. 
(Property (ii) applies to all densities of the metallic phase \cite{Carmelo_17}.)

The $\eta$-spins of projection $-1/2$ and $+1/2$ whose one-particle state configurations are shown in
this paper to be different for non-doped and doped Mott-Hubbard insulators, control important
physical effects associated with other types of excitations. Specifically, they
control third-harmonic generation spectroscopy, 
which plays a key role in the realization of all-optical switching, modulating, and computing devices
of modern optical technology \cite{Shinjo_18,Kishida_00}. 
Indeed, excited states containing a pair of $\eta$-spins with opposite projection $-1/2$ and $+1/2$
are behind the anomalously enhanced third-order nonlinear optical susceptibility $\chi^{(3)}$,
as observed in the 1D Mott-Hubbard insulator Sr$_2$CuO$_3$ \cite{Kishida_00}.
Within a description in terms of the 1D half-filled Hubbard model, this follows from
odd- and even-parity states being nearly degenerate with a large transition dipole moment between them.
That degeneracy is due to the spin-charge separation \cite{Mizuno_00}. For large $u$, rotated fermions become 
fermions and $\eta$-spins of projection $-1/2$ and $+1/2$ refer 
to specific degrees of freedom of sites doubly occupied and unoccupied by 
unrotated fermions, which have been called doublons and holons, respectively \cite{Zawadzki_19}.
A simplified $u\gg1 $ holon-doublon model reproduces very well the characteristic behaviors of the experimental
$\chi^{(3)}$ including data from the third-harmonic generation spectroscopy \cite{Mizuno_00}.

Conversely, $\eta$-spins of projection $-1/2$ and $+1/2$ are a generalization of doublons and holons, respectively,
for the whole $u>0$ range. This involves the replacement of the $u\gg 1$ fermion's doubly occupied and unoccupied
sites by rotated-fermion's doubly occupied and unoccupied sites, respectively, for $u>0$. 
In the present case of one-particle excited states, our studies account for 
the qualitative different $\eta$-spins configurations of non-doped and doped Mott-Hubbard insulators.
They rely on a corresponding extension to the gapped subspace associated with
the one-particle excited energy eigenstates of the 1D Hubbard model with one fermion per site
of the dynamical theory for the quantum metallic
phases of that model introduced in Refs. \onlinecite{Carmelo_05,Carmelo_08}. 
That theory was used in Ref. \onlinecite{Carmelo_17} to calculate expressions 
for the line shape at and near up- and down-spin one-particle spectral functions's cusp singularities 
in the case of the 1D Hubbard model in a magnetic field's metallic phase. 
For the up- and down-spin one-particle spectral functions of the half-filled 1D Hubbard model, 
whose ground state refers to the Mott-Hubbard insulating quantum phase,
the dynamical theory used in this paper refers to a limiting case of that used in Ref. \onlinecite{Carmelo_17}.

On the other hand, in Refs. \onlinecite{Carmelo_21,Carmelo_16} the dynamical theory was used 
for a subspace of the half-filled 1D Hubbard model spanned by 
spin excited energy eigenstates. The Mott-Hubbard gap is not seen by
spin excitations, which are gapless at half filling. The corresponding gapless spin quantum problem is
qualitatively different from the gapped one-particle problem studied in this paper.
Indeed, for it the Bethe-ansatz charge $c$ branch of excitations does
not contribute to the dynamical theory. In contrast, for the one-particle problem both the
charge $c$ and spin $s$ branch of excitations contribute to the dynamical theory
and this applies {\it both} to the metallic and Mott-Hubbard insulator quantum phases.

In the case of integrable models, the general dynamical theories of Refs.
\onlinecite{Carmelo_21,Carmelo_16,Carmelo_17,Carmelo_05,Carmelo_08} are equivalent to and account for the
same microscopic processes \cite{Carmelo_18} as the mobile quantum impurity model scheme 
of Refs. \onlinecite{Imambekov_09,Imambekov_12}. Momentum dependent exponents in the expressions
of spectral functions have also been obtained in Refs. \onlinecite{Sorella_96,Sorella_98}.

The dynamical theory of Refs. \onlinecite{Carmelo_17,Carmelo_05,Carmelo_08} is a generalization to
the whole $u=U/4t>0$ range of the approach used in the $u\rightarrow\infty$ limit in Refs.
\onlinecite{Karlo_96,Karlo_97}. In that limit the Bethe-ansatz solution simplifies, which 
makes possible the derivation of several one-particle quantities \cite{Ogata_90,Sorella_92}.

The paper is organized as follows. In Sec. \ref{SECII} the model and its up- and down-spin 
one-particle spectral functions are introduced and the spin-density and $u$ dependencies of the
Mott-Hubbard gap are studied. The rotated fermions, corresponding fractionalized 
particles, and the differences relative to the doped Mott-Hubbard insulator
are the issues addressed in Sec. \ref{SECIII}. In Sec. \ref{SECIV} the spectra and the 
general expressions of the spectral functions at and near their cusp singularities are introduced.
The zero spin-density spectra and exponents are studied in Sec. \ref{SECV}.
In Sec. \ref{SECVI} the specific line shapes at and in the vicinity of the branch and boundary lines of the up- and down-spin 
one-particle spectral functions are calculated for finite spin density. The effects of the charge-spin separation and 
charge-spin recombination are the issues discussed in Sec. \ref{SECVII} Finally, the discussion of the results and 
the concluding remarks are presented in Sec. \ref{SECVIII}. Some useful side results and 
derivations needed for our studies are presented in three Appendices.

\section{The model, the spectral functions, and the $m$
dependence of the Mott-Hubbard gap}
\label{SECII}

The Hubbard model with one fermion per site in a magnetic field $h$ 
under periodic boundary conditions on a 1D lattice with an even number $N_a\rightarrow\infty$ of 
sites is given by,
\begin{equation}
{\hat{H}} = {\hat{H}}_H + 2\mu_B h\,{\hat{S}}_s^{z} 
\hspace{0.20cm}{\rm where}\hspace{0.20cm}
{\hat{H}}_H = t\,\hat{T}+U\,\hat{V}_D \, ,
\label{H}
\end{equation}
and
\begin{eqnarray}
\hat{T} & = & -\sum_{\sigma=\uparrow,\downarrow }\sum_{j=1}^{L}\left(c_{j,\sigma}^{\dag}\,
c_{j+1,\sigma} + c_{j+1,\sigma}^{\dag}\,c_{j,\sigma}\right)\hspace{0.20cm}{\rm and}
\nonumber \\
\hat{V}_D & = & \sum_{j=1}^{L}\hat{\rho}_{j,\uparrow}\hat{\rho}_{j,\downarrow} \, ;
\hspace{0.50cm}
\hat{\rho}_{j,\sigma} = c_{j,\sigma}^{\dag}\,c_{j,\sigma} -1/2 \, ,
\label{HH}
\end{eqnarray}
are the kinetic-energy operator in units of $t$ and the on-site repulsion operator in units of $U$, 
respectively. The operator $c_{j,\sigma}^{\dagger}$ (and $c_{j,\sigma}$)
creates (and annihilates) a $\sigma=\uparrow ,\downarrow$ fermion (electron or atom) 
at lattice site $j=1,...,N_a$. The fermion number operators read
${\hat{N}}=\sum_{\sigma=\uparrow ,\downarrow}\,\hat{N}_{\sigma}$ and
${\hat{N}}_{\sigma}=\sum_{j=1}^{N_a}\hat{n}_{j,\sigma}= \sum_{j=1}^{N_a}c_{j,\sigma}^{\dag}\,c_{j,\sigma}$.
The model, Eq. (\ref{H}), describes $N=N_{\uparrow}+N_{\downarrow}=N_a$ 
interacting spin-$1/2$ fermions in a lattice with $N_a\rightarrow\infty$ sites
for spin densities $m = (N_{\uparrow}-N_{\downarrow})/N_a\in [0,1[$.

We use in general units of lattice constant and Planck constant one, the number of lattice sites 
$N_a=N_{\uparrow}+N_{\downarrow}$ equaling the lattice length $L$. 
In Eq. (\ref{H}), $\mu_B$ is the Bohr magneton, for simplicity in $g\mu_B$ we have taken the Land\'e factor 
to read $g=2$, and the operator ${\hat{S}}_s^{z}= -{1\over 2}\sum_{j=1}^{N_a}({\hat{n}}_{j,\uparrow}-{\hat{n}}_{j,\downarrow})$
is the diagonal generator of the model global spin $SU(2)$ symmetry, where
$\hat{n}_{j}=\sum_{\sigma}\hat{n}_{j,\sigma}$.
The model has also a global $\eta$-spin $SU(2)$ symmetry with diagonal generator
${\hat{S}}_{\eta}^{z}=-{1\over 2}\sum_{j=1}^{N_a}(1-\hat{n}_{j})$. 
We denote by $S_s$ and $S_s^z$ the energy eigenstates's spin
and spin projection and by $S_{\eta}$ and $S_{\eta}^z$ such states's $\eta$-spin
and $\eta$-spin projection.

The global symmetry of $\hat{H}_H$ for 1D and any bipartite lattice \cite{Carmelo_10} 
is at $U\neq 0$ given by $[SO(4)\otimes U(1)]/Z_2=
[SU(2)\otimes SU(2)\otimes U(1)]/Z_2^2$. 
Here $1/Z_2^2$ refers to the number $4^{N_a}$ of its irreducible  representations,
which refer to configurations of fractionalized particles considered in Appendix \ref{A}
and correspond to the $4^{N_a}$ energy eigenstates of $\hat{H}$,
being four times smaller than the dimension $4^{N_a+1}$ of the group $SU(2)\otimes SU(2)\otimes U(1)$. 
The global $c$ lattice $U(1)$ symmetry beyond $SO(4)=[SU(2)\otimes SU(2)]/Z_2$ is associated with the 
lattice degrees of freedom and does not exist at $U=0$, emerging at any arbitrarily small $u=U/4t$ value 
\cite{Carmelo_17,Carmelo_10,Carmelo_18A}.

The lowest weight states (LWSs) and highest weight states (HWSs) of the $\eta$-spin $(\alpha =\eta)$ and spin 
$(\alpha =s)$ $SU(2)$  symmetry algebras have numbers $S_{\alpha} = - S_{\alpha}^{z}$ and $S_{\alpha} = S_{\alpha}^{z}$, respectively. 

Our studies involve one-particle excited states of ground states with vanishing chemical potential $\mu=0$,
so that their zero energy reference level lies in the middle of the Mott-Hubbard gap $2\Delta_{MH}$. 
The up- and down-spin one-particle spectral functions $B_{\sigma,\gamma} (k,\,\omega)$ 
of the 1D half-filled Fermionic Hubbard model in a magnetic field $h$ where
$\gamma=-1$ (and $\gamma=+1$) for one-particle removal (and addition) then read,
\begin{eqnarray}
B_{\sigma,-1} (k,\,\omega) & = &  \sum_{\nu^-}
\vert\langle\nu^-\vert\, c_{k,\sigma} \vert \,GS\rangle\vert^2 \,\delta (\omega
+ \epsilon_{\sigma,\nu^-} (k)) 
\nonumber \\
{\rm for} && \omega \leq - \Delta_{MH} 
\nonumber \\
B_{\sigma,+1} (k,\,\omega) & = & \sum_{\nu^+}
\vert\langle\nu^+\vert\, c^{\dagger}_{k,\sigma} \vert
\,GS\rangle\vert^2 \,\delta (\omega - \epsilon_{\sigma,\nu^+} (k))  
\nonumber \\
{\rm for} && \omega \geq \Delta_{MH}\hspace{0.20cm}{\rm where}\hspace{0.20cm}
\sigma=\uparrow ,\downarrow \, .
\label{Bkomega}
\end{eqnarray}

Here $c_{k,\sigma}$ and $c^{\dagger}_{k,\sigma}$ are fermion
annihilation and creation operators, respectively, of momentum $k$. $\vert GS\rangle$ denotes the
initial $N_{\uparrow},N_{\downarrow}$ particle ground state of energy $E_{GS}^{N_{\sigma},N_{\bar{\sigma}}}$.
The $\nu^-$ and $\nu^+$ summations run over the $N_{\sigma}-1$ and $N_{\sigma}+1$-particle excited 
energy eigenstates, respectively. $E_{\nu^-}^{N_{\sigma}-1,N_{\bar{\sigma}}}$ and 
$E_{\nu^+}^{N_{\sigma}+1,N_{\bar{\sigma}}}$ are in $\epsilon_{\sigma,\nu^-} (k) = 
(E_{\nu^-}^{N_{\sigma}-1,N_{\bar{\sigma}}}-E_{GS}^{N_{\sigma},N_{\bar{\sigma}}})$
and $\epsilon_{\sigma,\nu^+} (k) = (E_{\nu^+}^{N_{\sigma}+1,N_{\bar{\sigma}}}-E_{GS}^{N_{\sigma},N_{\bar{\sigma}}})$
the corresponding energies.

Since the chemical potential, $\mu =0$, lies at the middle of the 
Mott-Hubbard gap, the following exact symmetry,
\begin{eqnarray}
B_{\sigma,+1} (k,\omega) & = & B_{\bar{\sigma},-1} (\pi-k,-\omega)
\hspace{0.20cm}{\rm for}\hspace{0.20cm} \sigma = \uparrow, \downarrow
\nonumber \\
{\rm with} && \bar{\uparrow} =\downarrow
\hspace{0.20cm}{\rm and}\hspace{0.20cm}
\bar{\downarrow} =\uparrow \, ,
\label{BB}
\end{eqnarray}
holds. We also rely on the following symmetry that refers to the spin density intervals 
$m\in [-1,0]$ and $m\in [0,1]$,
\begin{equation}
B_{\sigma,-1} (k,\omega) \vert_m = B_{\bar{\sigma},-1} (k,\omega)\vert_{-m}
\hspace{0.20cm}{\rm for}\hspace{0.20cm}m \in [0,1] \,  .
\label{SPMMPmm}
\end{equation}
We thus only consider explicitly the removal function 
$B_{\sigma,-1} (k,\omega)$ for the spin density interval $m\in [0,1]$.

The Mott-Hubbard gap $2\Delta_{MH}$
plays an important role in the one-particle spectral properties. 
Its general expression for general spin densities, $m\in [0,1]$, has not been given explicitly
in the literature. Relying on the Bethe-ansatz representation reported in Appendix \ref{B}, 
we have derived it in Appendix \ref{C} with the result,
\begin{eqnarray}
2\Delta_{MH} & = & U - 4t + \int_{-B}^{B}d\Lambda\,2t\,\eta_s (\Lambda)\,\Phi (\Lambda) + 2\mu_B\, h 
\nonumber \\
& = & U - 4t + \int_{-B}^{\infty}d\Lambda\,2t\,\eta_s (\Lambda)\,\Phi (\Lambda) 
\hspace{0.20cm}{\rm where}
\nonumber \\
\Phi (\Lambda) & = & \frac{2}{\pi}\arctan \left({\Lambda\over u}\right)\hspace{0.40cm}{\rm for}\hspace{0.40cm}\Lambda < B
\nonumber \\
& = & 1\hspace{0.40cm}{\rm for}\hspace{0.40cm}\Lambda > B \, .
\label{2DeltaMHallm}
\end{eqnarray}
Here the distribution $2t\,\eta_{s} (\Lambda)$ is solution of the coupled integral equations, Eq. (\ref{equA5}) 
of Appendix \ref{B}, and the parameter $B$ is defined in Eq. (\ref{QB-r0rs})
of that appendix. 
\begin{figure}
\begin{center}
\centerline{\includegraphics[width=8.75cm]{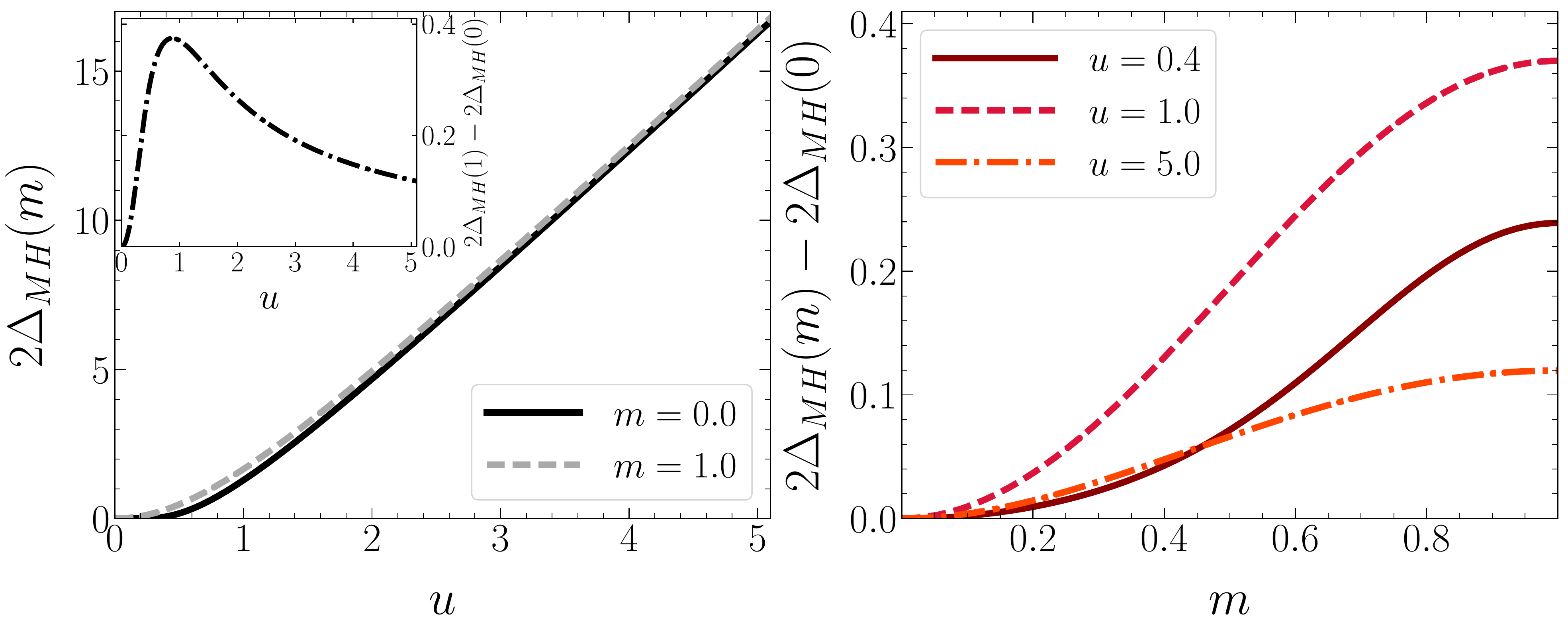}}
\caption{(a) The Mott-Hubbard gap $2\Delta_{MH}$, Eq. (\ref{2DeltaMHallm}), in units
of $t$ as a function of $u=U/4t$ for spin densities $m=0$ and $m=1$ and (b) the small 
deviation $2\Delta_{MH} (m) - 2\Delta_{MH} (0)$ in units of $t$ where $2\Delta_{MH} (0)$
corresponds to $m=0$ as a function of the spin density $m$ for
$u = 0.4, 1.0, 5.0$. The inset in (a) shows $2\Delta_{MH} (1) - 2\Delta_{MH} (0)$
as a function of $u$.}
\label{figure1}
\end{center}
\end{figure}

At $m= 0$, its expression simplifies and is well known\cite{Lieb_68,Lieb_03,Ovchinnikov_70}.
At $m= 0$ and in the $m\rightarrow 1$ limit, it reads,
\begin{eqnarray}
2\Delta_{MH} & = & U - 4t + 8t\int_0^{\infty}d\omega {J_1 (\omega)\over\omega\,(1+e^{2\omega u})} 
\nonumber \\
& = & {16\,t^2\over U}\int_1^{\infty}d\omega {\sqrt{\omega^2-1}\over\sinh\left({2\pi t\omega\over U}\right)} 
\hspace{0.20cm}{\rm at}\hspace{0.20cm}m = 0 \hspace{0.20cm}{\rm and}
\nonumber \\
& = & \sqrt{(4t)^2+U^2} - 4t \hspace{0.20cm}{\rm for}\hspace{0.20cm}m\rightarrow 1 \, ,
\label{2mu0}
\end{eqnarray}
respectively, where $J_1 (\omega)$ is a Bessel function. This gives,
\begin{eqnarray}
2\Delta_{MH} & \approx  & {8\sqrt{t\,U}\over \pi}\,e^{-2\pi \left({t\over U}\right)}
\hspace{0.20cm}{\rm for}\hspace{0.20cm}u\ll 1\hspace{0.20cm}{\rm and}\hspace{0.20cm}m=0
\nonumber \\
& \approx  & {U^2\over 8t}\hspace{0.20cm}{\rm for}\hspace{0.20cm}u\ll 1\hspace{0.20cm}{\rm and}\hspace{0.20cm}m\rightarrow 1
\nonumber \\
& \approx  & U - 4t\hspace{0.20cm}{\rm for}\hspace{0.20cm}u\gg 1\hspace{0.20cm}{\rm and}\hspace{0.20cm}
m \in [0,1[ \, .
\label{2mu0LB}
\end{eqnarray}

The Mott-Hubbard gap, Eqs. (\ref{2DeltaMHallm}) and (\ref{2mu0}), is plotted in Fig. \ref{figure1}\,(a) as a function of
$u=U/4t$ for spin densities $m=0$ and $m=1$. Although it is very little $u$ dependent, note that 
as shown in the insert of that figure, there is a small 
deviation $2\Delta_{MH} (1) - 2\Delta_{MH} (0)$ as a function of $u = U/4t$ for spin densities $m =
0$ and $m = 1$. That deviation reaches a maximum value at $u = U/4t = u_* \sim 1$.
The Mott-Hubbard gap $2\Delta_{MH} (m)$ is for all spin densities $m\in [0,1]$ an increasing
function of $u$. In order to compare its dependence on the spin density 
$m$ conveniently, we present that gap deviation from its $m = 0$ 
value in Fig. \ref{figure1}\,(b) for three representative $u$ values. As the intersect of the lines 
for $u=0.4$ and $u=5.0$ reveals, for smaller $m$ the deviation $2\Delta_{MH} (m) - 2\Delta_{MH} (0)$ is larger
for $u=5.0$ than for $u=0.4$ and the opposite occurs for $m$ larger than approximately
$0.45$. However, we recall that $2\Delta_{MH} (m)\vert_{u=5.0}>2\Delta_{MH} (m)\vert_{u=1}>2\Delta_{MH} (m)\vert_{u=0.4}$
for all spin densities $m\in [0,1]$.

That the Mott-Hubbard small $m$ dependence is more pronounced at intermediate
$u=U/4t\sim 1$ values is consistent with that gap having the same values for all spin densities $m\in [0,1]$ in the
$u\rightarrow 0$ and $u\gg 1$ limits in which it reads $2\Delta_{MH}=0$ and
$2\Delta_{MH}=U - 4t$, respectively. $2\Delta_{MH}$ is associated with
charge degrees of freedom, its little dependence on $m$ being
consistent with the $u>0$ charge - spin separation being strongest at half filling.

The interval  $m\in [0,1[$ refers to fields
in the range $h \in [0,h_c[$ where $h_c$ is the critical field for $m\rightarrow 1$
fully polarized ferromagnetism. The spin density curve $h (m) \in  [0,h_c[$
for $m\in [0,1[$ and 
$h_c$ are given by,
\begin{equation}
h (m) =  - {\varepsilon_s^0 (k_{F\downarrow})\over 2\mu_B} 
\hspace{0.20cm}{\rm and}\hspace{0.20cm}h_c = {\sqrt{(4t)^2+U^2} - U\over 2\mu_B} \, ,
\label{magcurve}
\end{equation}
respectively. Here the bare $s$-band energy dispersion $\varepsilon_s^0 (q)$ 
is defined in Eq. (\ref{equA112}) of Appendix \ref{C}. In the thermodynamic limit, one has that the 
momentum $k_{F\downarrow}$ in $\varepsilon_s^0 (k_{F\downarrow})$ and $k_{F\uparrow} = \pi - k_{F\downarrow}$ 
are for $m\in [0,1[$ given by,
\begin{equation}
k_{F\downarrow} = {\pi\over 2}(1-m)\hspace{0.20cm}{\rm and}\hspace{0.20cm}k_{F\uparrow} = {\pi\over 2}(1+m) \, .
\label{kkk}
\end{equation}

\section{Different one-particle processes of the doped and undoped Mott-Hubbard insulator}
\label{SECIII}

The dynamical theory used in our studies 
relies on a representation in terms of fractionalized particles that naturally
emerge from the rotated-fermion degrees of freedom separation 
\cite{Carmelo_21,Carmelo_16,Carmelo_05,Carmelo_08,Carmelo_17}.
It is briefly described for the whole Hilbert space in Appendix \ref{A}. This is useful for the introduction of 
its simplified form for the Mott-Hubbard insulator subspace of our study, which
is spanned by energy eigenstates described only by real Bethe-ansatz rapidities.
They are populated by $c$ particles, $s$ particles, unpaired spins $1/2$
of projection $+1/2$, and zero or one unpaired $\eta$-spin $1/2$ of projection $+1/2$.
Only the $c$- and $s$-bands whose Bethe-ansatz quantum numbers $q_j = {2\pi\over L}I_j^c$
and $q_j = {2\pi\over L}I_j^s$, respectively, are given in Eqs. (\ref{q-j}) and (\ref{Ic-an}) of Appendix \ref{B}
have finite occupancy. 

In contrast to the doped Mott-Hubbard insulator
and general metallic phase, the states that span that subspace are 
not populated by the $\eta$ particles considered in Appendix \ref{B}, which also refer to real Bethe-ansatz 
rapidities. The Bethe-ansatz equations and quantum numbers are given in Eqs. (\ref{TapcoS})-(\ref{Ic-an}) 
of Appendix \ref{B}. In the thermodynamic limit for $u>0$ and $m\geq 0$, the ground-state $s$-band Fermi points 
and limiting momentum values read $\pm k_{F\downarrow}$ and $\pm k_{F\uparrow}$, respectively, Eq. (\ref{kkk}).

Mott-Hubbard insulator ground states are for $m\in [0,1[$ not populated 
by $\eta$-spin $1/2$'s whose number is denoted by $M_{\eta,\pm 1/2}$
for projections $\pm 1/2$, are populated by a number $N_c = N = N_a$ of $c$ particles without
internal degrees of freedom, a number $N_s =\Pi_s = N_{\downarrow}$ of $s$ particles whose 
internal degrees of freedom refer to one unbound spin-singlet pair of spins $1/2$'s, 
and a number $M_{s,+1/2}= N_{\uparrow}-N_{\downarrow}$ of unpaired spin $1/2$'s of projection $+1/2$.
The translational degrees of freedom of such unpaired spin $1/2$'s are described by the $N_s^h = 
2S_s = N_{\uparrow}-N_{\downarrow}$ $s$-band holes. The $m=0$ ground state for which 
$M_s=M_{s,+1/2}+M_{s,-1/2} =2S_s =0$
is not populated by unpaired spin $1/2$'s and thus has no $s$-band holes.

Ground states of a doped Mott-Hubbard insulator with concentration $\delta = \vert 1-n_f\vert$ 
very small yet finite are populated by a number $N_c$ of $c$ particles and $N_s$ of $s$ particles 
that for spin LWSs for which $m\geq 0$ are given by,
\begin{eqnarray}
N_c & = & N \hspace{0.20cm}{\rm and}\hspace{0.20cm}N_s = N_{\downarrow} 
\hspace{0.20cm}{\rm for}\hspace{0.20cm}n_f = N/N_a \in [0,1]
\nonumber \\
N_c & = & N^h \hspace{0.20cm}{\rm and}\hspace{0.20cm}N_s = N_{\downarrow}^h 
\hspace{0.20cm}{\rm for}\hspace{0.20cm}n_f \in [1,2] \, .
\label{NcNs}
\end{eqnarray}
Here,
\begin{eqnarray}
N^h = 2N_a - N \, ,\hspace{0.15cm}
N_{\uparrow}^h = N_a - N_{\downarrow} \, , \hspace{0.15cm} N_{\downarrow}^h = N_a - N_{\uparrow} \, ,
\label{Nholes}
\end{eqnarray}
are the numbers of fermionic holes that are more suitable to describe the quantum problem for $n_f\in [1,2]$. 
Indeed, the corresponding fermionic hole density $n_f^h = N^h/L = N^h/N_a$ varies in the range $n_f^h\in [0,1]$.
(Each site has two orbitals, $N^h = 2N_a - N$ being the number of orbitals that are not occupied by fermions
and thus refer to fermionic holes.) The spin densities of such ground states are $m\in [0,n_f[$ for $n_f \in [0,1]$ 
and $m\in [0,n_f^h[$ for  $n_f \in [1,2]$ where $n_f^h = N^h/N_a$. The present representation 
$L_s = 2S_s + 2\Pi_s$ spin $1/2$'s are for $n_f \in [1,2]$ those of the $N^h$ rotated fermionic holes that single
occupy sites. For $n_f \in [0,2]$ there is are $M_s=2S_s = N_{\uparrow}-N_{\downarrow}=N_{\uparrow}^h-N_{\downarrow}^h$ 
unpaired spin $1/2$'s whose translational degrees of freedom 
are described by $N_s^h = 2S_s = N_{\uparrow}-N_{\downarrow}=N_{\uparrow}^h-N_{\downarrow}^h$
$s$-band holes. In the ground states of the doped case, there are
$M_{\eta} = 2S_{\eta} = N_a - N$ unpaired $\eta$-spins of projection $+1/2$
for $n_f \in [0,1]$ and $M_{\eta} = 2S_{\eta} = N - N_a$ unpaired $\eta$-spins of projection $-1/2$
for $n_f \in [1,2]$. 

To address the qualitative differences of the one-particle 
problem for the Mott-Hubbard insulator and the corresponding doped insulator, respectively,
we start by considering fermionic densities $n_f\in [0,2]$ and
an Hamiltonian ${\hat{H}} + 2\mu\,{\hat{S}}_{\eta}^{z}$. Here ${\hat{H}}$ is 
given in Eq. (\ref{H}) and $\mu$ is the chemical potential.
The ranges $n_f\in [0,1]$ and $n_f\in [1,2]$ refer to the $\eta$-spin LWS and HWS Bethe 
ansatzes and thus to subspaces spanned by energy eigenstates
that are $S_{\eta}^z = - S_{\eta}$ $\eta$-spin LWSs and $S_{\eta}^z = S_{\eta}$ $\eta$-spin HWSs,
respectively. An useful symmetry relation that connects the $\sigma = \uparrow,\downarrow$ one-particle spectral functions for
fermionic densities $n_f<1$ and $n_f'=2-n_f>1$, respectively, is,
\begin{equation}
B_{\sigma,\gamma} (k,\,\omega)\vert_{n<1} 
= B_{\bar{\sigma},-\gamma} (\pi - k,\,-\omega)\vert_{n'=2-n>1} \, .
\label{relationBBnn}
\end{equation}
For $n_f\rightarrow 1$ and thus $n_f^h=2-n_f\rightarrow 1$
this general relation gives that provided in Eq. (\ref{BB}).

The $\sigma=\uparrow,\downarrow$ one-particle spectral functions obey the following sum rules,
\begin{eqnarray}
\sum_k\int_{-\infty}^{\infty}d\omega\,B_{\sigma,-1} (k,\,\omega) & = & N_{\sigma} 
\nonumber \\
\sum_k\int_{-\infty}^{\infty}d\omega\,B_{\sigma,+1} (k,\,\omega) & = & N_a - N_{\sigma} 
\nonumber \\
\sum_k\int_{-\infty}^{\infty}d\omega\,B^{\rm LHB}_{\sigma,+1} (k,\,\omega) & = & N_a - N 
\nonumber \\
\sum_k\int_{-\infty}^{\infty}d\omega\,B^{\rm UHB}_{\sigma,+1} (k,\,\omega) & = & N - N_{\sigma} 
= N_{\bar{\sigma}} \, ,
\label{sumrulesplus}
\end{eqnarray}
for $n_f\in [0,1]$ and,
\begin{eqnarray}
\sum_k\int_{-\infty}^{\infty}d\omega\,B_{\sigma,-1} (k,\,\omega) & = & N_a - N_{\sigma}^h 
\nonumber \\
\sum_k\int_{-\infty}^{\infty}d\omega\,B_{\sigma,+1} (k,\,\omega) & = & N_{\sigma}^h 
\nonumber \\
\sum_k\int_{-\infty}^{\infty}d\omega\,B^{\rm -LHB}_{\sigma,-1} (k,\,\omega) & = & N - N_a 
\nonumber \\
\sum_k\int_{-\infty}^{\infty}d\omega\,B^{\rm -UHB}_{\sigma,-1} (k,\,\omega) & = & 
N^h - N_{\sigma}^h = N_{\bar{\sigma}}^h \, ,
\label{sumrulesminus}
\end{eqnarray}
for $n_f\in [1,2]$. Here the symbols lower Hubbard band (LHB) and upper Hubbard band (UHB)
(and -LHB and -UHB) refer for $n_f\in [0,1[$ (and $n_f\in ]1,2]$) 
to one-particle addition (and removal) for $\omega >0$ (and $\omega <0$.) 

The LHB (and -LHB) is generated by transitions to addition (and removal) excited energy eigenstates
that are not populated by $\eta$-spins of projection $-1/2$ (and $+1/2$). 
The UHB (and -UHB) is generated by transitions to such excited states 
that are populated by $\eta$-spins of projection $-1/2$ (and $+1/2$). Nearly all UHB 
(and -UHB) spectral weight stems from transitions to addition (and removal) excited energy eigenstates that are 
populated by a single $\eta$-spin of projection $-1/2$ \cite{Carmelo_17} (and $+1/2$.) 

The first two sum rules in Eqs. (\ref{sumrulesplus}) and (\ref{sumrulesminus}) are well known and exact for all $u$ values. 
The (i) $B^{\rm LHB}_{\sigma,+1} (k,\,\omega)$ and $B^{\rm UHB}_{\sigma,+1} (k,\,\omega)$ sum rules 
and (ii) $B^{\rm -LHB}_{\sigma,-1} (k,\,\omega)$ and $B^{\rm -UHB}_{\sigma,-1} (k,\,\omega)$ sum rules 
are for $u>0$ found to be exact in the (i) $n_f\rightarrow 0$ and $n_f\rightarrow 1$ limits and
(ii) $n_f^h\rightarrow 0$ and $n_f^h\rightarrow 1$ limits, respectively.
Both in the $u\ll 1$ and $u\gg 1$ limits they are exact for all fermionic densities 
$n_f \in [0,2]$ and all corresponding spin densities.
They are exact or a very good approximation also for intermediate $u$ values.
Clarification of this issue is not needed for our studies, since
for the Mott Hubbard insulator and the doped Mott Hubbard insulator 
for which $\delta = \vert 1-n_f\vert\ll 1$ is finite but very small all such sum rules apply.

For simplicity, in general in this paper we consider spin densities $m\geq 0$ associated with
spin LWSs such that $\delta M_{s,-1/2}=0$, yet similar results are obtained for $m\leq 0$.
The processes associated with one-particle removal (REM) for $n_f\leq 1$ and 
one-particle addition for $n_f\geq 1$ are similar for the doped Mott-Hubbard insulator
and the present $n_f = n_f^h =1$ Mott-Hubbard insulator. (See the REM fractionalized particles numbers deviations in 
Table \ref{table1} for $n_f\in [0,1[$. For how they relate to those of addition 
for $n_f\in ]1,2]$, see that table caption.)

A first qualitative difference is that the one-particle addition LHB for $n_f\in [0,1[$ and 
the one-particle removal -LHB for $n_f\in ]1,2]$ do not
exist for the Mott-Hubbard insulator. As justified in the following, the 
processes associated with the UHB one-particle addition for $n_f\leq 1$
and -UHB one-particle removal for $n_f\geq 1$ are also qualitatively different for the doped Mott-Hubbard insulator
and the Mott-Hubbard insulator. 

Let $\vert \nu^{+,0},N\rangle$ and $\vert \nu^{-,0},N\rangle$ denote energy 
eigenstates of the Hamiltonian ${\hat{H}} + 2\mu\,{\hat{S}}_{\eta}^{z}$ where the upper index $0$ indicates they 
are $\eta$-spin LWSs and HWSs for upper indices $+$ and $-$, respectively. They thus refer to two 
corresponding Bethe-ansatz solutions for $n_f\in [0,1[$ and $n_f\in ]1,2]$, respectively, and $u>0$. 
Their label $\nu$ represents $u=U/4t$ and the set of all $u$-independent 
quantum numbers other than $N$ needed to uniquely specify an energy eigenstate. This refers to occupancy
configurations of Bethe-ansatz momentum quantum numbers $q_j = {2\pi\over L}\,I^{\beta}_j$. Here $I^{\beta}_j$ are 
successive integers, $I^{\beta}_j=0,\pm 1,\pm 2,...$, or half-odd integers ,$I^{\beta}_j=\pm 1/2,\pm 3/2,\pm 5/2,...$, 
according to well-defined boundary conditions. Their allowed Pauli-like occupancies are zero and one. The index
$\beta$ labels the Bethe-ansatz band $\beta=c,s,\eta,\eta n,s n$ where $n>1$ are $n$-strings lengths.
We denote by $\vert GS^{+,0},N\rangle$ and $\vert GS^{-,0},N\rangle$ the ground states that
are $\eta$-spin LWSs and HWSs for $n_f\in [0,1[$ and $n_f\in ]1,2]$, respectively. 
\begin{table}
\begin{center}
\begin{tabular}{|c|c|c|c|c|c|c|} 
\hline
& $\delta N_c$ & $\delta N_s$ & $\delta N_{\eta}$ & $\delta M_{\eta,-1/2}$ & $\delta M_{\eta,+1/2}$ & $\delta M_{s,1/2}$\\
\hline
I$_{\rm UHB}$ $\delta N_{\uparrow} = 1$ & $-1$ & $-1$ & $0$ & $1$ & $0$ & $1$ \\
\hline
II$_{\rm UHB}$ $\delta N_{\uparrow} = 1$ & $-1$ & $-1$ & $1$ & $0$ & $-1$ & $1$ \\
\hline
I$_{UHB}$ $\delta N_{\downarrow} = 1$ & $-1$ & $0$ & $0$ & $1$ & $0$ & $-1$ \\
\hline
II$_{\rm UHB}$ $\delta N_{\downarrow} = 1$ & $-1$ & $0$ & $1$ & $0$ & $-1$ & $-1$ \\
\hline
$_{\rm LHB}$ $\delta N_{\uparrow} = 1$ & $1$ & $0$ & $0$ & $0$ & $-1$ & $1$ \\
\hline
$_{\rm LHB}$ $\delta N_{\downarrow} = 1$ & $1$ & $1$ & $0$ & $0$ & $-1$ & $-1$ \\
\hline
$_{\rm REM}$ $\delta N_{\uparrow} = -1$ & $-1$ & $0$ & $0$ & $0$ & $1$ & $-1$ \\
\hline
$_{\rm REM}$ $\delta N_{\downarrow} = -1$ & $-1$ & $-1$ & $0$ & $0$ & $1$ & $1$ \\
\hline
\end{tabular}
\caption{Fractionalized particles numbers deviations for one-particle 
types I and II UHB addition, LWS addition, and removal for 
$n_f\in [0,1]$. Similar values apply for types I and II -UHB removal, -LWS removal, 
and addition for $n_f\in [1,2]$ provided that the numbers $\delta N_{\sigma}$
and $\delta M_{\eta,\mp 1/2}$ of rotated-fermion unoccupied $(+1/2)$ and
doubly occupied $(-1/2)$ sites
are replaced by those for $\delta N_{\sigma}^h$ 
and $\delta M_{\eta,\pm 1/2}$, respectively. (For excited states of $m=0$ ground states, the 
LHB deviations for $\delta N_{\downarrow} = 1$ and the REM deviations for $\delta N_{\uparrow} = -1$ 
should be replaced by those for $\delta N_{\uparrow} = 1$ and $\delta N_{\downarrow} = -1$,
respectively, with the number $\delta M_{s,-1/2}$ of rotated-fermion singly occupied
sites of spin projection $-1/2$ replaced by $\delta M_{s,1/2}$.)}
\label{table1}
\end{center}
\end{table} 

Most of the UHB and -UHB spectral weight results from transitions from the ground states $\vert GS^{+,0},N\rangle$ and 
$\vert GS^{-,0},N\rangle$ to energy eigenstates populated by a single $\eta$-spin of projection 
$-1/2$ and $+1/2$, respectively. There are three {\it qualitatively different} types of such states:\\ 

Type I: The single $\eta$-spin of projection $-1/2$ and $+1/2$ is unpaired and {\it is not generated} from a
$\eta$-spin of projection $+1/2$ and $-1/2$, respectively, by an $\eta$-spin flip process.\\

Type II: The single $\eta$-spin is paired within an $\eta$-spin singlet pair.\\  

Type III: The single $\eta$-spin of projection $-1/2$ and $+1/2$ is unpaired and {\it is generated} from a
$\eta$-spin of projection $+1/2$ and $-1/2$, respectively, by an $\eta$-spin flip process.\\

Our goal here is to confirm that the type I and II excited states are those that play an active role
in the cases of the Mott-Hubbard insulator and doped Mott-Hubbard insulator 
for which $\delta = \vert 1-n_f\vert\ll 1$ is finite but very small, respectively.
The I$_{\rm UHB}$ and II$_{\rm UHB}$ fractionalized particles numbers deviations 
are given in Table \ref{table1} for the UHB type I and II excited states, respectively. 
That table caption reports how they relate to those of the -UHB.

In the case of a doped Mott-Hubbard insulator for which $\delta = \vert 1-n_f\vert\ll 1$ is finite but very small,
creation of a single unpaired $\eta$-spin of projection $-1/2$ and $+1/2$ under
UHB creation of one fermion for $n_f <1$ and -UHB annihilation of one fermion for $n_f >1$,
respectively, can only occur through transitions from the ground state to type II or type III excited states. 
Indeed, it turns out that type I excited states are populated by a single $\eta$-spin $1/2$ 
whose projection is $-1/2$ and $+1/2$ for fermion creation and annihilation, respectively.
Type I excited states thus do not exist in the subspace of the doped Mott-Hubbard insulator
under consideration. Our following analysis thus involves matrix elements between that 
insulator's ground state and type II and type III excited states, respectively.

We start by considering transitions to the latter type III UHB and -UHB excited energy eigenstates 
of the doped Mott-Hubbard insulator, which we denote by $\vert\nu^+_{uhb},N+1\rangle$ and $\vert\nu^-_{uhb},N-1\rangle$,
respectively. The absence of the upper label $0$ means they are not $\eta$-spin LWSs and 
HWSs, respectively. They can be written as,
\begin{eqnarray}
\vert \nu^+_{uhb},N+1\rangle & = & {1\over\sqrt{N_a-N+1}}\,{\hat{S}}^{+}_{\eta}\vert \nu^{+,0},N-1\rangle
\nonumber \\
\vert \nu^-_{uhb},N-1\rangle & = & {1\over\sqrt{N-N_a+1}}\,{\hat{S}}^{-}_{\eta}\vert \nu^{-,0},N+1\rangle \, .
\label{twostates}
\end{eqnarray}
These type III excited states are $\eta$-spin non-LWSs (+) (and non-HWSs (-)) of the 
doped Mott-Hubbard insulator that are generated from corresponding 
$\eta$-spin LWSs (+) $\vert  \nu^{+,0},N-1\rangle$ (and HWSs (-) $\vert  \nu^{-,0},N+1\rangle$.)
The latter states $\eta$-spin is such that $2S_{\eta} = N_a-N+1>0$ (and $2S_{\eta}=N-N_a+1>0$)
and thus do not exist for the $S_{\eta}=0$ Mott-Hubbard insulator. They
are populated by $N-1$ (and $N+1$) fermions and by $M_{\eta}=2S_{\eta}=\vert N_a-N+1\vert$ unpaired 
$\eta$-spins of projection $+1/2$ (and $-1/2$) and {\it are not} populated by $\eta$-spins of projection $-1/2$ 
(and $+1/2$.) 

The off-diagonal generators of the $\eta$-spin $SU(2)$ 
symmetry algebra appearing in Eq. (\ref{twostates})
are given by \cite{Carmelo_17,Carmelo_18A,Essler_91},
\begin{equation}
{\hat{S}}^{+}_{\eta} = \sum_{j=1}^{N_a}(-1)^j\,c_{j,\downarrow}^{\dag}\,c_{j,\uparrow}^{\dag} 
\hspace{0.20cm}{\rm and}\hspace{0.20cm}{\hat{S}}^{-}_{\eta} = \sum_{j=1}^{N_a}(-1)^j\,c_{j,\uparrow}\,c_{j,\downarrow} \, ,
\label{LadderOp}
\end{equation}
or equivalently in terms of momentum $k$,
\begin{equation}
{\hat{S}}^{+}_{\eta} = \sum_{k}\,c_{\pi -k,\downarrow}^{\dag}\,c_{k,\uparrow}^{\dag} 
\hspace{0.20cm}{\rm and}\hspace{0.20cm}{\hat{S}}^{-}_{\eta} = \sum_{k}\,c_{k,\uparrow}\,c_{\pi-k,\downarrow} \, .
\label{LadderOpk}
\end{equation}
The effect of ${\hat{S}}^{\pm}_{\eta}$ in Eq. (\ref{twostates}) is to produce an $\eta$-spin flip that transforms 
one unpaired $\eta$-spin of projection $\pm 1/2$ into one unpaired $\eta$-spin of projection $\mp 1/2$.

We consider the following matrix elements associated with transitions from the $N$-fermion
ground states to the type III excited states $\vert\nu^+_{uhb},N+1\rangle$ and $\vert\nu^-_{uhb},N-1\rangle$,
\begin{eqnarray}
&& \langle \nu^+_{uhb},N+1\vert c_{k,\sigma}^{\dag}\vert GS^{+,0},N\rangle =
\nonumber \\
&& {1\over\sqrt{N_a-N+1}}\,\langle \nu^{+,0},N-1\vert\,{\hat{S}}^{-}_{\eta}\,c_{k,\sigma}^{\dag}\vert GS^{+,0},N\rangle =
\nonumber \\
&& {1\over\sqrt{N_a-N+1}}\,\langle \nu^{+,0},N-1\vert\,[{\hat{S}}^{-}_{\eta},c_{k,\sigma}^{\dag}]\vert GS^{+,0},N\rangle =
\nonumber \\
&& {- \gamma_{\sigma}\over\sqrt{N_a-N+1}}\,\langle \nu^{+,0},N-1\vert\,c_{\pi-k,{\bar{\sigma}}}\vert GS^{+,0},N\rangle \, ,
\label{matrixplus}
\end{eqnarray}
and
\begin{eqnarray}
&& \langle \nu^-_{uhb},N-1\vert c_{k,\sigma}\vert GS^{-,0},N\rangle =
\nonumber \\
&& {1\over\sqrt{N-N_a+1}}\,\langle \nu^{-,0},N+1\vert\,{\hat{S}}^{+}_{\eta}\,c_{k,\sigma}\vert GS^{-,0},N\rangle =
\nonumber \\
&& {1\over\sqrt{N-N_a+1}}\,\langle \nu^{-,0},N+1\vert\,[{\hat{S}}^{+}_{\eta},c_{k,\sigma}]\vert GS^{-,0},N\rangle =
\nonumber \\
&& {\gamma_{\sigma}\over\sqrt{N-N_a+1}}\,\langle \nu^{-,0},N+1\vert\,c_{\pi-k,{\bar{\sigma}}}^{\dag}\vert GS^{-,0},N\rangle \, ,
\label{matrixminus}
\end{eqnarray}
where the bra representation of the states, Eq. (\ref{twostates}), was used and the
coefficient $\gamma_{\sigma}$ is for $\sigma =\uparrow,\downarrow$ given by,
\begin{equation}
\gamma_{\uparrow} = + 1\hspace{0.20cm}{\rm and}\hspace{0.20cm}
\gamma_{\downarrow} = - 1 \, ,
\label{cssigma}
\end{equation}
respectively. From the use of the relations, Eqs. (\ref{matrixplus}) and (\ref{matrixminus}),
one finds for a doped Mott-Hubbard insulator with very small 
yet finite concentration $\delta = \vert(1-n_f)\vert$ for which $1/(N_a-N+1)\approx 1/N_a\,\delta$ and $1/(N-N_a+1)\approx 1/N_a\,\delta$ the following matrix elements square ratios,
\begin{eqnarray}
{\vert\langle \nu^+_{uhb},N+1\vert c_{k,\sigma}^{\dag}\vert GS^{+,0},N\rangle\vert^2\over 
\vert\langle \nu^{+,0},N-1\vert\,c_{\pi-k,{\bar{\sigma}}}\vert GS^{+,0},N\rangle\vert^2} & = & {1\over N_a\,\delta}
\nonumber \\
{\rm and} & &
\nonumber \\
{\vert\langle \nu^-_{uhb},N-1\vert c_{k,\sigma}\vert GS^{-,0},N\rangle\vert^2\over 
\vert\langle \nu^{-,0},N+1\vert\,c_{\pi-k,{\bar{\sigma}}}^{\dag}\vert GS^{-,0},N\rangle\vert^2} & = & {1\over N_a\,\delta} \, .
\label{matrixboth}
\end{eqnarray}
Such ratios then vanish in the thermodynamic limit.

On the other hand, the type II UHB and -UHB excited energy eigenstates 
 $\vert \nu^{+,0}_{uhb},N+1\rangle$ and $\vert \nu^{-,0}_{uhb},N-1\rangle$
of such a doped insulator are $\eta$-spin LWSs and HWSs, respectively. In Eq. (\ref{sumrulesplus}) 
both the sum rules of $B_{\bar{\sigma},-1} (k,\,\omega)$ (with $\sigma$ replaced by 
$\bar{\sigma}$) and $B^{\rm UHB}_{\sigma,+1} (k,\,\omega)$
give exactly the same number value $N_{\bar{\sigma}}$. In Eq. (\ref{sumrulesminus}) 
the sum rules of $B_{\bar{\sigma},+1} (k,\,\omega)$ (again with $\sigma$ replaced by
$\bar{\sigma}$) and $B^{\rm -UHB}_{\sigma,-1} (k,\,\omega)$
also give the same number value $N_{\bar{\sigma}}^h$. One then finds that,
\begin{eqnarray}
{\vert\langle \nu^{+,0},N-1\vert\,c_{\pi-k,{\bar{\sigma}}}\vert GS^{+,0},N\rangle\vert^2\over 
\vert\langle \nu^{+,0}_{uhb},N+1\vert c_{k,\sigma}^{\dag}\vert GS^{+,0},N\rangle\vert^2} 
& \approx & 1
\nonumber \\
{\rm and} & &
\nonumber \\
{\vert\langle \nu^{-,0},N+1\vert\,c_{\pi-k,{\bar{\sigma}}}^{\dag}\vert GS^{-,0},N\rangle\vert^2\over 
\vert\langle \nu^{-,0}_{uhb},N-1\vert c_{k,\sigma}\vert GS^{-,0},N\rangle\vert^2} 
& \approx & 1 \, .
\label{ratius}
\end{eqnarray}
That such $k$ dependent ratios are finite in the thermodynamic limit is what matters for our analysis.
In average they are given approximately by one, as given here.

Combining the relations, Eq. (\ref{ratius}), with those given in Eqs. (\ref{matrixplus}) 
and (\ref{matrixminus}), one then finds that,
\begin{eqnarray}
{\vert\langle \nu^+_{uhb},N+1\vert c_{k,\sigma}^{\dag}\vert GS^{+,0},N\rangle\vert^2\over 
\vert\langle \nu^{+,0}_{uhb},N+1\vert c_{k,\sigma}^{\dag}\vert GS^{+,0},N\rangle\vert^2} 
& \approx &{1\over N_a\,\delta}
\nonumber \\
{\rm and} & &
\nonumber \\
{\vert \langle \nu^-_{uhb},N-1\vert c_{k,\sigma}\vert GS^{-,0},N\rangle\vert^2\over 
\vert\langle \nu^{-,0}_{uhb},N-1\vert c_{k,\sigma}\vert GS^{-,0},N\rangle\vert^2} 
& \approx & {1\over N_a\,\delta} \, ,
\label{ratius2}
\end{eqnarray}
for very small yet finite concentration $\delta = \vert(1-n_f)\vert$.
That the ratios in this equation vanish in the thermodynamic limit, 
confirms that for the doped insulator the type III UHB and -UHB excited 
states do not contribute to the one-particle spectral-weight
distributions. 

Since no type-I excited states exist for the doped insulator,
the dominant processes that are associated with the UHB
creation of one fermion for $n_f <1$ and with the -UHB annihilation of one fermion for $n_f >1$ 
refer to type II excited states.

On the other hand, both type II and type III excited states are not allowed
for the Mott-Hubbard insulator whose ground states are not populated by $\eta$-spins $1/2$.
Its UHB and -UHB are rather generated by transitions to type I excited states that are populated by one 
unpaired $\eta$-spin with projection $-1/2$ and $+1/2$, respectively.
They are not created by $\eta$-spin flips, as in Eq. (\ref{twostates}) for the 
type III excited states. For the Mott-Hubbard insulator there exist only the UHB and -UHB 
that refer to the addition and removal spectral functions,
respectively, which are related as given in Eq. (\ref{BB}).

Only for excited states of ground states for which $\vert N_a -N\vert$ is finite 
and $\delta = \vert(1-n_f)\vert$ is of order $1/N_a$ and thus vanishes in the thermodynamic limit,
the UHB for $N<N_a$ and the -UHB for $N>N_a$ result from transitions to both
excited states of type II and type III.

The UHB one-particle processes reported here for a doped Mott-Hubbard insulator
with $n_f<1$ are the same as those of the metallic phase 
already studied in Ref. \onlinecite{Carmelo_17}. This is why
in this paper we limit our study to the up- and down-spin one-particle
spectral functions, Eq. (\ref{Bkomega}), of the Mott-Hubbard insulator. 

An important qualitative difference of the dynamical theory version suitable to the Mott-Hubbard insulator
used in the studies of this paper relative to that of the metallic phase used in the studies
of Ref. \onlinecite{Carmelo_17} is thus that the excited energy eigenstates populated by $\eta $ particles do not 
contribute to the line shape near the $(k,\omega)$-plane cusp singularities under study.
As reported in the following, also the dynamical theory's spectral parameters and
$c$ and $s$ particle's phase shifts have specific values and $u$ and $m$ dependencies
for the Mott-Hubbard insulator, different from those of its version suitable to the
metallic phase.

Due to the symmetry, Eq. (\ref{BB}), our study refers explicitly to one-particle removal. As given
in Table \ref{table1}, for both cases of up- and down-spin one-particle removal, 
one $c$ particle is removed and one unpaired $\eta$-spin $1/2$ of projection $+1/2$ is created. Within
removal of one down-spin fermion for $m>0$, the unbound spin-singlet pair of 
one $s$ particle is broken along with its annihilation and one
unpaired spin $1/2$ of projection $+1/2$ is created. The spin $1/2$ of projection $-1/2$
that also originates from the $s$ particle annihilation recombines with the annihilated $c$ particle
within the removed down-spin fermion.
For removal of one up-spin fermion for $m>0$, the number of $s$ particles remains unchanged and one
unpaired spin $1/2$ of projection $+1/2$ is removed. It
recombines with the annihilated $c$ particle within the removed up-spin fermion. 

As reported in the caption of Table \ref{table1}, some of the numbers 
provided in it are different
for excited states of $m=0$ ground states. Then the unbound spin-singlet pair of 
one $s$ particle is broken along with its annihilation and one
unpaired spin $1/2$ of projection $+1/2$ and $-1/2$ is created under
removal of one down-spin fermion and one up-spin fermion, respectively.
The spin $1/2$ of projection $-1/2$ and $+1/2$ left over
that also originates from the $s$ particle annihilation recombines with the annihilated $c$ particle
within the removed down-spin fermion and up-spin fermion, respectively.

\section{The spectra and the spectral functions near their singularities}
\label{SECIV}

For simplicity, we do not provide here the details of the present dynamical theory 
that are common to those already given in Ref. \onlinecite{Carmelo_17} for the metallic case.
The theory applies to several dynamical correlation functions.
In the case of the present quantum problem, it
provides the line shape of the up- and down-spin one-particle spectral functions, Eq. (\ref{Bkomega}),
at and in the vicinity of two types of singular spectral features called branch lines 
and boundary lines, respectively. Those are the most important spectral features
of such functions. 
\begin{figure}
\begin{center}
\centerline{\includegraphics[width=8.75cm]{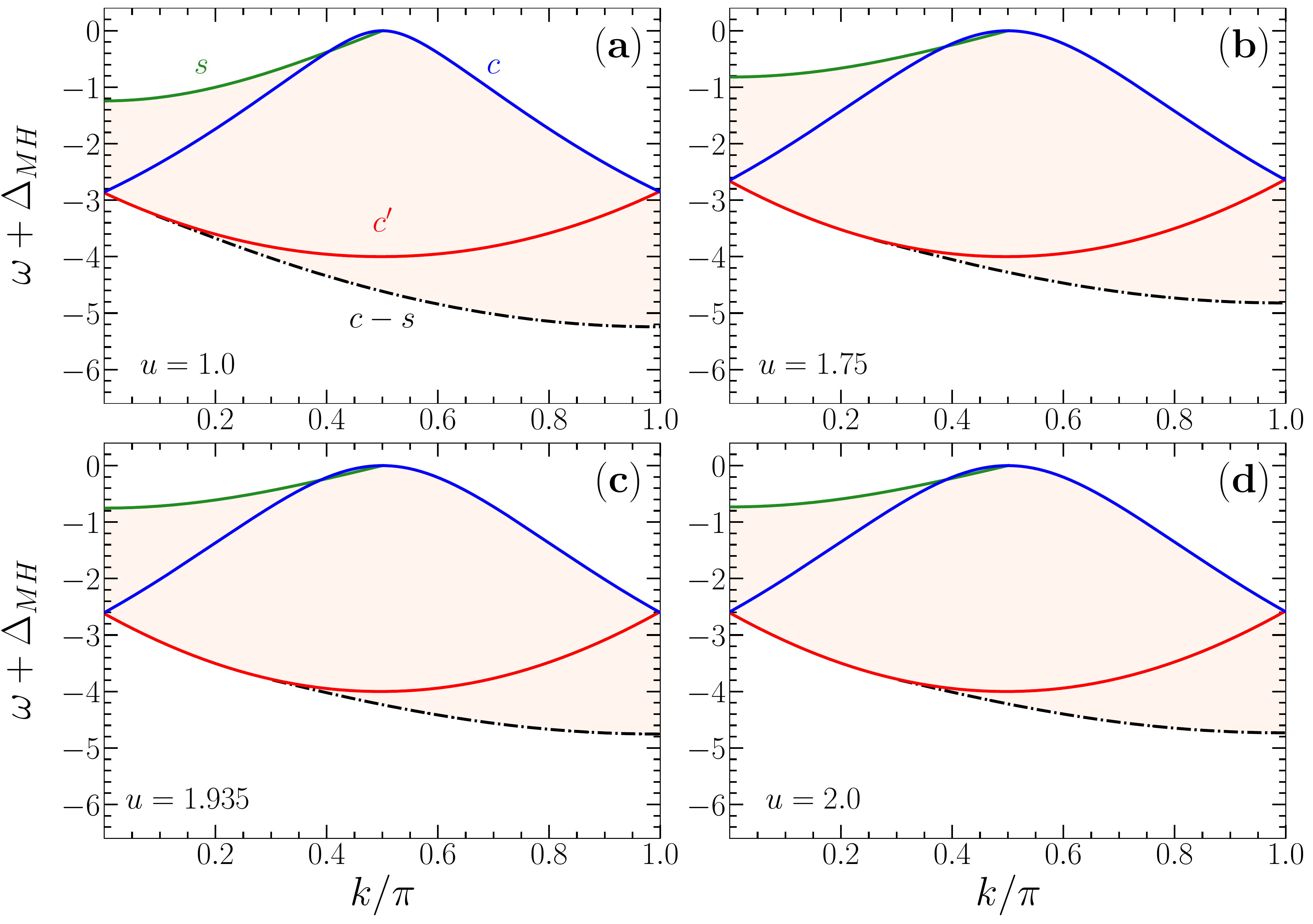}}
\caption{The $(k,\omega)$-plane regions defined by the spectrum
$\epsilon (k)$, Eq. (\ref{SpElremom0}), where for spin density $m=0$ and 
(a) $u=1.0$, (b) $u=1.75$, (c) $u=1.935$, and (d) $u=2.0$ there is in 
the thermodynamic limit more spectral weight in the removal one-particle spectral function
of the 1D Hubbard model with one fermion per site. The $c$, $c'$, and $s$ branch lines 
whose spectra are given in Eq. (\ref{OkudRLAccxm0})
are represented by solid lines and the $c-s$ boundary line by a dashed-dotted line.
The $(k,\omega)$-plane distributions presented here and in Figs. \ref{figure3} and \ref{figure6}-\ref{figure12} 
do not provide information on the relative amount of spectral weight contained within each spectrum's 
colored continuum, most weight being actually located in the vicinity of the branch lines and boundary line.
For the above intermediate $u$ values, the present one-particle spectra were studied by other methods
and other authors. They are to be compared with those plotted 
in (a) Fig. 3(a) of Ref. \onlinecite{Nocera_18}
for $U/t = 4.0$ (and $u=U/4t=1.0$) calculated with tDMRG, (b)
Fig. 7(b) of Ref. \onlinecite{Lante_09} for $U/t = 7.0$ (and $u=1.75$) derived 
combining Bethe-ansatz results, Lanczos diagonalizations, and field theoretical approaches,
(c) Fig. 10 of Ref. \onlinecite{Benthien_07}
for $U/t = 7.74$ (and $u=1.935$) calculated with DDMRG, and
(d) Fig. 1(a) of Ref. \onlinecite{Nocera_18}
for $U/t = 8.0$ (and $u=2.0$) obtained with tDMRG.}
\label{figure2}
\end{center}
\end{figure}

\subsection{The spectra of the excitations behind most one-particle spectral weight}
\label{SECIVA}

For an initial insulator ground state, any transition under which
the deviation $\delta N_c$ is finite refers to a gapped excitation spectrum, consistently with
the term $- \Delta_{MH}$ in the $c$-band energy dispersion in Eq. (\ref{equA4}) of Appendix \ref{B}.
This applies to the removal up- and down-spin one-particle spectra for which $\delta N_c = -1$,
as given in Table \ref{table1} for one-particle removal (REM) excited energy eigenstates. 
Those are the states that span the present subspaces for which
$\delta M_{\eta,+1/2}=\delta N_c^h = 1$ and $\delta M_{s,+1/2}=\delta N_s^h = 1$ and $=-1$ 
for $\delta N_{\downarrow} = -1$ and $\delta N_{\uparrow} = -1$, respectively. 
Due to the perturbative character
of the present quantum problem in terms of the number of elementary fractionalized-particle 
processes \cite{Carmelo_17}, transitions 
to one-particle removal excited energy eigenstates associated with $\delta N_c <0$ deviations 
such that $\vert \delta N_c\vert >1$ lead to very little spectral weight and do not contribute
to the line-shape at and near the one-particle spectral functions's singularities studied in
this paper.
\begin{figure}
\begin{center}
\centerline{\includegraphics[width=8.75cm]{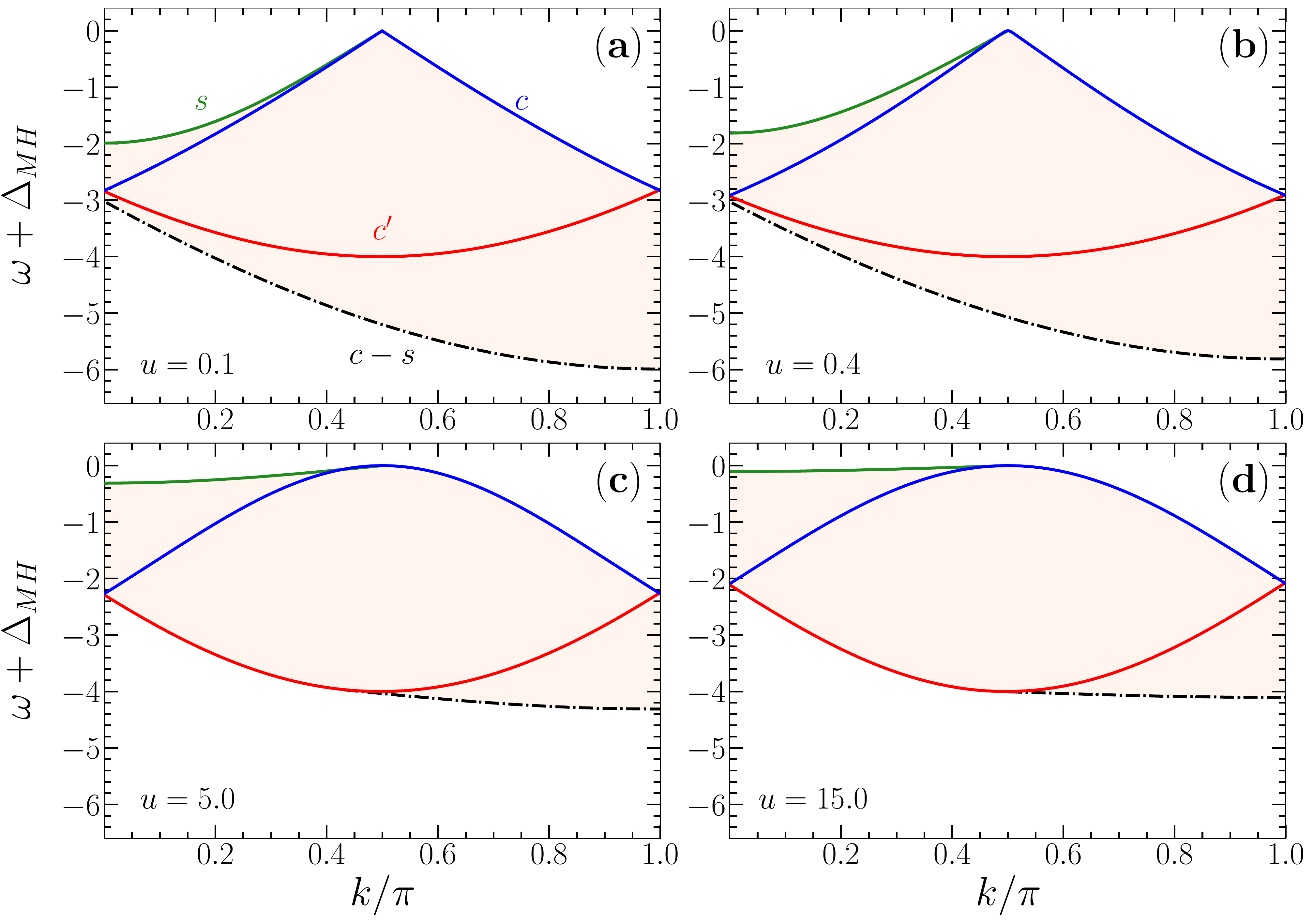}}
\caption{The $(k,\omega)$-plane regions defined by the spectrum
$\epsilon_{\downarrow} (k)$, Eq. (\ref{SpElremom0}), where for spin density $m=0$ and 
(a) $u=0.1$, (b) $u=0.4$, (c) $u=5.0$, and (d) $u=15.0$ there is in 
the thermodynamic limit more spectral weight in the removal one-particle spectral function
of the 1D Hubbard model with one fermion per site. Most spectral weight is located in the vicinity 
of the $c$, $c'$, and $s$ branch lines, Eq. (\ref{OkudRLAccxm0}), and of the $c-s$ boundary line
represented here by solid lines and a dashed-dotted line, respectively.}
\label{figure3}
\end{center}
\end{figure}

Within a $k$ extended zone scheme, the $-\epsilon_{\downarrow} (k) <0$ and $-\epsilon_{\uparrow} (k)<0$
spectra of such excited states that generate most of the down- and up-spin
one-particle removal spectral weight, respectively, has the general form,
\begin{eqnarray}
&& \epsilon_{\downarrow} (k) = -\varepsilon_c (q) - \varepsilon_{s} (q') 
\nonumber \\
& & {\rm where}\hspace{0.20cm}k = \iota\pi - q - q' \hspace{0.20cm}
{\rm and}\hspace{0.20cm}\iota = \pm 1
\nonumber \\
&& {\rm for} \hspace{0.2cm}q \in [-\pi,\pi]\hspace{0.20cm}{\rm and}\hspace{0.20cm} q' \in [-k_{F\downarrow},k_{F\downarrow}] \, ,
\label{SpdownElremo}
\end{eqnarray}
and
\begin{eqnarray}
&& \epsilon_{\uparrow} (k) = -\varepsilon_c (q) + \varepsilon_{s} (q')
\nonumber \\
& & {\rm where}\hspace{0.20cm}k = - q + q' 
\nonumber \\
&& {\rm for} \hspace{0.2cm}q \in [-\pi,\pi]\hspace{0.20cm}{\rm and}\hspace{0.20cm}
\vert q'\vert \in [k_{F\downarrow},k_{F\uparrow}] \, ,
\label{SpupElremo}
\end{eqnarray}
respectively. The energy dispersions $\varepsilon_c (q) $ and
$\varepsilon_{s} (q')$ appearing here are defined in Eq. (\ref{equA4}) of Appendix \ref{B}.
The number of $s$ band holes is in the present subspace given by $N_s^h = N_c - 2N_s$.
This leads to $\delta N_s^h = -1$ for up-spin one-particle removal for which 
$\delta N_c = -1$ and $\delta N_s = 0$, as given in Table \ref{table1}
($\delta N_{\uparrow} = -1$ REM deviations.) The annihilation of one $s$
band hole is behind the $q'$ plus sign in the excitation
momentum $k = - q + q'$,  Eq. (\ref{SpupElremo}).

The $(k,\omega)$-plane continuum associated with the two-parametric spectrum 
in Eq. (\ref{SpdownElremo}) that has two overlapping $\iota=\pm 1$ branches
is shown for several $u$ values in Figs. \ref{figure2}-\ref{figure3} for $m=0$
and in Figs. \ref{figure7}-(\ref{figure9} for a set of $m$ values. That 
associated with the two-parametric spectrum in Eq. (\ref{SpupElremo}) is shown
in Figs. \ref{figure10}-\ref{figure12}, respectively, for a set of spin density $m$ and $u$ values.
The zero energy of all such spectra is that of the deviation $\omega + \Delta_{MH}$
where $\omega \leq - \Delta_{MH}$. The energy level $\omega =0$ refers to
the middle of the Mott-Hubbard gap $2\Delta_{MH} $, Eqs. (\ref{2DeltaMHallm})-(\ref{2mu0LB}).
The $u$ and $m$ dependencies of that gap amplitude are illustrated in Figs. \ref{figure1} (a) and (b).

The $(k,\omega)$-plane distributions shown in Figs. \ref{figure2}-\ref{figure3}
and (\ref{figure7})-(\ref{figure12}) do not provide information on the relative 
amount of spectral weight contained within each spectrum's colored continuum. Most
one-particle spectral weight is located in the vicinity of the branch lines and boundary line
shown in such figures. 
\begin{figure}
\begin{center}
\centerline{\includegraphics[width=8.75cm]{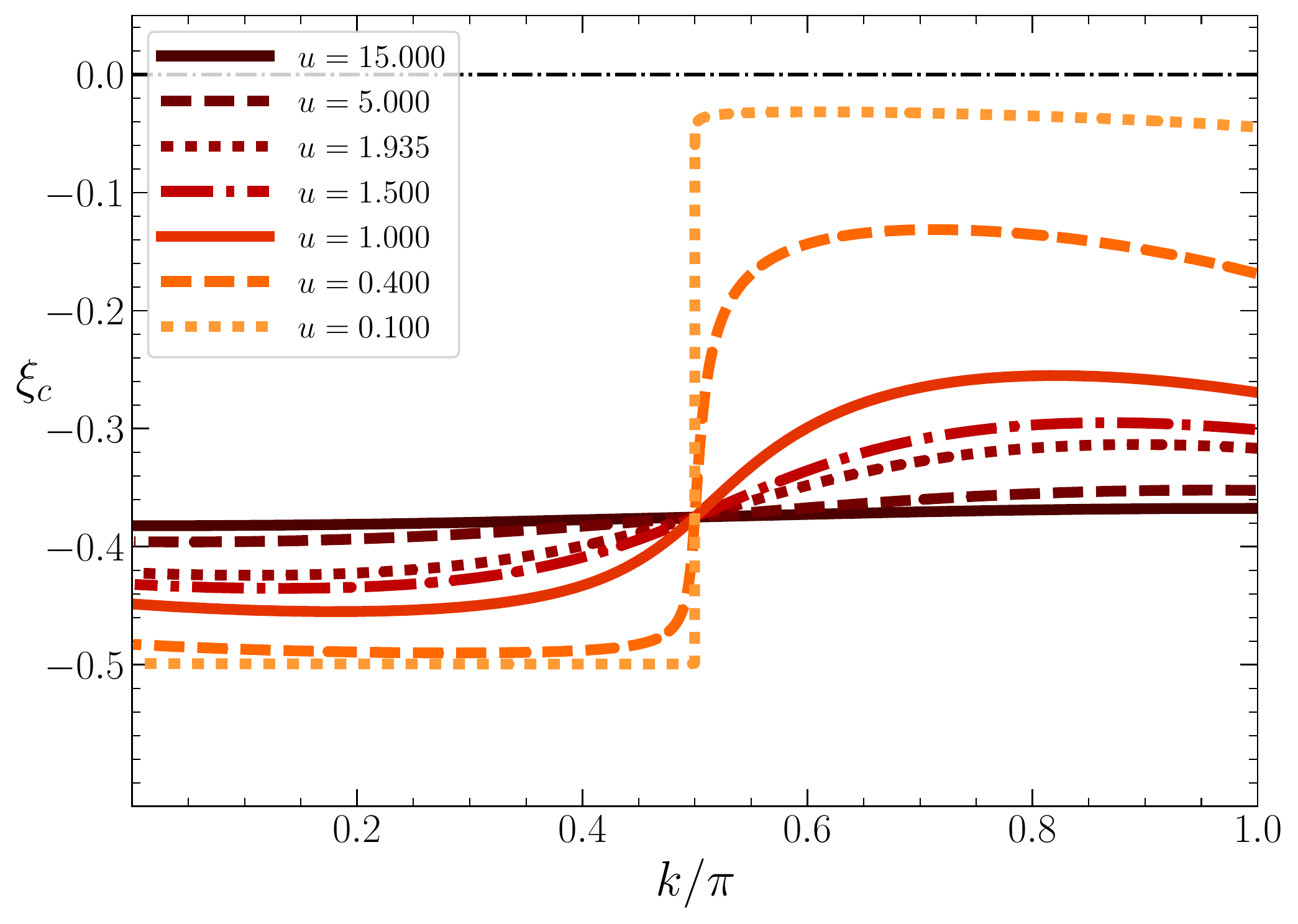}}
\caption{The negative $c$-branch line exponent, Eq. (\ref{xiccpskm0}), as a function
of the excitation momentum $k$ for spin density $m=0$
and a set of $u$ values.}
\label{figure4}
\end{center}
\end{figure}
\begin{figure}
\begin{center}
\centerline{\includegraphics[width=8.75cm]{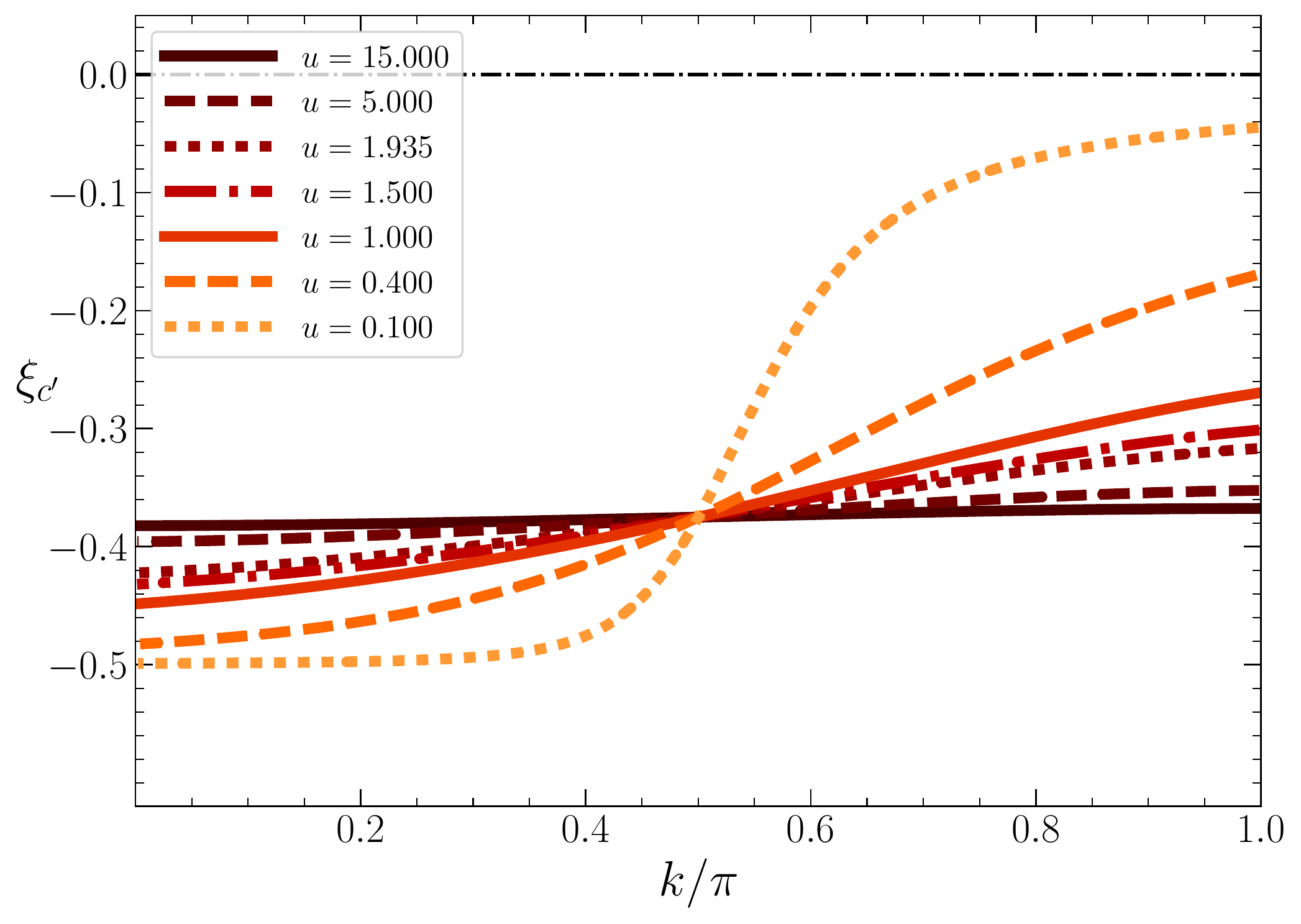}}
\caption{The same as in Fig. \ref{figure4} for the $c'$-branch line exponent, Eq. (\ref{xiccpskm0}).}
\label{figure5}
\end{center}
\end{figure}

\subsection{The spectral functions near their singularities}
\label{SECIVB}

The following number and current number deviations under transitions to excited states 
play an important role in the dynamical theory used in our studies,
\begin{eqnarray}
&& \delta N_{c,\iota}^F \hspace{0.20cm}{\rm and}\hspace{0.20cm}\delta N_{s,\iota}^F
\hspace{0.20cm}{\rm for}\hspace{0.20cm}\iota =1,-1\hspace{0.20cm}{\rm (right,left)}\hspace{0.20cm}{\rm particles}
\nonumber \\
&& \delta N_{c}^F = \sum_{\iota = \pm 1}\delta N_{c,\iota}^F
\hspace{0.20cm}{\rm and}\hspace{0.20cm}\delta N_{s}^F = \sum_{\iota = \pm 1}\delta N_{s,\iota}^F
\nonumber \\
&& \delta J_{c}^F = {1\over 2}\sum_{\iota = \pm 1}\iota\,\delta N_{c,\iota}^F 
\hspace{0.20cm}{\rm and}\hspace{0.20cm}
\delta J_{s}^F =  {1\over 2}\sum_{\iota = \pm 1}\iota\,\delta N_{s,\iota}^F 
\nonumber \\
&& \delta N_{\beta}=\delta N_{\beta}^{F}+\delta N_{\beta}^{NF}
 \hspace{0.20cm}{\rm for}\hspace{0.20cm}\beta = c,s \, .
\label{NcFNcFJcFJsF}
\end{eqnarray}
The $\beta =c,s$ deviations $\delta N_{\beta,\iota}^F$ refer to changes in the 
number of $\beta$ particles at the $\iota=+1$ right and $\iota=-1$ left Fermi points
$q_{F\beta}^{\iota}$, respectively. Except for $1/L$ corrections,
$q_{F\beta}^{\iota}\approx \iota{\pi\over L}N_{\beta}$, so that
$q_{Fc}^{\iota}\approx \iota\pi$ and $q_{Fc}^{\iota}\approx \iota k_{F\downarrow}$.
Rather than $\delta N_{\beta,\iota}^F$, one uses in general
$\beta =c,s$ number deviations $\delta N_{\beta}^F$ and number current
deviations $\delta J_{\beta}^F$, Eq. (\ref{NcFNcFJcFJsF}), which contain exactly the same information.

Removal of $\beta =c,s$ particles at $\beta$-band momenta $\vert q\vert < q_{F\beta}$
away from the Fermi points $q_{F\beta}^{\pm 1}$ and addition of $s$ particles at $s$-band 
momenta $k_{F\downarrow}<\vert q'\vert<k_{F\uparrow}$ are associated with number
deviations denoted by $\delta N_{\beta}^{NF}$, so that $\delta N_{\beta}=\delta N_{\beta}^{F}+\delta N_{\beta}^{NF}$,
Eq. (\ref{NcFNcFJcFJsF}).

The index $\bar{\beta}=c,c',s$ labels the branch lines that run within the $(k,\omega)$-plane continua associated 
with the spectra, Eqs. (\ref{SpdownElremo}) and (\ref{SpupElremo}).
An index $\beta_c=c,c'$ is used only for the $c$ and $c'$ branch lines.
The branch-line processes lead to a one-parametric $(k,\omega)$-plane $\bar{\beta} =c,c',s$ branch line 
spectrum of the general form,
\begin{eqnarray}
\epsilon_{\sigma,\bar{\beta}} (k) & = & \delta_{\bar{\beta},s}\Delta_{MH} + c_{\bar{\beta}}\,\varepsilon_{\bar{\beta}} (q)\ \geq 0
\hspace{0.20cm}{\rm where}
\nonumber \\
k & = & k_0 + c_{\bar{\beta}}\,q\hspace{0.20cm}{\rm for}\hspace{0.20cm}\bar{\beta} = c,c',s \, . 
\label{dE-dP-bl}
\end{eqnarray}
The convention is that when as here $\bar{\beta}$ labels an energy dispersion $\varepsilon_{\bar{\beta}} (q)$, 
Eq. (\ref{equA4}) of Appendix \ref{B}, it reads $\bar{\beta}=c$ both for the $c$ and $c'$ branch lines.
In Eq. (\ref{dE-dP-bl}), $\sigma =\uparrow,\downarrow$ refers to the one-particle spectral function under
consideration and the constants $c_{\bar{\beta}}$ are given by,
\begin{eqnarray}
c_{c} & = & c_{c'} = -1
\hspace{0.20cm}{\rm for}\hspace{0.20cm}q \in]-\pi,\pi[
\nonumber \\
& & {\rm and}\hspace{0.20cm}\sigma=\uparrow,\downarrow\hspace{0.20cm}{\rm fermion}\hspace{0.20cm}
c,c'\hspace{0.20cm}{\rm branch}\hspace{0.20cm}{\rm lines}
\nonumber \\
c_{s} & = & -1
\hspace{0.20cm}{\rm for}\hspace{0.20cm}
q = q' \in]-k_{F\downarrow},k_{F\downarrow}[
\nonumber \\
& & {\rm and}\hspace{0.20cm}{\rm the}\hspace{0.20cm}\downarrow\hspace{0.20cm}{\rm fermion}\hspace{0.20cm}
s\hspace{0.20cm}{\rm branch}\hspace{0.20cm}{\rm line}
\nonumber \\
& = & 1
\hspace{0.20cm}{\rm for}\hspace{0.20cm}
\vert q\vert = \vert q'\vert \in ]k_{F\downarrow},k_{F\uparrow}]
\nonumber \\
& & {\rm and}\hspace{0.20cm}{\rm the}\hspace{0.20cm}\uparrow\hspace{0.20cm}{\rm fermion}\hspace{0.20cm}
s\hspace{0.20cm}{\rm branch}\hspace{0.20cm}{\rm line} \, .
\label{AcAs}
\end{eqnarray}
The momentum $k_0$ in Eq. (\ref{dE-dP-bl}) is determined by the current number deviations as follows,
\begin{equation}
k_0 = 2\pi\,\delta J_{c}^F + 2k_{F\downarrow}\,\delta J_{s}^F  \, .
\label{1el-omega0}
\end{equation}
\begin{figure}
\begin{center}
\centerline{\includegraphics[width=8.75cm]{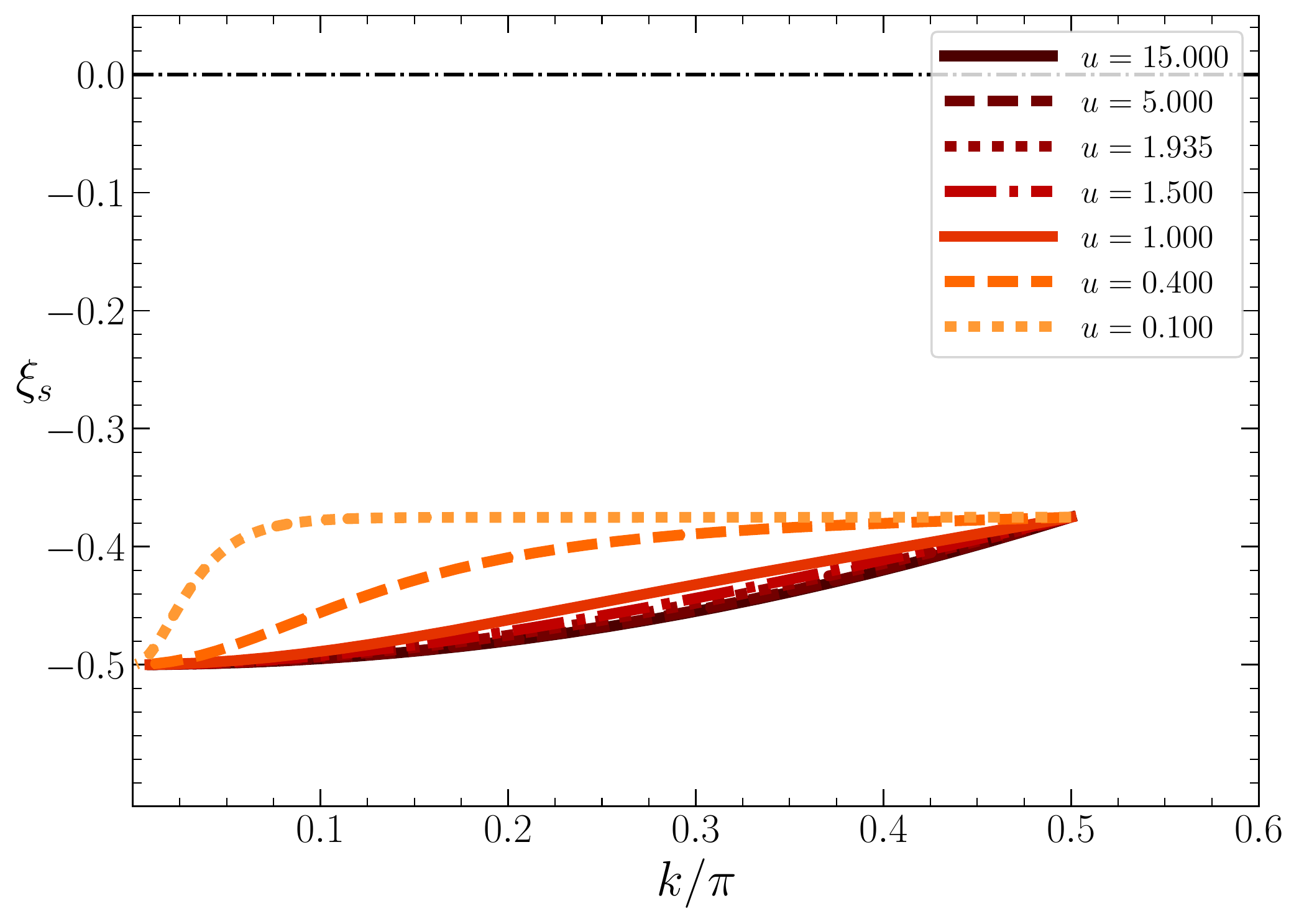}}
\caption{The same as in Fig. \ref{figure4} for the $s$-branch line exponent, Eq. (\ref{xiccpskm0}).}
\label{figure6}
\end{center}
\end{figure}

Within the dynamical theory used in our studies, the up- and down-spin spectral 
functions have for $\omega<0$ and small values of the deviation $(\omega + \epsilon_{\sigma,\bar{\beta}} (k))\leq 0$
at and near a $\bar{\beta} = c,c',s$ branch line the following general behavior,
\begin{eqnarray}
B_{\sigma,-1} (k,\omega) & = & C_{\sigma,\bar{\beta}}\,
\Bigl(\omega + \epsilon_{\sigma,\bar{\beta}} (k)\Bigr)^{\xi_{\bar{\beta}}^{\sigma} (k)}
\nonumber \\
{\rm for} && {\rm small}\hspace{0.20cm}(\omega + \epsilon_{\sigma,\bar{\beta}} (k)) \leq 0 \, .
\label{branch-l}
\end{eqnarray}
Here the constant $C_{\sigma,\bar{\beta}}$ is a $m$ and $u$ dependent function that has a fixed value 
for the $k$ and $\omega$ ranges corresponding to small values of the energy deviation 
$(\omega + \epsilon_{\sigma,\bar{\beta}} (k))\leq 0$ for which this expression is valid. 
The momentum dependent exponents in it have the general form,
\begin{equation}
\xi_{\bar{\beta}}^{\sigma} (k) = -1 + \sum_{\iota =\pm1}\sum_{\beta=c,s} \Phi_{\bar{\beta},\beta,\iota}^2 (q) \, ,
\label{expbsk}
\end{equation}
where the spectral functionals $\Phi_{\bar{\beta},\beta,\iota} (q)$ are defined below.
The spectral function expression, Eq. (\ref{branch-l}), is exact when there is no spectral weight just 
above the corresponding $\bar{\beta}$ branch line. In the present case of one-particle spectral functions,
either there is no weight above the branch lines or the amount of that weight 
is small. In the latter case, the very weak coupling to it leads to a higher order contribution to the line shape 
expression given in Eq. (\ref{branch-l}) that can be neglected in the present thermodynamic limit. 
Indeed, such a contribution only slightly changes the value of the momentum-dependent exponent,
preserving its negativity or positivity.

The spectral functionals $\Phi_{\bar{\beta},\beta,\iota} (q)$ in Eq. (\ref{expbsk})
depend on the $\bar{\beta} = c,c',s$ branch lines's
excitation momentum spectrum $k = k_0 \pm q$, Eq. (\ref{dE-dP-bl}). 
(The index $\bar{\beta}$ that labels such functionals also reads
$\bar{\beta}=c$ both for the $c$ and $c'$ branch lines.) In the present case of the Mott-Hubbard insulator,
the phase-shift related parameters $\xi_{\beta,\beta'}$ 
in the general expression of such spectral functionals\cite{Carmelo_17} have the specific form
given in Eq. (\ref{ZZ-gen}) of Appendix \ref{B}. Their use leads to the following expressions,
\begin{eqnarray}
\Phi_{\bar{\beta},c,\iota} (q) & = & {\iota\,\delta N^F_{c}\over 2} + \delta J^F_{c} + \xi_{s\,c}\,\delta J^F_{s}
\nonumber \\
& & + c_{\bar{\beta}}\,\Phi_{c,\bar{\beta}}(\iota\pi,q) \, ,
\label{cfunDc}
\end{eqnarray}
and
\begin{eqnarray}
\Phi_{\bar{\beta},s,\iota} (q) & = & -\iota {\xi_{c\,s}\over 2\xi_{s\,s}}\,\delta N^F_{c} 
+ {\iota\,\delta N^F_{s}\over 2\xi_{s\,s}} + \xi_{s\,s}\,\delta J^F_{s}
\nonumber \\
& & + c_{\bar{\beta}}\,\Phi_{s,\bar{\beta}}(\iota k_{F\downarrow},q) \, .
\label{cfunDs}
\end{eqnarray}
Here the $\bar{\beta}$-band momentum $q$ is outside the $\bar{\beta} =c,s$ Fermi points and
$c_{\bar{\beta}}$ are the constants, Eq. (\ref{AcAs}).
The $\beta=c,s$ particle phase shifts $\Phi_{\beta,\bar{\beta}}(q,q')$ in units of $2\pi$ are defined by
Eqs. (\ref{Phi-barPhi})-(\ref{Phis-all-qq}) of Appendix \ref{B}.
Physically, $\mp 2\pi\Phi_{\beta,\bar{\beta}}(q,q')$ is the phase shift acquired by a $\beta$ particle
of momentum $q$ upon creation of one $\bar{\beta}$ hole 
($-2\pi\Phi_{\beta,\bar{\beta}}$) and one $\bar{\beta}$ particle ($+2\pi\Phi_{\beta,\bar{\beta}}$) at a momentum $q'$.

The down- and up-spin one-particle removal excited states may generate
an overall $c$-band momentum shift and/or a $s$-band momentum shift.
Their possible values are $0,\pm \pi/L$. Such shifts are behind the possibility of the 
$\beta =c,s$ deviations $\delta N_{\beta,\iota}^F$ in 
Eq. (\ref{NcFNcFJcFJsF}) being half-odd integer numbers
and are implicitly accounted for by the functionals, Eqs. (\ref{cfunDc}) and (\ref{cfunDs}).

A coefficient $b_{\beta_c}$ is used for $\beta_c =c,c'$ in some of the expressions given in the following
that refer to the $c$ and $c'$ branch lines, respectively. It reads,
\begin{equation}
b_{c} = 1 \hspace{0.20cm}{\rm and}\hspace{0.20cm}b_{c'} = - 1 \, .
\label{acc}
\end{equation}

There is a second type of $(k,\omega)$-plane feature near which the present
dynamical theory provides an analytical expression of 
the one-particle spectral functions. It is generated by processes where one $c$ particle 
is removed at a $c$-band momentum value $q$ and one $s$ particle is created or removed at 
a $s$-band momentum value $q'$. 

Those are one-parametric features called $c-s$ boundary lines. Indeed, they are part of the limiting boundary 
lines of the $(k,\omega)$-plane continua associated with the two-parametric spectra, Eqs. (\ref{SpdownElremo}) 
and (\ref{SpupElremo}). A $c-s$ boundary line $(k,\omega)$-plane spectrum has the following 
general form,
\begin{eqnarray}
\epsilon_{\sigma,c-s} (k) & = & \left(-\varepsilon_{c}(q) + c_s\,\varepsilon_{s}(q')\right)\,\delta_{v_{c}(q) ,\,v_{s}(q')}
\hspace{0.20cm}{\rm where}
\nonumber \\
k & = & k_0 - q + c_s\,q' \, .
\label{dE-dP-c-s1}
\end{eqnarray}
Here $c_s=-1$ or $c_s=1$, Eq. (\ref{AcAs}), and several 
$q$ and $q'$ limiting values that obey the equality $v_{c}(q)=v_{s}(q')$
are defined by the relations, Eq. (\ref{cfirstsector}).

Near such a $c-s$ boundary line, the up- and down-spin one-particle spectral functions 
behave as \cite{Carmelo_08},
\begin{eqnarray}
B_{\sigma,-1} (k,\omega) & \propto & \Bigl(\omega +\epsilon_{\sigma,c-s} (k)\Bigr)^{-1/2} 
\nonumber \\
{\rm for} && {\rm small}\hspace{0.20cm}
(\omega +\epsilon_{\sigma,c-s} (k)) \leq 0 \, .
\label{B-bol}
\end{eqnarray}

The branch- and boundary-line spectra studied in this paper are defined 
in the momentum interval $k\in [0,\pi]$.

\section{The $m=0$ zero-magnetization spectra and exponents}
\label{SECV}

For vanishing spin density and magnetic field of interest for ARPES
there are previous studies on the present Mott-Hubbard insulator's spectral function
\cite{Nocera_18,Lante_09,Benthien_07}. At $h=0$ and $m=0$ the spectral function $B_{-1} (k,\omega)$ equals the spectral functions 
$B_{\downarrow,-1} (k,\omega)$ and $B_{\uparrow,-1} (k,\omega)$ obtained in the $m\rightarrow 0$ limit 
from $m>0$ and $m<0$ values, respectively. 
In the $m\rightarrow 0$ limit from $m>0$ values, the up-spin one-particle 
removal spectrum $\epsilon_{\uparrow} (k)$, Eq. (\ref{SpupElremo}), becomes one-parametric and coincides 
with the $c$ and $c'$ branch lines of the down-spin one-particle removal spectrum $\epsilon_{\downarrow} (k)=\epsilon (k)$,
Eq. (\ref{SpdownElremo}). In that limit from $m<0$ values, the situation is
the opposite, the down-spin one-particle removal spectrum becoming one-parametric and the 
up-spin one-particle removal spectrum giving $\epsilon (k)$.
\begin{figure}
\begin{center}
\centerline{\includegraphics[width=8.75cm]{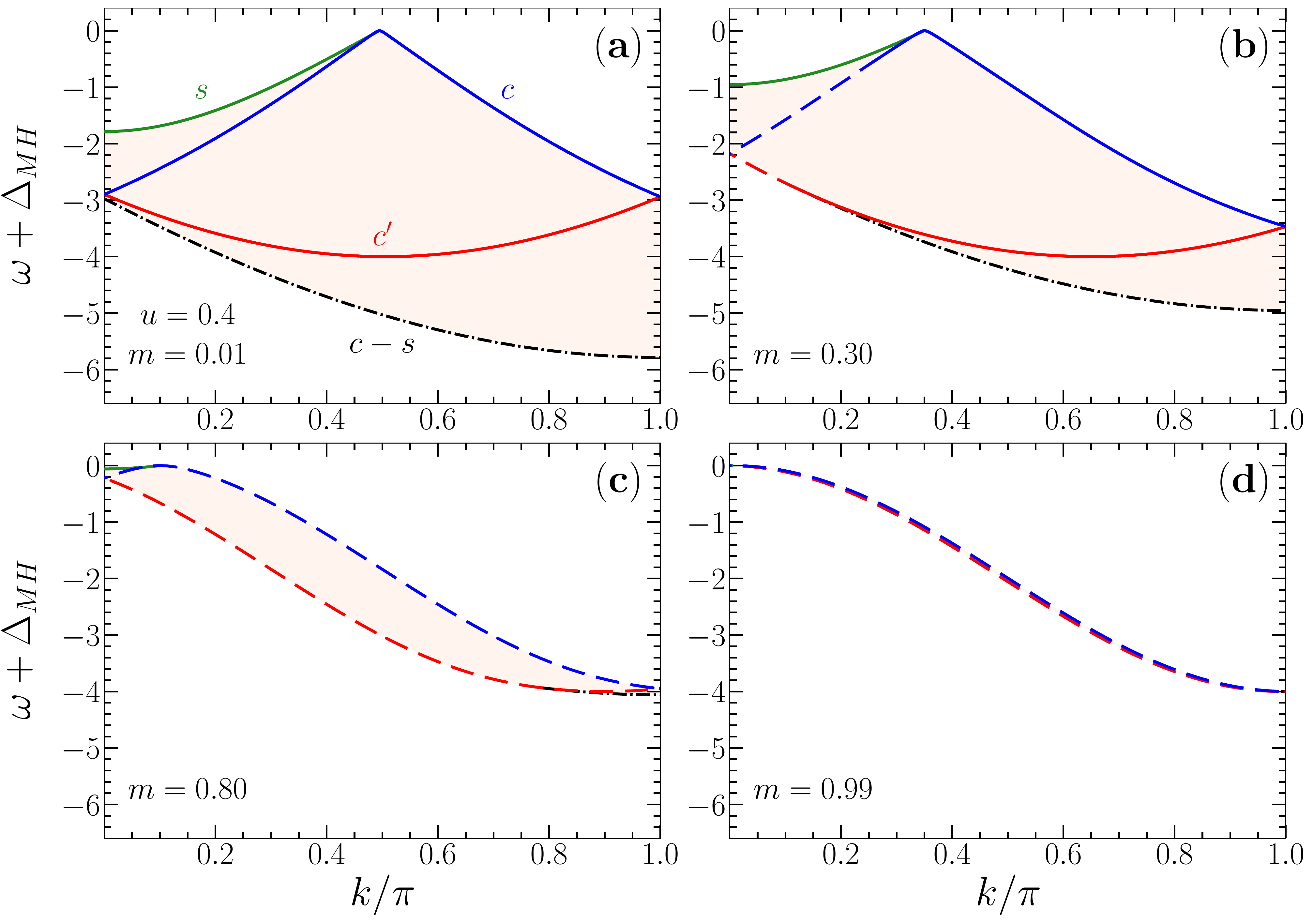}}
\caption{The $(k,\omega)$-plane regions 
corresponding to the spectrum $\epsilon_{\downarrow} (k)$, Eq. (\ref{SpdownElremo}),
where for spin densities
(a) $m=0.01$, (b) $m=0.30$, (c) $m=0.80$, and (d) $m=0.99$ and
$u=0.4$ there is in the thermodynamic limit more spectral weight in the 
removal down-spin one-particle spectral function. Most such a weight is located in the vicinity 
of the $c$, $c'$, and $s$ branch lines and of the $c-s$ boundary line.
Here and in Figs. \ref{figure8}-\ref{figure12}
the branch lines $k$ intervals for which the corresponding exponents 
are negative and positive are represented by solid and dashed lines,
respectively, and the boundary line by a dashed-dotted line.}
\label{figure7}
\end{center}
\end{figure}

Within a $k$ extended zone scheme, the $-\epsilon (k) <0$ 
spectrum of the excitations that contain most of the one-particle removal spectral weight 
is given by,
\begin{eqnarray}
&& \epsilon (k) = -\varepsilon_c (q) - \varepsilon_{s} (q') 
\nonumber \\
& & {\rm where}\hspace{0.20cm}k = \iota\pi - q - q' \hspace{0.20cm}
{\rm and}\hspace{0.20cm}\iota = \pm 1
\nonumber \\
&& {\rm for} \hspace{0.2cm}q \in [-\pi,\pi]\hspace{0.20cm}{\rm and}\hspace{0.20cm} q' \in [-\pi/2,\pi/2] \, .
\label{SpElremom0}
\end{eqnarray}
The $c$- and $s$-band energy dispersions $\varepsilon_c (q) $ and
$\varepsilon_{s} (q')$ appearing here have at $m=0$ simplified expressions,
defined by Eq. (\ref{varesm0}) of Appendix \ref{B}.
\begin{figure}
\begin{center}
\centerline{\includegraphics[width=8.75cm]{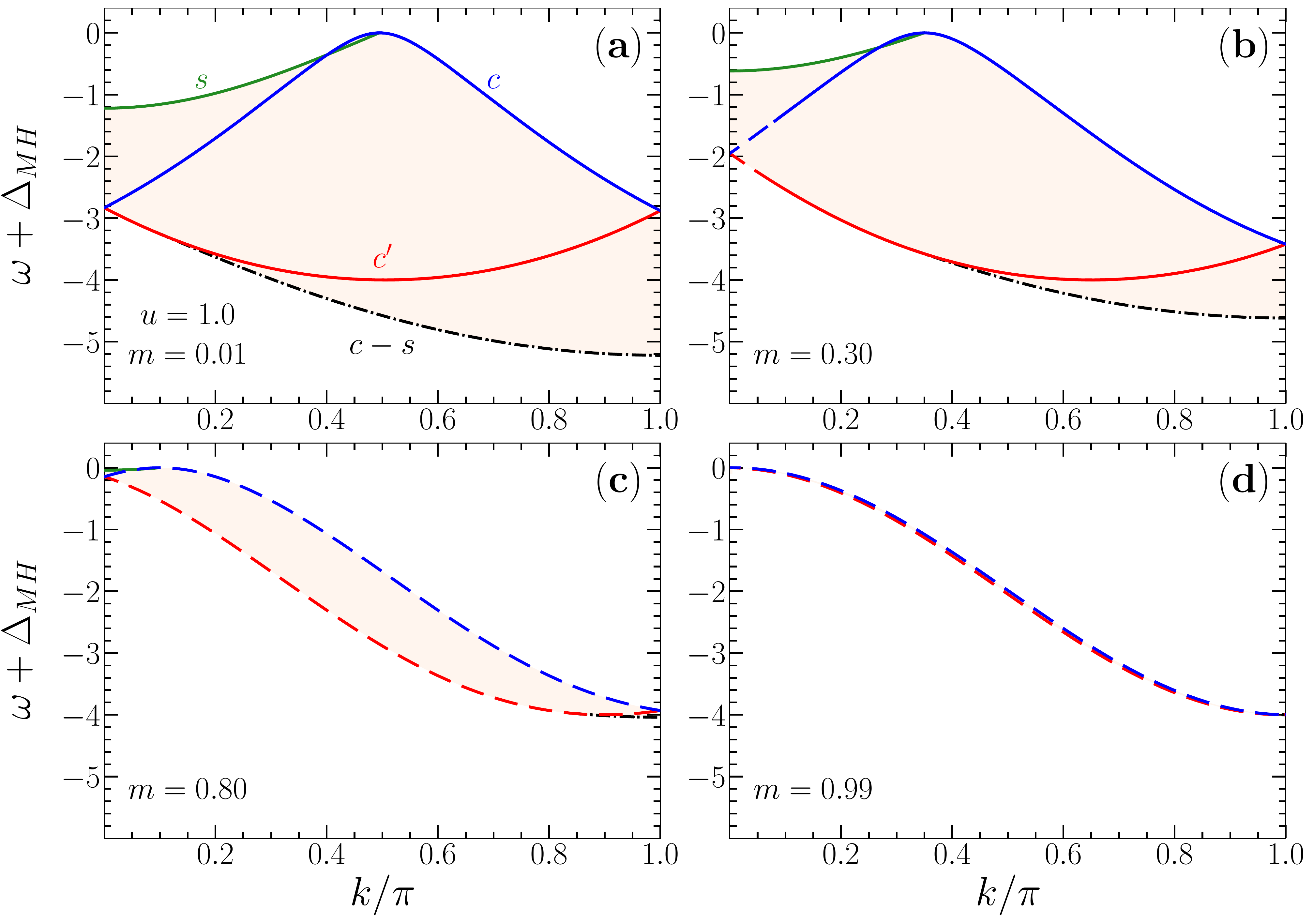}}
\caption{The same as Fig. \ref{figure7} for a larger $u$ value, $u=1.0$.}
\label{figure8}
\end{center}
\end{figure}

The $(k,\omega)$-plane continuum shown in Figs. \ref{figure2} and \ref{figure3} for several $u$ values 
in the range $u\in [0.1,15.0]$ refers to the two overlapping $\iota =\pm 1$ branches two-parametric spectrum, 
Eq. (\ref{SpElremom0}). The continuum shown in Fig. \ref{figure2} is to be compared with those plotted 
in (a) Fig. 3(a) of Ref. \onlinecite{Nocera_18} for $U/t = 4.0$ (and $u=U/4t=1.0$) calculated with tDMRG, (b)
Fig. 7(b) of Ref. \onlinecite{Lante_09} for $U/t = 7.0$ (and $u=1.75$) derived 
combining Bethe-ansatz results, Lanczos diagonalizations, and field theoretical approaches,
(c) Fig. 10 of Ref. \onlinecite{Benthien_07} for $U/t = 7.74$ (and $u=1.935$) calculated with DDMRG, and
(d) Fig. 1(a) of Ref. \onlinecite{Nocera_18} for $U/t = 8.0$ (and $u=2.0$) derived with tDMRG.

For excitation momentum $k>0$, the two-parametric spectrum, Eq. (\ref{SpElremom0}), contains 
the $c$, $c'$, and $s$ branch lines represented by solid lines in Figs. \ref{figure2} and \ref{figure3}
and a $c-s$ boundary line represented by a dashed-dotted line. 
 
The one-particle removal $c$, $c'$, and $s$ branch-line spectra have
the general form, Eq. (\ref{branch-l}), and are given by,
\begin{eqnarray}
B_{-1} (k,\omega) & = & C_{\bar{\beta}}\,
\Bigl(\omega + \epsilon_{\bar{\beta}} (k)\Bigr)^{\xi_{\bar{\beta}} (k)}
\nonumber \\
{\rm for} && {\rm small}\hspace{0.20cm}(\omega + \epsilon_{\bar{\beta}} (k)) \leq 0 \, .
\label{branch-l-0}
\end{eqnarray}
Here the $\beta_c = c,c'$ and $s$ branch-line spectra $\epsilon_{\bar{\beta}} (k)$ read,
\begin{eqnarray}
\epsilon_{\beta_c} (k) & = & - \varepsilon_c (q)\hspace{0.20cm}{\rm for}\hspace{0.20cm}
\bar{\beta} = \beta_c = c,c'\hspace{0.20cm}{\rm and}
\nonumber \\
\epsilon_{s} (k) & = & \Delta_{MH} - \varepsilon_s (q')
\hspace{0.20cm}{\rm for}\hspace{0.20cm}\bar{\beta} = s \, ,
\label{OkudRLAccxm0}
\end{eqnarray}
where $\Delta_{MH}$ is half the $m=0$ Mott-Hubbard gap, Eqs. (\ref{2mu0})-(\ref{2mu0LB}),
and the expressions of $k$ are given below.
\begin{figure}
\begin{center}
\centerline{\includegraphics[width=8.75cm]{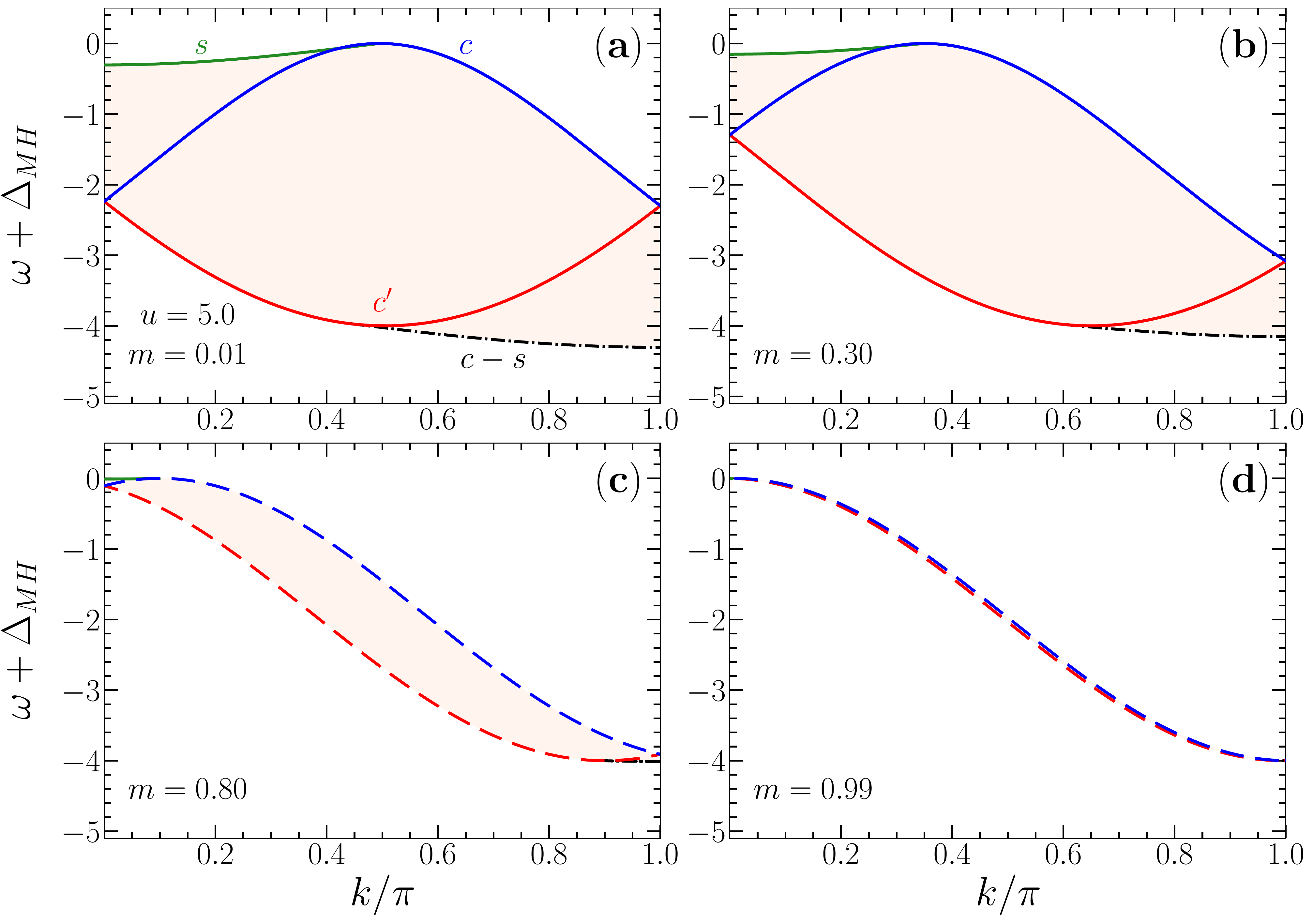}}
\caption{The same as Fig. \ref{figure7} for a larger $u$ value, $u=5.0$.}
\label{figure9}
\end{center}
\end{figure}

The functionals, Eqs. (\ref{cfunDc})-(\ref{cfunDs}), simplify at $m=0$ to,
\begin{eqnarray}
\Phi_{\bar{\beta},c,\iota} (q) & = & {1\over 2}\left(\iota\,\delta N^F_{c} + 2\,\delta J^F_{c} + \delta J^F_{s}\right)
\nonumber \\
& & + c_{\bar{\beta}}\,\Phi_{c,\bar{\beta}}(\iota\pi,q) 
\nonumber \\
\Phi_{\bar{\beta},s,\iota} (q) & = & {1\over\sqrt{2}}\left(- {\iota \,\delta N^F_{c}\over 2}
+ \iota\,\delta N^F_{s} + \delta J^F_{s}\right)
\nonumber \\
& & + c_{\bar{\beta}}\,\Phi_{s,\bar{\beta}}(\iota\pi/2,q) \, .
\label{cfunDcsm0}
\end{eqnarray}
The use of the $m=0$ related phase-shift parameters and phase-shift's expressions, 
Eqs. (\ref{ZZ-m0}) and (\ref{Phicccs}) of Appendix \ref{B}, respectively, leads
for $u>0$ to the following simplified expressions for the exponents in Eq. (\ref{branch-l-0}),
\begin{eqnarray}
\xi_c (k) & = & - {1\over 2} + {1\over 8}\left(4\Psi \left({\sin k_c (q)\over u}\right) + 1\right)^2
\hspace{0.20cm}{\rm where}
\nonumber \\
 k & = & - {\pi\over 2} - q \in \left[0,{\pi\over 2}\right]
\hspace{0.20cm}{\rm and}\hspace{0.20cm}q \in \left[-\pi,-{\pi\over 2}\right] 
\nonumber \\
 k & = & {3\pi\over 2} - q \in \left[{\pi\over 2},\pi\right]
\nonumber \\
\xi_{c'} (k) & = & - {1\over 2} + {1\over 8}\left(4\Psi \left({\sin k_c (q)\over u}\right) - 1\right)^2
\hspace{0.20cm}{\rm where}
\nonumber \\
 k & = & {\pi\over 2} - q \in \left[0,\pi\right]
\hspace{0.20cm}{\rm and}\hspace{0.20cm}q \in \left[-{\pi\over 2},{\pi\over 2}\right] 
\nonumber \\
\xi_s (k) & = & - {1\over 2} + {1\over 2\pi^2}\left(\arctan\left(\sinh\left({\pi\over 2}{\Lambda_s (q)\over u}\right)\right)\right)^2
\hspace{0.20cm}{\rm where}
\nonumber \\
k & = & - q \in \left[0,{\pi\over 2}\right]
\hspace{0.20cm}{\rm and}\hspace{0.20cm}q \in \left[-{\pi\over 2},0\right] \, .
 \label{xiccpskm0}
\end{eqnarray}
The rapidity functions $k_c (q)$ and $\Lambda_s (q)$ in these expressions 
are defined in terms of their inverse functions $q = q_c (k)$ for $k \in [-\pi,\pi]$ and
$q = q_s (\Lambda)$ for $\Lambda\in [-\infty,\infty]$ in Eq.
(\ref{qLambdam0}) of Appendix \ref{B}, respectively, and $\Psi (x)$ is the function given in 
Eq. (\ref{Psix}) of that appendix.
\begin{figure}
\begin{center}
\centerline{\includegraphics[width=8.75cm]{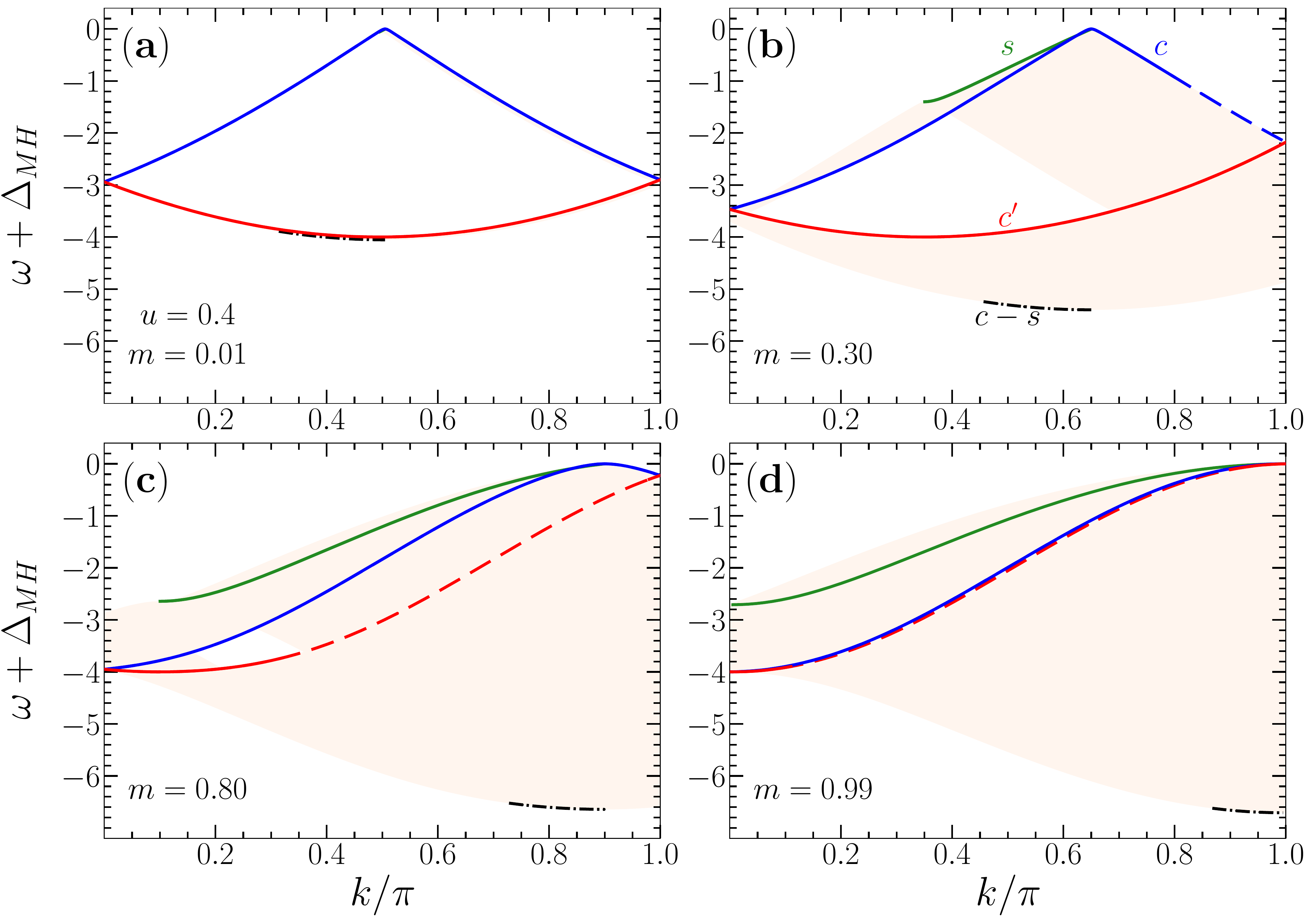}}
\caption{The $(k,\omega)$-plane regions 
corresponding to the spectrum $\epsilon_{\uparrow} (k)$, Eq. (\ref{SpupElremo}), where for spin densities
(a) $m=0.01$, (b) $m=0.30$, (c) $m=0.80$, and (d) $m=0.99$ and
$u=0.4$ there is in the thermodynamic limit more spectral weight in the 
removal up-spin one-particle spectral function. Most such a weight is located in the vicinity 
of the $c$, $c'$, and $s$ branch lines and of the $c-s$ boundary line.
The lines notations are the same as in Fig. \ref{figure7}.}
\label{figure10}
\end{center}
\end{figure}

The exponents in Eq. (\ref{xiccpskm0}) are plotted as
a function of the momentum $k$ for several $u$ values in Figs. \ref{figure4}, \ref{figure5}, and
\ref{figure6}. They are negative for their $k$ intervals and for all $u>0$ values.
Hence, consistently with results obtained by other methods \cite{Nocera_18,Lante_09,Benthien_07},
there are line-shape $(k,\omega)$-plane cusp singularities 
in the one-particle removal spectral function, Eq. (\ref{branch-l-0}), 
at and in the vicinity of the $c$, $c'$, and $s$ branch lines shown in Figs. \ref{figure2} and \ref{figure3}.

The spectrum of the one-particle removal $c-s$ boundary line represented
in these figures by a dashed-dotted line has the general
form, Eq. (\ref{dE-dP-c-s1}). It
reads $\epsilon_{c-s} (k) = -\varepsilon_c (q) - \varepsilon_{s} (q')$
for $v_c (q)=v_s (q')$ where $k = \pi - q - q'\in [\pi/2 - k_{c-s},\pi]$.
Here $k_{c-s}=\pi/2$ for $u\ll 1$ and $k_{c-s}=\pi/4u$ for $u\gg 1$. The
spectrum $-\epsilon_{c-s} (k)$ reaches a minimum value at $k=\pi$, $\partial\epsilon_{c-s} (k)/\partial k\vert_{k=\pi} = 0$.
Again consistently with results obtained by other methods \cite{Nocera_18,Lante_09,Benthien_07},
at and near this $c-s$ boundary line the spectral function
displays singular behavior, $B_{-1} (k,\omega) \propto \Bigl(\omega +\epsilon_{c-s} (k)\Bigr)^{-1/2}$. 

\section{The branch and boundary lines of the one-particle spectral functions for $m>0$}
\label{SECVI}

\begin{figure}
\begin{center}
\centerline{\includegraphics[width=8.75cm]{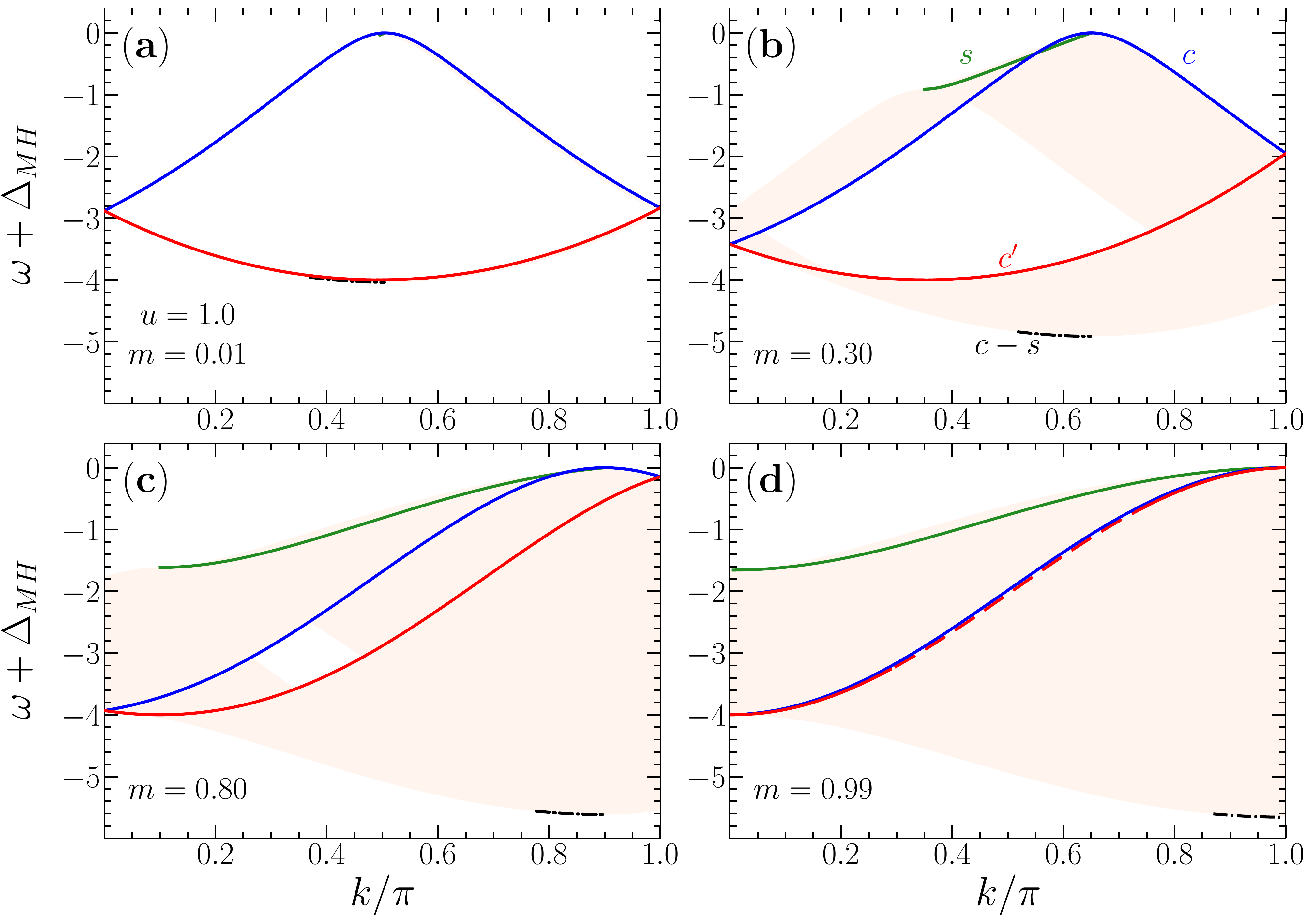}}
\caption{The same as Fig. \ref{figure10} for a larger $u$ value, $u=1.0$.}
\label{figure11}
\end{center}
\end{figure}

\begin{figure}
\begin{center}
\centerline{\includegraphics[width=8.75cm]{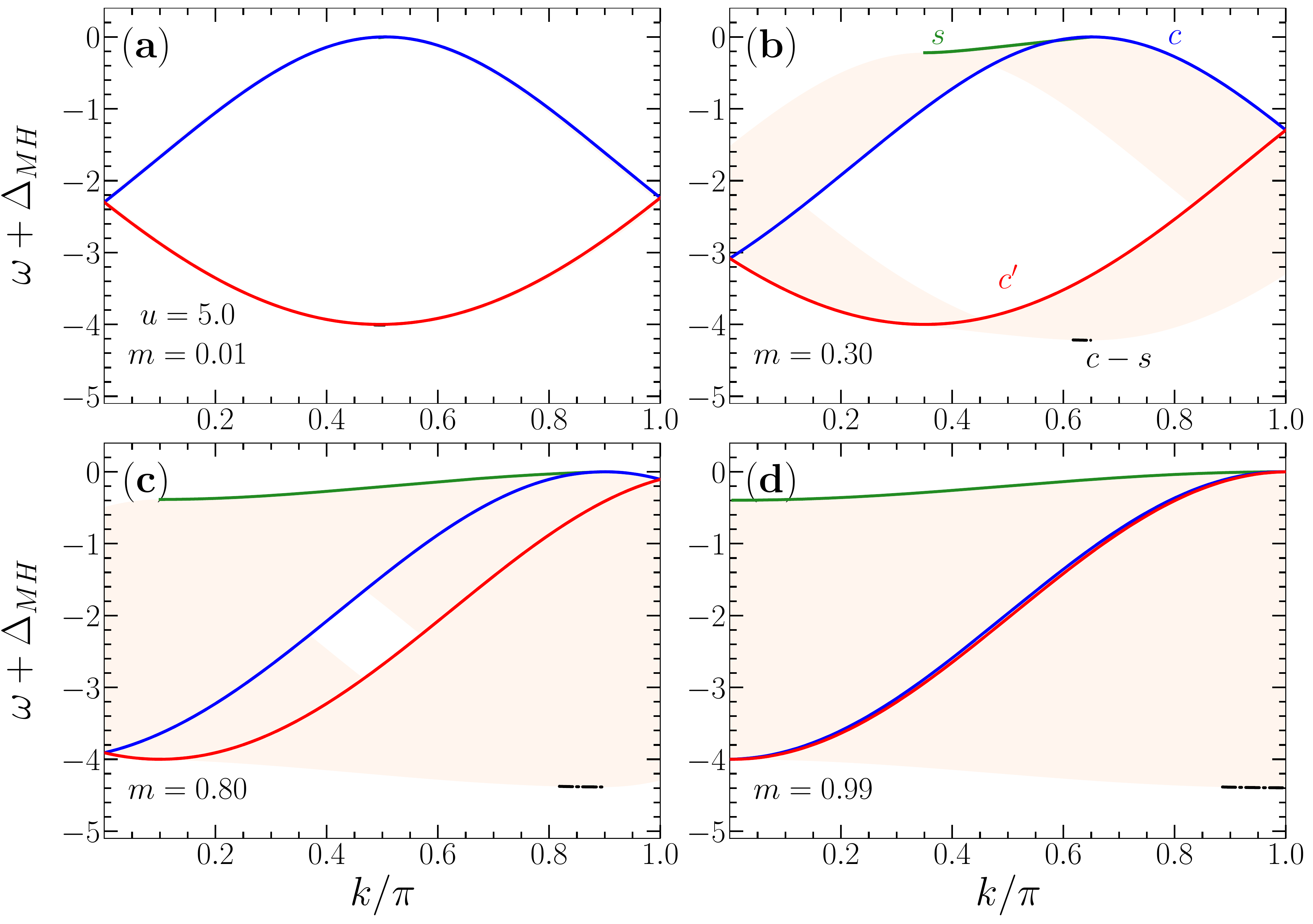}}
\caption{The same as Fig. \ref{figure10} for a larger $u$ value, $u=5.0$.}
\label{figure12}
\end{center}
\end{figure}

In the $(k,\omega)$-plane distributions shown in Figs. \ref{figure7}-\ref{figure12}, 
the $m>0$ branch lines $k$ intervals for which the corresponding exponents, Eq. (\ref{expbsk}),
are negative and positive are represented by solid and dashed lines,
respectively, and the boundary line by a dashed-dotted line. Expressions
of the spectra and exponents specific to each branch line are provided 
in the following.

\subsection{The down-spin $\bar{\beta} = c,c',s$ branch lines}
\label{SECVIA}

The $c$ and $c'$ branch lines one-parametric spectrum is contained in that defined by Eq. (\ref{SpdownElremo}) 
and reads,
\begin{equation}
\epsilon_{\beta_c,\downarrow} (k) = - \varepsilon_c (q) 
\hspace{0.20cm}{\rm for}\hspace{0.20cm}\beta_c=c,c' \, ,
\label{OkudRLAcc}
\end{equation}
where $\varepsilon_c (q)$ is the $c$-band energy dispersion defined in Eq. (\ref{equA4}) of Appendix \ref{B}.
This and all branch-line spectra given in the following have the general form, Eq. (\ref{dE-dP-bl}),
and all the spectral weight near them has been transferred to the first Brillouin zone for $k>0$.. 

The $c$ and $c'$ branch lines refer to excited states with deviations $\delta N_c^F =0$,  $\delta N_s^F = -1$, $\delta N_c^{NF} =-1$,
$\delta J_c^F = - b_{\beta_c}b_{cs}/2$, and $\delta J_{s}^F = b_{\beta_c}/2$.
Here the coefficient $b_{cs}=\pm 1$ refers to two spectral weight contributions from different extended-zone $k$ regions
that in the first Brillouin zone refer to $c$ and $c'$ branch lines with the same energy
spectrum but different weight distributions. In the following the value $b_{cs}=1$ is that used
because it refers to the processes that determine the line shape.

The $c$ branch has for $k>0$ two subdomains,
\begin{eqnarray}
k & = & - k_{F\uparrow} - q \in [0,k_{F\downarrow}]\hspace{0.2cm}{\rm for}\hspace{0.2cm}
q \in [-\pi,-k_{F\uparrow}]\hspace{0.2cm}{\rm and}
\nonumber \\
k & = & \pi + k_{F\downarrow} - q \in [k_{F\downarrow},\pi]\hspace{0.2cm}{\rm for}\hspace{0.2cm}
q \in [k_{F\downarrow},\pi] \, ,
\label{OkudRLAc}
\end{eqnarray}
whereas the $k>0$ interval of the $c'$ branch line is,
\begin{equation}
k = k_{F\uparrow} - q \in [0,\pi]\hspace{0.2cm}{\rm for}\hspace{0.2cm}
q \in [-k_{F\downarrow},k_{F\uparrow}] \, .
\label{OkudRLAcprime}
\end{equation}

Near the present branch lines $B_{\downarrow,-1} (k,\omega)$ reads,
\begin{eqnarray}
B_{\downarrow,-1} (k,\omega) & = & C_{\downarrow,\beta_c} 
\Bigl(\omega + \epsilon_{\beta_c,\downarrow} (k)\Bigr)^{\xi_{\beta_c}^{\downarrow} (k)}  
\nonumber \\
& & {\rm for}\hspace{0.20cm} (\omega + \epsilon_{\beta_c,\downarrow} (k)) \leq 0  \, .
\label{cpm-branchccd}
\end{eqnarray}
The $\beta_{c}=c,c'$ constants $C_{\downarrow,\beta_c}$ and all multiplicative constants in
the branch-line spectral function expressions of the general form, Eq. (\ref{branch-l}), provided in the following 
have a fixed value for the $k$ and $\omega$ ranges corresponding to small values of the energy deviation 
for which such expressions are valid, which here is given by $(\omega + \epsilon_{\beta_c,\downarrow} (k))$.

The use of the above deviations in the spectral functionals, Eqs. (\ref{cfunDc}) and (\ref{cfunDs})
for $\bar{\beta}=c,c'$, leads to the following expression for the exponents in Eq. (\ref{cpm-branchccd}),
\begin{eqnarray}
&& \xi_{\beta_c}^{\downarrow} (k) = -1
+ \sum_{\iota}\left(- b_{\beta_c}{(1-\xi_{c\,s})\over 2} - \Phi_{c,c}(\iota\pi,q)\right)^2 
\nonumber \\
&& + \sum_{\iota}\left(- {\iota\over 2\xi_{s\,s}} + b_{\beta_c}{\xi_{s\,s}\over 2} 
- \Phi_{s,c}(\iota k_{F\downarrow},q)\right)^2 \, ,
\label{expccdne1}
\end{eqnarray}
of the general form, Eq. (\ref{expbsk}), where $k=k (q)$ is given 
in Eqs. (\ref{OkudRLAc}) and (\ref{OkudRLAcprime}) 
for the $c$ and $c'$ branch lines, respectively.

\begin{figure}
\begin{center}
\centerline{\includegraphics[width=8.75cm]{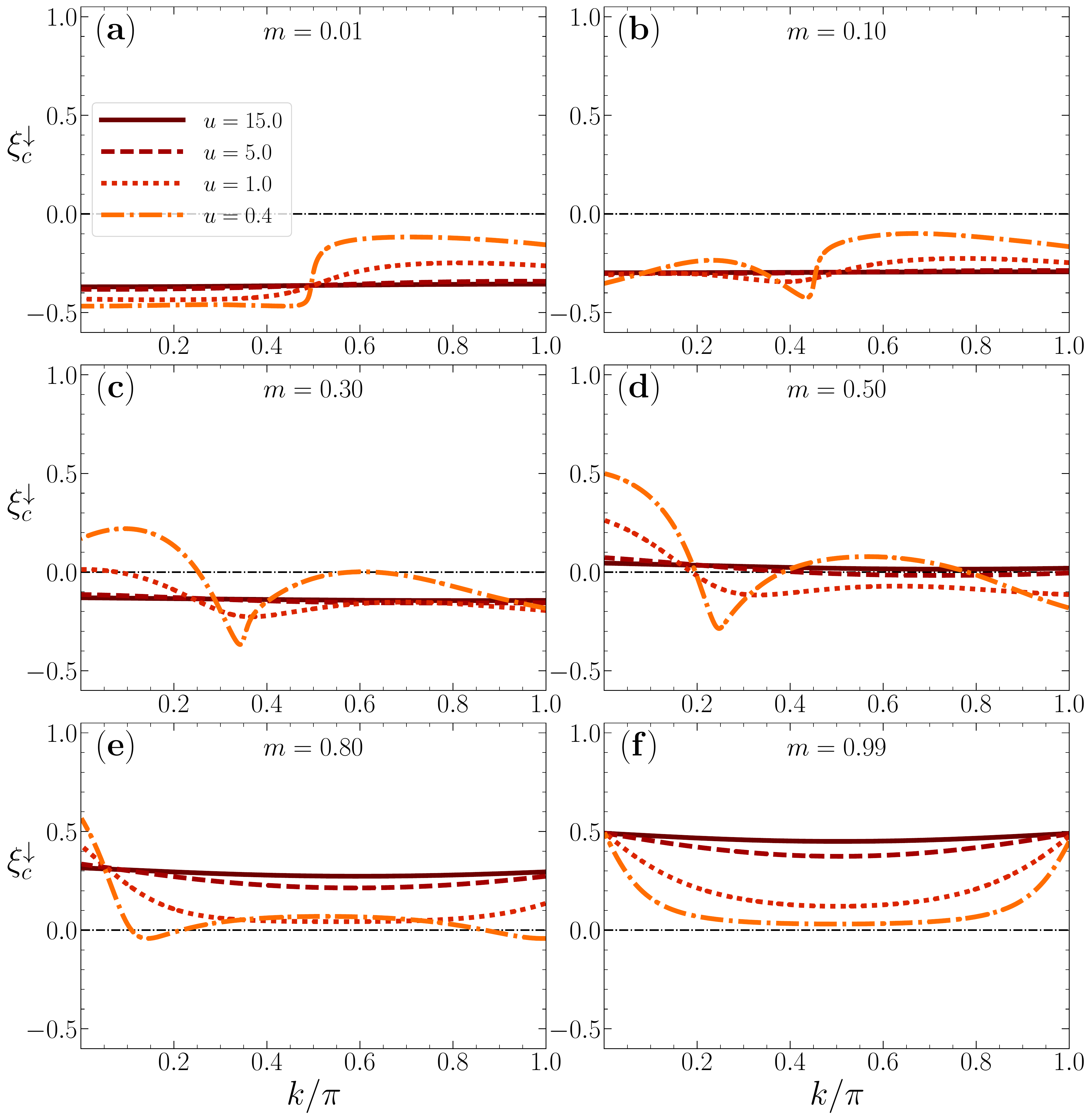}}
\caption{The down-spin $c$-branch line exponent as a function
of the excitation momentum $k$ for spin densities
(a) $m=0.01$, (b) $m=0.10$, (c) $m=0.30$, (d) $m=0.50$, (e) $m=0.80$, and (f) $m=0.99$ 
and a set of $u$ values. For small $m$ this exponent is negative. It acquires
$k$ intervals where it becomes positive whose momentum width increases upon increasing $m$.}
\label{figure13}
\end{center}
\end{figure}

\begin{figure}
\begin{center}
\centerline{\includegraphics[width=8.75cm]{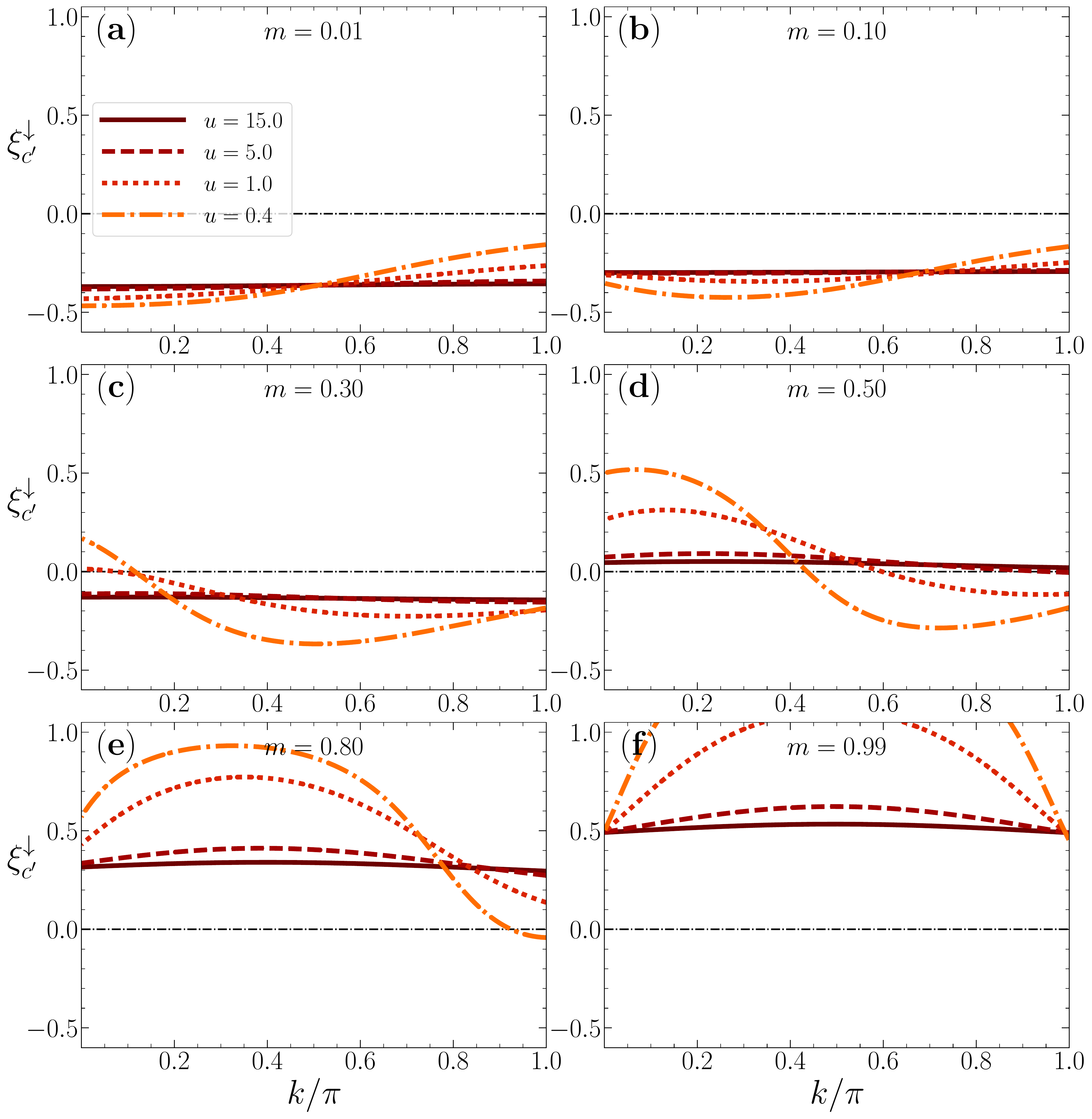}}
\caption{The $k$ dependence of the down-spin $c'$-branch line exponent 
for the same spin densities and $u$ values as Fig. \ref{figure13}. 
For small $m$ this exponent is negative. It becomes positive in $k$ intervals 
whose momentum width increases upon increasing $m$.}
\label{figure14}
\end{center}
\end{figure}

The down-spin $c$ and $c'$ branch-line exponents, Eq. (\ref{expccdne1}),
are plotted in Figs. \ref{figure13} and \ref{figure14}, respectively, as a function of 
$k$ for a set of $m$ and $u$ values. For the $k$ intervals for
which they are negative, there are $(k,\omega)$-plane cusp singularities 
in the down-spin spectral function, Eq. (\ref{cpm-branchccd}), at in the vicinity of the corresponding branch lines.
In contrast to their $m=0$ behavior, Figs. \ref{figure4} and \ref{figure5}, for
some $k$ intervals these exponents are positive.

The spectrum of the down-spin $s$ branch line reads,
\begin{eqnarray}
\epsilon_{s,\downarrow} (k) & = & \Delta_{MH} - \varepsilon_s (q') \hspace{0.2cm}{\rm where}
\nonumber \\
k & = & - q' \in [0,k_{F\downarrow}]
\hspace{0.2cm}{\rm for}\hspace{0.2cm}
q' \in [-k_{F\downarrow},0] \, .
\label{OkudRLAs}
\end{eqnarray}
Here $\Delta_{MH}$ is half the Mott-Hubbard gap, Eq. (\ref{2mu0}), and
$\varepsilon_s (q')$ is the $s$-band energy dispersion defined in Eq. (\ref{equA4}) of Appendix \ref{B}.
This branch line corresponds to excited energy eigenstates with deviations 
$\delta N_c^F = -1$,  $\delta N_s^F = 1$, $\delta N_s^{NF} =-1$,
and $\delta J_c^F = \delta J_s^F = 0$.

Near the $s$ branch line the expression of $B_{\downarrow,-1} (k,\omega)$ is,
\begin{eqnarray}
B_{\downarrow,-1} (k,\omega) & = & C_{\downarrow,s} 
\Bigl(\omega + \epsilon_{s,\downarrow} (k)\Bigr)^{\xi_{s}^{\downarrow} (k)}  
\nonumber \\
& & {\rm for}\hspace{0.20cm} (\omega + \epsilon_{s,\downarrow} (k)) \leq 0  \, .
\label{cpm-branchsd}
\end{eqnarray}
The exponent obtained by the use in the functionals, Eqs. (\ref{cfunDc}) and (\ref{cfunDs}), 
of the above deviations is given by,
\begin{eqnarray}
\xi_{s}^{\downarrow} (k) & = & -1 + \sum_{\iota}\left(- {\iota\over 2} - \Phi_{c,s}(\iota\pi,-k)\right)^2 
\nonumber \\
& + & \sum_{\iota}\left({\iota\,\xi_{c\,c}\over 2\xi_{s\,s}} 
- \Phi_{s,s}(\iota k_{F\downarrow},-k)\right)^2 \, .
\label{exps3dne1}
\end{eqnarray}

\begin{figure}
\begin{center}
\centerline{\includegraphics[width=8.75cm]{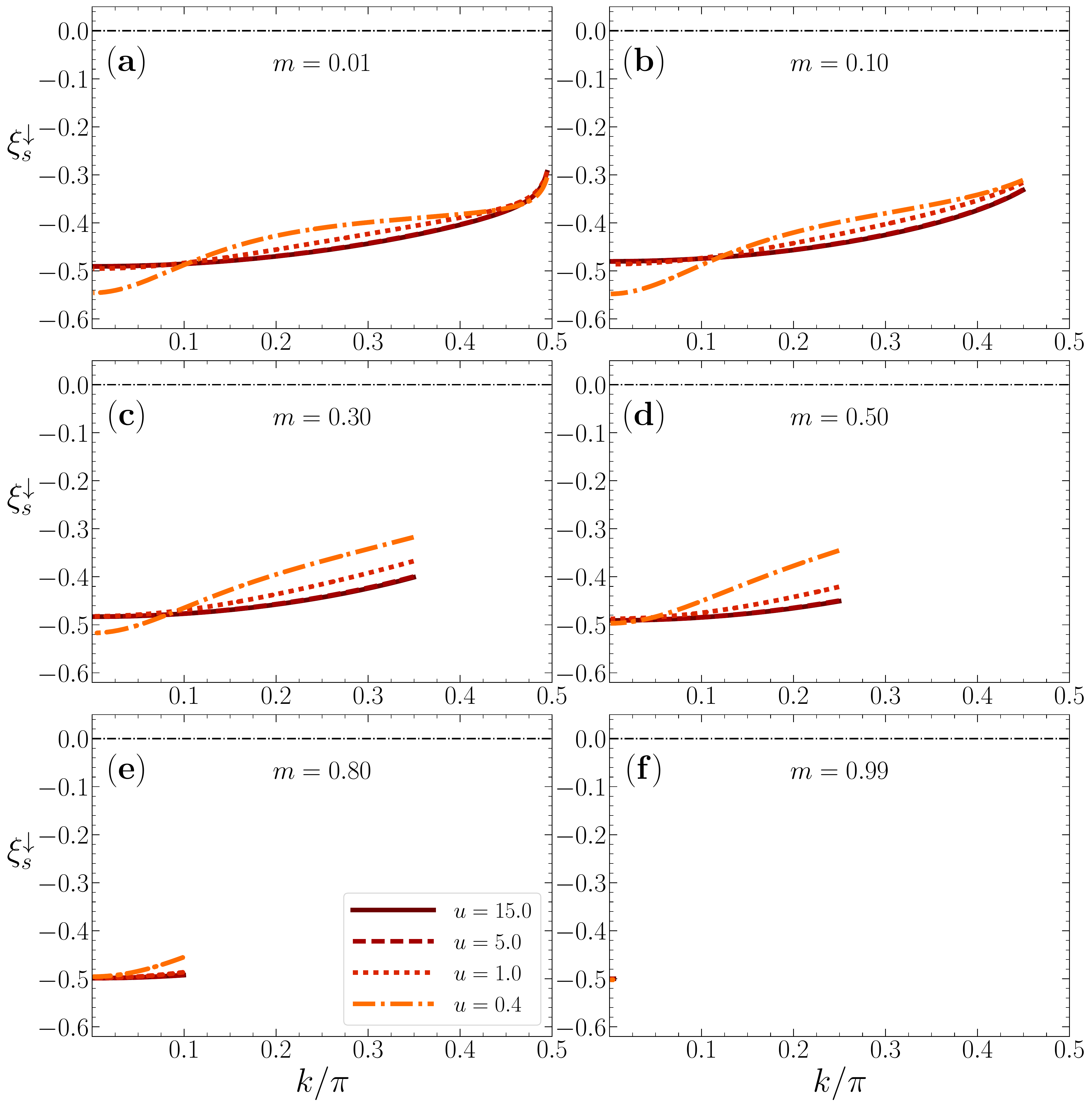}}
\caption{The $k$ dependence of the negative down-spin $s$-branch line exponent 
for the same spin densities and $u$ values as Fig. \ref{figure13}.}
\label{figure15}
\end{center}
\end{figure}

The down-spin $s$ branch-line exponent, Eq. (\ref{exps3dne1}),
is plotted in Fig. \ref{figure15} as a function of the excitation momentum $k$ 
for a set of spin density $m$ and $u$ values. It is negative in
all its $k$ intervals, so that there are $(k,\omega)$-plane cusp singularities 
in the down-spin spectral function, Eq. (\ref{cpm-branchsd}), at and in the vicinity of 
this branch line.

\subsection{The up-spin $\bar{\beta} = c,c',s$ branch lines}
\label{SECVIB}

The $\beta_c =c,c'$ branch lines refer to excited energy 
eigenstates with deviations $\delta N_c^F = \delta N_{s}^F =0$, $\delta N_c^{NF} = -1$,
$\delta J_c^F = 0$, and $\delta J_{s}^F = - b_{\beta_c}/2$.
Their spectra read,
\begin{eqnarray}
\epsilon_{\beta_c,\uparrow} (k) & = & - \varepsilon_c (q)\hspace{0.2cm}{\rm where}
\nonumber \\
k & = & - k_{F\downarrow} - q \in [0,k_{F\uparrow}]
\hspace{0.2cm}{\rm for}\hspace{0.2cm}q \in [-\pi,-k_{F\downarrow}]
\nonumber \\
k & = & 2\pi - k_{F\downarrow} - q \in [k_{F\uparrow},\pi]
\hspace{0.2cm}{\rm for}\hspace{0.2cm}
q \in [k_{F\uparrow},\pi]\hspace{0.2cm}
\nonumber \\
&& {\rm and}\hspace{0.2cm}\beta_c = c
\nonumber \\
k & = & k_{F\downarrow} - q \in [0,\pi]\hspace{0.2cm}{\rm for}\hspace{0.2cm}
q \in [-k_{F\uparrow},k_{F\downarrow}]
\nonumber \\
&& {\rm and}\hspace{0.2cm}\beta_c = c' \, .
\label{OkuuRLAcc}
\end{eqnarray}

Near the $\beta_{c}=c,c'$ branch lines $B_{\uparrow,-1} (k,\omega)$ is given by,
\begin{eqnarray}
B_{\uparrow,-1} (k,\omega) & = & C_{\uparrow,\beta_{c}} 
\Bigl(\omega + \epsilon_{\beta_{c}}^{\uparrow} (k)\Bigr)^{\xi_{\beta_{c}}^{\uparrow} (k)}  
\nonumber \\
& & {\rm for}\hspace{0.20cm} (\omega + \epsilon_{\beta_{c}}^{\uparrow} (k)) \leq 0  \, .
\label{cpm-branchcu}
\end{eqnarray}
The exponent obtained from the use in the functionals, Eqs. (\ref{cfunDc}) and (\ref{cfunDs}),
of the above deviations reads,
\begin{eqnarray}
\xi_{\beta_c}^{\uparrow} (k) & = & -1 + \sum_{\iota}\left(- b_{\beta_c}{\xi_{c\,s}\over 2} - \Phi_{c,c}(\iota\pi,q)\right)^2 
\nonumber \\
& + & \sum_{\iota}\left(- b_{\beta_c}{\xi_{s\,s}\over 2} 
- \Phi_{s,c}(\iota k_{F\downarrow},q)\right)^2 \, .
\label{expccune1}
\end{eqnarray}

\begin{figure}
\begin{center}
\centerline{\includegraphics[width=8.75cm]{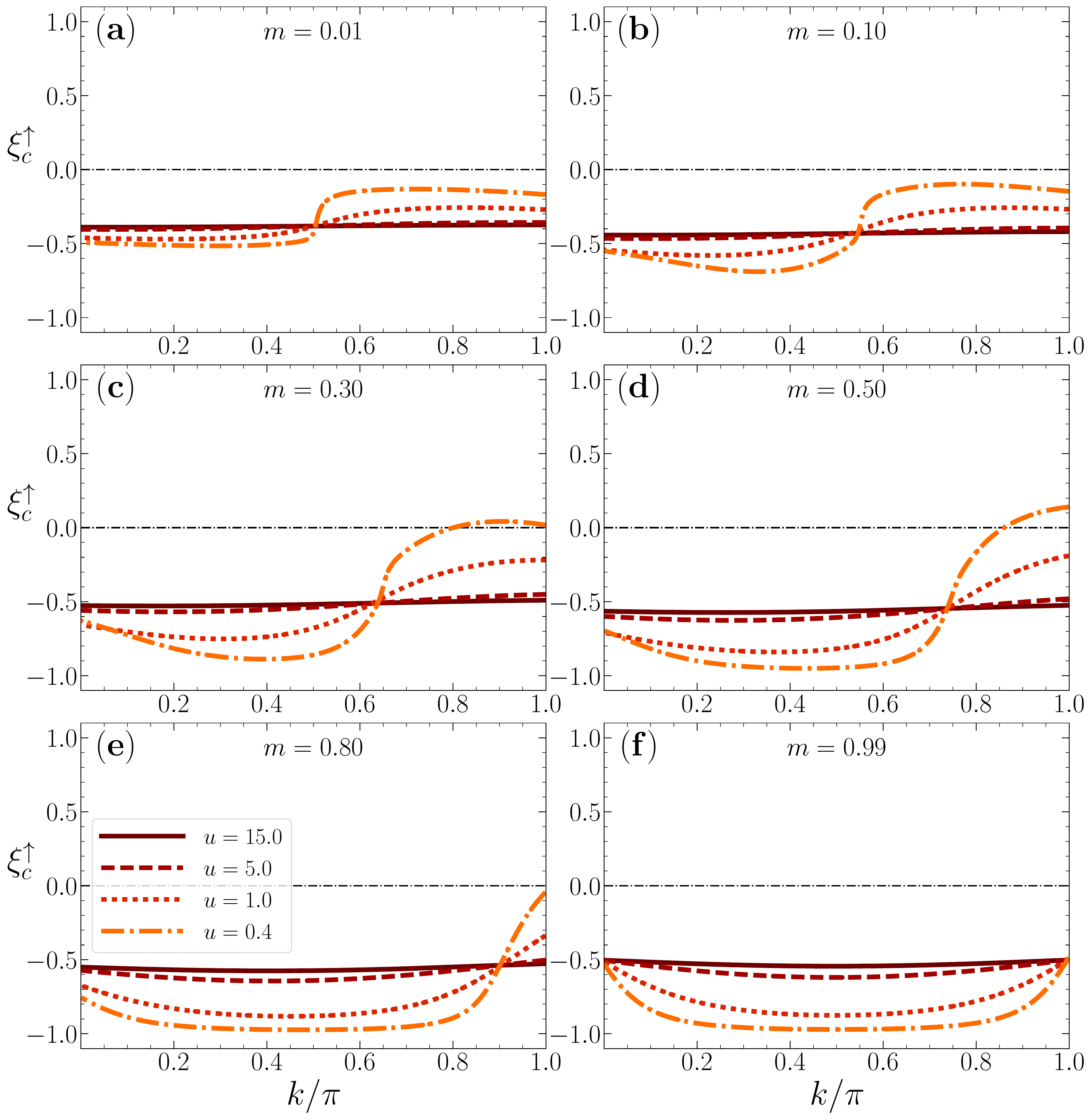}}
\caption{The $k$ dependence of the up-spin $c$-branch line exponent 
for the same spin densities and $u$ values as Fig. \ref{figure13}.
Except for a small $k$ interval that emerges for small $u$ values
and intermediate $m$ values, this exponent is negative.}
\label{figure16}
\end{center}
\end{figure}

These $c$ and $c'$ branch-line exponents are plotted in Figs. \ref{figure16} and \ref{figure17}, 
respectively, as a function of the excitation 
momentum $k$ for a set of spin density $m$ and $u$ values. They are mostly negative
yet upon decreasing $u$ and increasing $m$ they become positive for some $k$ intervals.
For those for which such exponents are negative, there are cusp singularities 
in the up-spin spectral function, Eq. (\ref{cpm-branchcu}), at and near the $\beta_{c}=c,c'$ branch lines.

The up-spin one-particle $s$ branch line refers to excited states with 
deviations $\delta N_c^F = \delta N_{s}^F = -1$,
$\delta N_s^{NF} = 1$, $\delta J_c^F = 1/2$, and $\delta J_{s}^F = 0$.
Its spectrum is given by,
\begin{eqnarray}
\epsilon_{s,\uparrow} (k) & = & \Delta_{MH} + \varepsilon_s (q') 
\hspace{0.2cm}{\rm where}\hspace{0.2cm}
k = \pi + q' \in [k_{F\downarrow},k_{F\uparrow}]
\nonumber \\
&&{\rm for}\hspace{0.2cm}q' \in [-k_{F\uparrow},-k_{F\downarrow}]
\label{OkuuRLAs}
\end{eqnarray}
Near the present $s$ branch line, $B_{\uparrow,-1} (k,\omega)$ reads,
\begin{eqnarray}
B_{\uparrow,-1} (k,\omega) & = & C_{\uparrow,s} 
\Bigl(\omega + \epsilon_{s,\uparrow} (k)\Bigr)^{\xi_{s}^{\uparrow} (k)}  
\nonumber \\
& & {\rm for}\hspace{0.20cm} (\omega + \epsilon_{s,\uparrow} (k)) \leq 0  \, .
\label{cpm-branchsu}
\end{eqnarray}
The exponent obtained from the use of the above deviations in
the functionals, Eqs. (\ref{cfunDc}) and (\ref{cfunDs}), reads,
\begin{eqnarray}
\xi_{s}^{\uparrow} (k) & = & -1
+ \sum_{\iota}\left({(1-\iota)\over 2} + \Phi_{c,s}(\iota\pi,q')\right)^2 
\nonumber \\
& + & \sum_{\iota}\left( - {\iota\,(1 - \xi_{c\,s})\over 2\xi_{s\,s}} 
+ \Phi_{s,s}(\iota k_{F\downarrow},q')\right)^2 \, .
\label{exps3une1}
\end{eqnarray}

This exponent is plotted in Fig. \ref{figure18} as a function of the excitation momentum $k$ 
for a set of spin density $m$ and $u$ values. It is negative for smaller spin densities and
becomes positive upon increasing $m$. For the $k$ intervals, spin densities, and $u$ values for
which it is negative, there are $(k,\omega)$-plane cusp singularities 
in the up-spin spectral function, Eq. (\ref{cpm-branchsu}), at and near this branch line.

\subsection{Branch lines for $h=h_c$ and $m=1$}
\label{SECVIC}

For the fully polarized limit at $h=h_c$ and $m=1$ one has that
$B_{\downarrow,-1} (k,\omega) =0$. Indeed, all spins are up and thus there are
no spin-down electrons to be annihilated. In spite of the onsite interaction playing no
role when all spins are up, annihilation of one up-spin electron generates energy
eigenstates with one rotated-electron unoccupied site and thus one $\eta$-spin of
projection $1/2$ (see Appendix \ref{A}.) This is a high-energy process that
involves the many-electron interactions. 

Similarly, $B_{\uparrow,+1} (k,\omega) =0$ at $h=-h_c$ and $m=-1$ and
creation of one down-spin electron generates energy
eigenstates with one rotated-electron doubly occupied site and thus one $\eta$-spin of
projection $-1/2$. Again this is a high-energy process that involves 
the electronic correlations. 

In the present case of $h=h_c$ and $m=1$, the form of the $-\epsilon_{\uparrow} (k)<0$ spectrum, Eq. (\ref{SpupElremo}), 
simplifies. Within a $k$ extended zone scheme, it reads,
\begin{eqnarray}
&& \epsilon_{\uparrow} (k) = 2t\cos q + {1\over 2}\sqrt{(4t)^2 + U^2} 
\nonumber \\
&& - {2t\over\pi} \int_{-\pi}^{\pi} dq''\,\sin q''\arctan\left({\sin q'' - \Lambda_s (q')\over u}\right)
\hspace{0.2cm}{\rm where}
\nonumber \\
&& k = - q + q'\hspace{0.20cm}{\rm for}\hspace{0.20cm}
q \in [-\pi,\pi]\hspace{0.20cm}{\rm and}\hspace{0.20cm}q'\in [-\pi,\pi] \, .
\label{SpeUPm1}
\end{eqnarray}
Here the $s$ band rapidity function $\Lambda_s (q')$ is defined by its inverse function as,
\begin{eqnarray}
q' & = & {1\over\pi} \int_{-\pi}^{\pi} dq''\,\arctan\left({\Lambda_s (q') - \sin q''\over u}\right)\in [-\pi,\pi] 
\nonumber \\
&& {\rm for}\hspace{0.20cm}\Lambda\in [-\infty,\infty] \, .
\label{qsUPm1}
\end{eqnarray}
The $m=1$ spectrum, Eq. (\ref{SpeUPm1}), has nearly the same form as that plotted in Fig. \ref{figure10}\,(d)
for $m=0.99$.
\begin{figure}
\begin{center}
\centerline{\includegraphics[width=8.75cm]{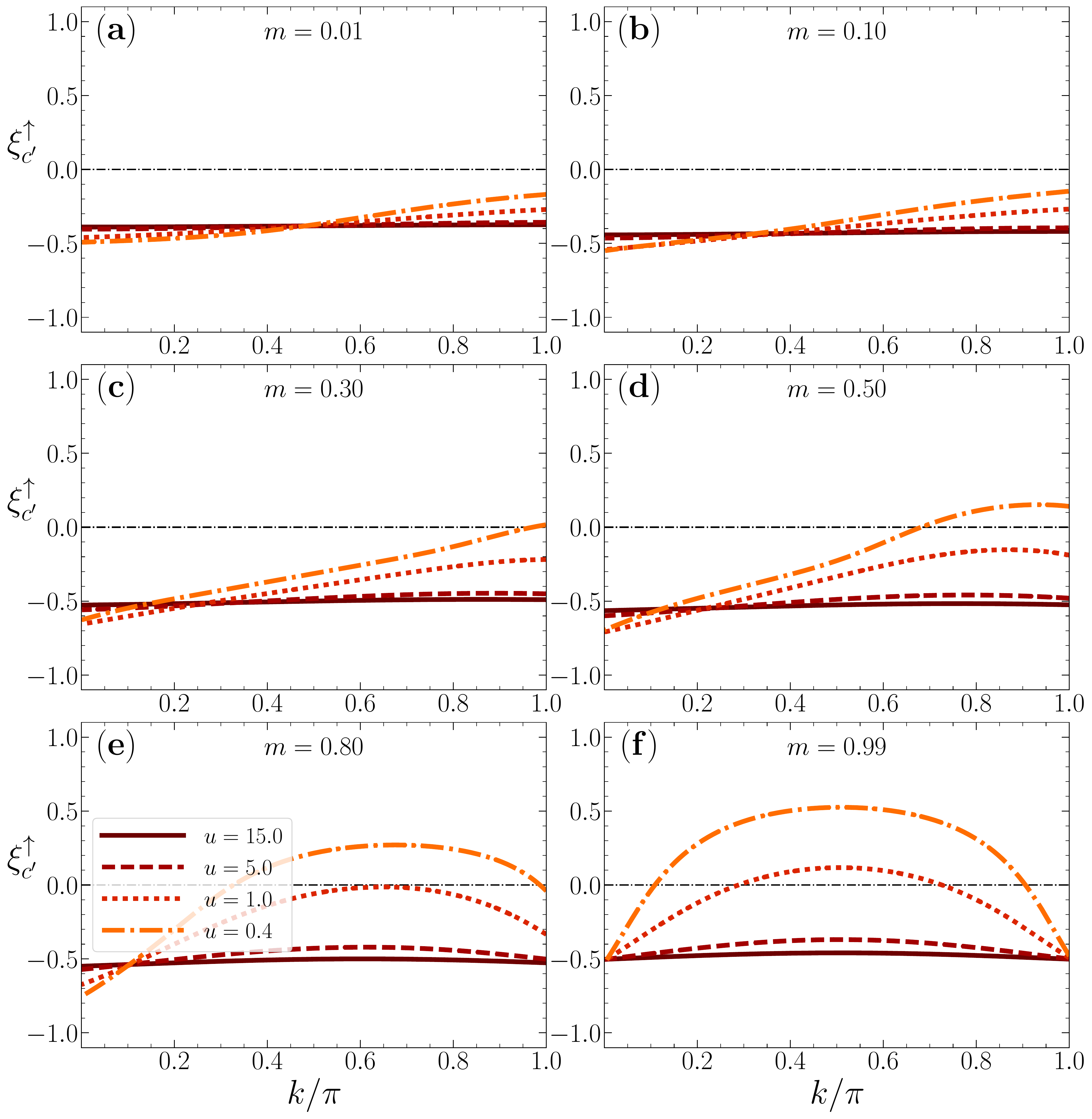}}
\caption{The $k$ dependence of the up-spin $c'$-branch line exponent 
for the same spin densities and $u$ values as Fig. \ref{figure13}.
Upon increasing $m$, it acquires for small $u$ values $k$ intervals 
where it becomes positive.}
\label{figure17}
\end{center}
\end{figure}
\begin{figure}
\begin{center}
\centerline{\includegraphics[width=8.75cm]{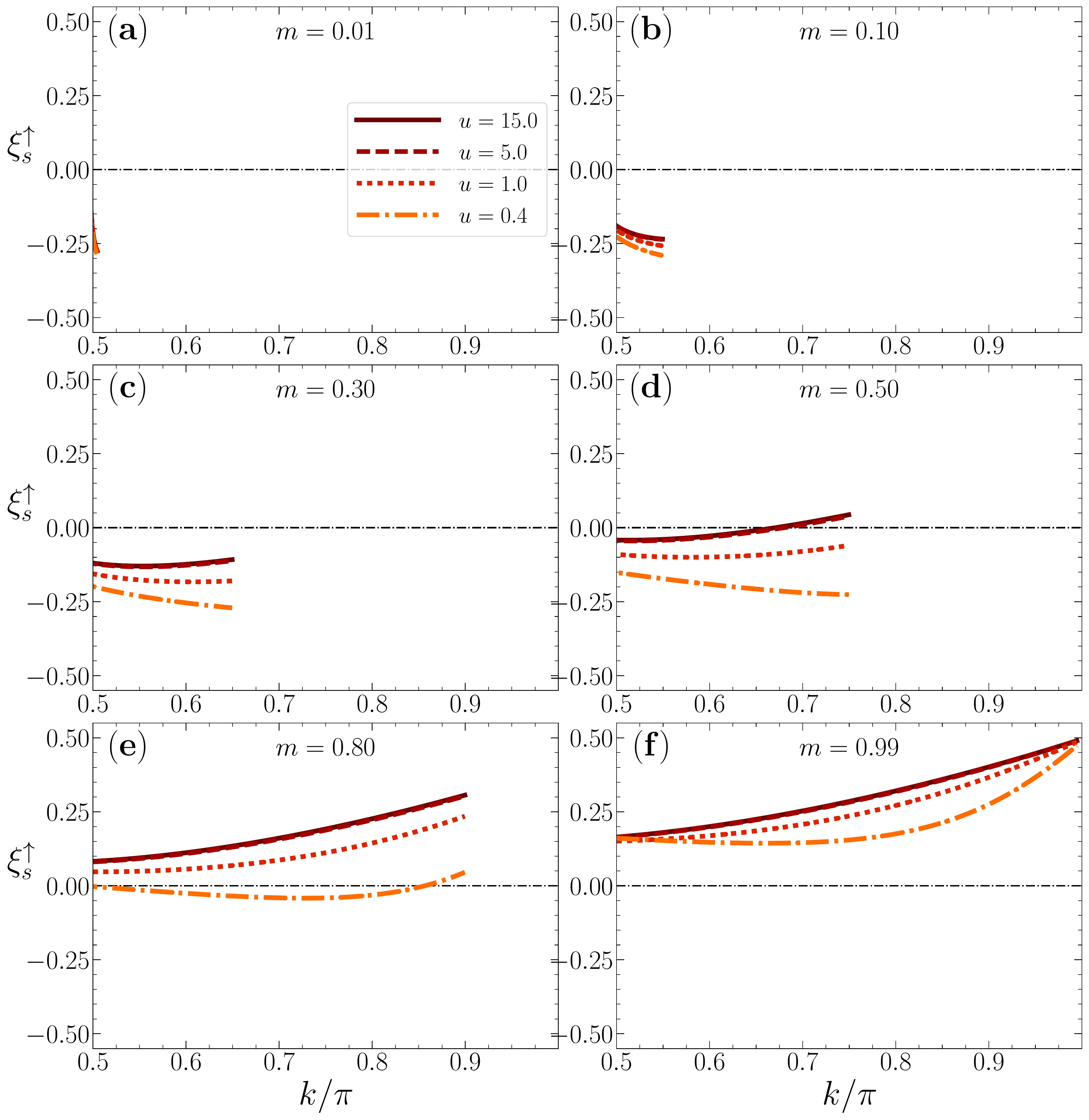}}
\caption{The $k$ dependence of the up-spin $s$-branch line exponent 
for the same spin densities and $u$ values as Fig. \ref{figure13}.
The general trend of this exponent for all $u$ values is that
it is negative at small $m$ and becomes positive upon increasing $m$.}
\label{figure18}
\end{center}
\end{figure}

The up-spin one-particle removal spectral function has in
the vicinity of the $\bar{\beta} = c,c',s$ branch lines 
the general form, Eq. (\ref{branch-l}). It is given by,
\begin{eqnarray}
B_{\uparrow,-1} (k,\omega) & = & C_{\uparrow,\bar{\beta}}\,
\Bigl(\omega + \epsilon_{\bar{\beta},\uparrow} (k)\Bigr)^{\xi_{\bar{\beta}}^{\uparrow} (k)}
\nonumber \\
{\rm for} && {\rm small}\hspace{0.20cm}(\omega + \epsilon_{\bar{\beta},\uparrow} (k)) \leq 0 \, .
\label{branch-l-m1}
\end{eqnarray}
The form of the $\beta_c = c,c'$ and $s$ branch-line spectra $-\epsilon_{\bar{\beta},\uparrow} (k)<0$ 
appearing in this expression also simplifies. It reads,
\begin{eqnarray}
\epsilon_{c,\uparrow} (k) & = & \epsilon_{c',\uparrow} (k) = 2t\cos k + {1\over 2}\sqrt{(4t)^2 + U^2} 
\nonumber \\
\epsilon_{s,\uparrow} (k) & = & - {2t\over\pi} \int_{-\pi}^{\pi} dq\,\sin q\arctan\left({\sin q - \Lambda_s (\pi - k)\over u}\right)
\nonumber \\
& + & \sqrt{(4t)^2 + U^2} - 4t 
\nonumber \\
&& {\rm both} \hspace{0.20cm}{\rm for}\hspace{0.20cm} k\in [0,\pi] \, .
\label{OkudRLAccxm1}
\end{eqnarray}
Here and in the following $\Lambda_s (\pi - k)$ is the $s$ band rapidity function defined by its inverse 
function in Eq. (\ref{qsUPm1}) for $q' = \pi - k \in [0,\pi]$.
The $c$, $c'$, and $s$ branch-line spectra, Eq. (\ref{OkudRLAccxm1}), have nearly the same form 
as those shown in Fig. \ref{figure10}\,(d) for $m=0.99$. Note that at $m=1$ the $c$ and $c'$ branch-line spectra
have exactly the same shape.

The form of the corresponding $\beta_c =c,c'$ and $s$ exponents in Eq. (\ref{branch-l-m1}) also 
simplifies for $h=h_c$ and $m=1$,
\begin{eqnarray}
\xi_{\beta_c}^{\uparrow} (k) & = & - {1\over 2} -  {2b_{\beta_c}\over\pi}\arctan\left({\sin k\over u}\right) 
\nonumber \\
& + & {2\over\pi^2}\left\{\arctan\left({\sin k\over u}\right)\right\}^2
\nonumber \\
\xi_s^{\uparrow} (k) & = &  {1\over 2} - {2\over\pi}\arctan\left({\Lambda_s (\pi - k)\over 2u}\right) 
\nonumber \\
& + & {2\over\pi^2}\Bigl(\left\{\arctan\left({\Lambda_s (\pi - k)\over u}\right)\right\}^2 
\nonumber \\
& + & \left\{\arctan\left({\Lambda_s (\pi - k)\over 2u}\right)\right\}^2 \Bigr)
\nonumber \\
&& {\rm both} \hspace{0.20cm}{\rm for}\hspace{0.20cm} k\in [0,\pi] \, .
 \label{xiccpskm1}
\end{eqnarray}
Here the coefficient $b_{\beta_c}$ used for $\beta_c =c,c'$ is defined in Eq. (\ref{acc}).
These $c$, $c'$, and $s$ exponents have nearly the same $k$ dependence
as those shown in Figs. \ref{figure16}\,(f), \ref{figure17}\,(f), and \ref{figure18}\,(f), respectively, 
for $m=0.99$. 

When two branch lines have exactly the same spectrum, as occurs here 
for $\epsilon_{c,\uparrow} (k) = \epsilon_{c',\uparrow} (k)$, the line shape
in their vicinity has the form shown in Eq. (\ref{branch-l-m1}) with the
exponent being the smallest of the corresponding two exponents.
As confirmed by comparison of the curves plotted in 
Figs. \ref{figure16}\,(f) and \ref{figure17}\,(f),
the smallest exponent is $\xi_{c}^{\uparrow} (k)$, Eq. (\ref{xiccpskm1})
for $\beta_c = c$ and $b_{\beta_c} = b_c =1$.

The values of the exponent curves plotted in  Figs. \ref{figure16}\,(f) and \ref{figure18}\,(f)
reveal that at $m=1$ the exponent $\xi_{c}^{\uparrow} (k)$ is negative and $\xi_{s}^{\uparrow} (k)$
is positive. Hence there are spectral peaks associated with the cusp singularities
near the $c$ branch line shown in Fig. \ref{figure10}\,(d).

\subsection{The line shape near the $c-s$ boundary lines}
\label{SECVII}

The limiting values of the $k$ intervals of the down- and up-spin $c-s$ boundary line spectra,
which have general form, Eq. (\ref{dE-dP-c-s1}), involve the
group velocities, Eq. (\ref{equA4B}) of Appendix \ref{B}. For $u> 0$ and $m> 0$ 
they are determined by the $c$-band values $q=\pm q_c^0$ and $q= \pm{\bar{q}}_c^0$ 
such that $q_c^0 < {\bar{q}}_c^0$
and the $s$-band values $q'= \pm k_{F\downarrow}$, $q' = \pm{\bar{q}}_s$, 
$q' = \pm{\tilde{q}}_s$, and $q' = k_{F\uparrow}$ such that
$k_{F\downarrow} < {\bar{q}}_s < {\tilde{q}}_s<k_{F\uparrow}$, which obey
the relations,
\begin{eqnarray}
v_c (0) & = & v_s (k_{F\uparrow}) 
\, ,\hspace{0.20cm}
v_c (q_c^0) = v_s (k_{F\downarrow}) = v_s ({\tilde{q}}_s)
\nonumber \\
v_c ({\bar{q}}_c^0) & = & v_s ({\bar{q}}_s) 
\, ,\hspace{0.20cm}
\vert v_s ({\bar{q}}_s)\vert \equiv {\rm max}\,\vert v_s (q')\vert \, .
\label{cfirstsector}
\end{eqnarray}

For $k>0$ and finite $u$ values the spectra of the down- and up-spin $c-s$ boundary lines 
represented in Figs. \ref{figure7}-\ref{figure9} and \ref{figure10}-\ref{figure12}, respectively,
by dashed-dotted lines are given by,
\begin{eqnarray}
\epsilon_{\downarrow,c-s} (k) & = & (-\varepsilon_c (q) - \varepsilon_{s} (q'))\,\delta_{v_c (q),v_s (q')}
\hspace{0.2cm}{\rm where}
\nonumber \\
k & = & \pm \pi - q - q' \in [k_{F\uparrow} - q_c^0,\pi]\hspace{0.20cm}{\rm for}
\nonumber \\
q & \in & [0,q_c^0]\hspace{0.20cm}{\rm and}\hspace{0.20cm}q'\in [0,k_{F\downarrow}] \, ,
\label{SpeBLdcs}
\end{eqnarray}
and
\begin{eqnarray}
\epsilon_{\uparrow,c-s} (k) & = & (-\varepsilon_c (q) +\varepsilon_{s} (q'))\delta_{v_c (q),v_s (q')} 
\hspace{0.2cm}{\rm where}
\nonumber \\
k & = & - q + q' \in [({\bar{q}}_s - {\bar{q}}_c^0),k_{F\uparrow}]\hspace{0.20cm}{\rm for}
\nonumber \\
q & \in & [0,{\bar{q}}_c^0]\hspace{0.20cm}{\rm and}\hspace{0.20cm}q'\in [{\bar{q}}_s,k_{F\uparrow}] \, ,
\label{SpeBLucs}
\end{eqnarray}
respectively. These spectra have a minimum value at $k=\pi$ and
$k=k_{F\uparrow}$, respectively, such that,
\begin{eqnarray}
{\partial\epsilon_{\downarrow,c-s} (k)\over\partial k}\vert_{k=\pi} & = & 0
\nonumber \\
- \epsilon_{\downarrow,c-s} (\pi) & = & - \Delta_{MH} - 4t + \varepsilon_{s} (0) 
\nonumber \\
{\partial\epsilon_{\uparrow,c-s} (k)\over\partial k}\vert_{k=k_{F\uparrow}} & = & 0
\nonumber \\
- \epsilon_{\uparrow,c-s} (k_{F\uparrow}) & = & - \Delta_{MH} - 4t - \varepsilon_{s} (k_{F\uparrow}) \, .
\label{BLpi}
\end{eqnarray}

In the $m\rightarrow 1$ limit, the down-spin $c-s$ boundary line does not exist.
At and near the $c-s$ boundary lines, the up- and down-spin spectral functions
show cusp singularities, $B_{\sigma,-1} (k,\omega) \propto \Bigl(\omega +\epsilon_{\sigma,c-s} (k)\Bigr)^{-1/2}$,
as given in Eq. (\ref{B-bol}).

\section{Effects of the charge-spin separation and charge-spin recombination}
\label{SECVII}

In contrast to Fermi liquids, 1D interacting systems are characterized by a breakdown of the 
basic quasiparticle picture. Indeed, no quasiparticles occur when the electrons range of motion is 
restricted to a single spatial dimension. In the quantum problem studied in this paper,
correlated electrons rather split into basic fractionalized $c$ charge-only and $s$ spin-only particles.
These particles can move with different speeds and even in different directions in the 1D 
many-electron system. 

In the case of both the metallic and Mott-Hubbard insulating phases of the 1D Hubbard model,
the usual quasiparticle Fermi momenta are
replaced by the $c$ particle Fermi points $\pm 2k_F$ and $s$ particle Fermi points
$\pm k_{F\downarrow}$ whose bands limits are $\pm\pi$ and $\pm k_{F\uparrow}$, respectively. 
Here $\pm 2k_F$ becomes $\pm \pi$ for the Mott-Hubbard insulator. Importantly and again in contrast 
to the spin $1/2$ quasiparticle Fermi momenta, for it the $c$ particle Fermi points read $\pm\pi$
and thus coincide with the limits of the Brillouin zone for the whole spin density interval $m\in [0,1]$.

Within studies of doped high-$T_c$ superconductors, a {\it shadow band} has appeared in angle-resolved photoemission spectra
that corresponds in two-dimensional antiferromagnetic Fermi liquids to an extra line of peaks that at low $\omega$ 
occurs at $k_F + \pi$ \cite{Kampf_90,Chubukov_95}. The non-perturbative and non-Fermi-liquid physics of the 
present 1D Hubbard model refers to a different quantum problem. 
However, at $h=0$ a line of peaks that runs in the interval $k\in [0,3k_F]$, 
which for the Mott-Hubbard insulator refers to $k\in [0,k_F+\pi]$ with an extreme at $k_F + \pi$ 
also corresponding to low $\omega$, occurs in the one-particle
removal spectral function. By analogy with the two-dimensional Fermi liquid,
it has also been named {\it shadow band} \cite{Karlo_96}. (See lower
figure in Fig. 1 of Ref. \onlinecite{Karlo_96} for the metallic phase at quarter filling.) 

The separation of the charge an spin degrees of freedom that gives rise to independent charge $c$ and $c'$ 
and spin $s$ branch line peaks in the one-particle removal spectral function occurs both in the metallic 
and Mott-Hubbard insulating phases of the 1D Hubbard model for all densities. 
Near such branch lines, the spectral functions line shape 
is of the form $B_{\sigma,-1} (k,\omega) = C_{\sigma,\bar{\beta}}\,
(\omega + \epsilon_{\sigma,\bar{\beta}} (k))^{\xi_{\bar{\beta}}^{\sigma} (k)}$,
Eq. (\ref{branch-l}), where $\bar{\beta} = c,c',s$ and the $k$, $u$, and densities 
dependence of the exponents $\xi_{\bar{\beta}}^{\sigma} (k)$ stems from 
the overlaps within the one-electron matrix elements.

In the general case of both the metallic and Mott-Hubbard insulating phases,
such a shadow band refers for the (i) down-spin and (ii) up-spin one-particle removal spectral functions 
to a charge branch line of peaks that runs within an extended scheme from 
$k=0$ at high negative values of $\omega$, reaches a minimum 
$\omega$ of largest absolute value at 
(i) $k=k_{F\uparrow}$ and (ii) $k=k_{F\downarrow}$, and 
at low $\omega$ values runs until (i) $k=2k_{F\uparrow}+k_{F\downarrow}$
and (ii) $k=2k_{F\downarrow}+k_{F\uparrow}$, respectively.
This shadow band comes from effects on the charge
degrees of freedom of spin fluctuations \cite{Karlo_96}.

Hence in the specific case of the 1D Mott-Hubbard insulator under study in this paper, 
the shadow band of the (i) down-spin and (ii) up-spin one-particle removal spectral functions 
runs within an extended zone scheme in the interval
(i) $k\in [0,k_{F\uparrow}+\pi]$ and (ii) $k\in [0,k_{F\downarrow}+\pi]$,
respectively. In (i) Figs. \ref{figure7}, \ref{figure8}, and \ref{figure9} and
(ii) Figs. \ref{figure10}, \ref{figure11}, and \ref{figure12} for $0<h<h_c$, it is the $c'$ branch line 
for $k\in [0,\pi]$ and its interval (i) $k\in [\pi,k_{F\uparrow}+\pi]$ and (ii)
$k\in [\pi,k_{F\downarrow}+\pi]$ was brought to the first Brillouin zone and corresponds to the
part of the $c$ branch line that runs in the interval (i) $k\in [k_{F\downarrow},\pi]$ and
(ii) $k\in [k_{F\uparrow},\pi]$, respectively.

As mentioned above, at $h=0$ the shadow band interval becomes $k\in [0,k_{F}+\pi]$. In Figs. \ref{figure2} and
\ref{figure3} for $h=0$ it is the $c'$ branch line for $k\in [0,\pi]$ whereas its interval
$k\in [\pi,k_{F}+\pi]$ was brought again to the first Brillouin zone. It corresponds to the
part of the $c$ branch line that runs in the interval $k\in [k_{F},\pi]$.

At spin density $m=1$, when the $c$ and $s$ bands have the same limiting momenta $\pm\pi$,
the $c$ and $c'$ branch lines merge. In the case of the down-spin one-particle removal 
spectral function all weight actually vanishes at $m=1$. The line shape near
the up-spin one-particle removal spectral function is controlled only by the
$c$ branch line exponent $\xi_{c}^{\uparrow} (k)$. As found in Sec. \ref{SECVIC},
$\xi_{c}^{\uparrow} (k)$  is smaller than the exponent $\xi_{c'}^{\uparrow} (k)$ of the $c'$ branch line,
Eq. (\ref{xiccpskm1}) for $\beta_c =c,c'$. Moreover, the non-shadowed-band interval
$k\in [0,k_{F\uparrow}]$ of the $c$ branch line extends to $k\in [0,\pi]$. The 
corresponding lack of the shadow band thus occurs only when the $c$ and $s$ 
bands have the same limiting momenta $\pm\pi$ and the $s$ band is empty.

Interestingly, the physics behind the other type of singular features called boundary lines,
is related to a charge-spin recombination that occurs in the $(k,\omega)$-plane 
{\it only} at and very near such lines. As shown in Sec. 3.2 of Ref. \onlinecite{Carmelo_08},
rather than from one-electron matrix elements overlaps,
the boundary-line power-law behavior $B_{\sigma,-1} (k,\omega) 
\propto (\omega +\epsilon_{\sigma,c-s} (k))^{-1/2}$,
Eq. (\ref{B-bol}), stems from singularities in the density of states of the two-parametric spectra 
$E (k) = -\varepsilon_{c}(q) + c_s\,\varepsilon_{s}(q')$ where $k = k_0 - q + c_s\,q'$.
Here $c_s=-1$ or $c_s=1$, Eq. (\ref{AcAs}).

The physical reason why such one-electron spectral-function singularities are
controlled by an exponent $-1/2$ that does not depend on the momentum $k$, interaction $u$, and spin
density $m$ is indeed a phenomenon of charge-spin recombination. 
As reported above, in the present non-perturbative 
many-electron problem there is a separation of the charge and spin degrees of freedom
such that the $c$ charge and $s$ spin excitations propagate in general with different group
velocities $v_{c}(q)$ and $v_{s}(q')$, respectively. Only at and very near a $(k,\omega)$-plane boundary
line does one have that $v_{c}(q)=v_{s}(q')$. This equality of the $c$ charge and $s$ spin
velocities is associated with a partial recombination of the charge and spin degrees at
and near such $(k,\omega)$-plane lines, consistently with the corresponding
Fermi-liquid like exponent $-1/2$, in that it does not depend on $k$, $u$, and $m$.

\section{Discussion of the results and concluding remarks}
\label{SECVIII}

In this paper we have studied the one-particle spectral properties
of the 1D Hubbard model with one fermion per site both at zero
and finite magnetic field. Specifically, the momentum and energy dependence of the 
exponents and energy spectra that control the line shape at and near the cusp singularities
of the up- and down-spin one-particle spectral functions, Eq. (\ref{Bkomega}), 
was derived.

An important qualitative difference of the Mott-Hubbard insulator relative
to the one-particle properties of the model metallic phase studied in Ref. \onlinecite{Carmelo_17}
refers to the values of the exponents $\xi_{\bar{\beta}}^{\sigma} (k)$ in  Eq. (\ref{branch-l})
as $u\rightarrow 0$ when $m<1$. In the metallic case, that for a given $\bar{\beta} =c,s$ 
branch line $k$ interval the exponent reads $\xi_{\bar{\beta}}^{\sigma} (k)=-1$
in the  $u\rightarrow 0$ limit means that the exact expression of the spectral function is not that given in 
Eq. (\ref{branch-l}) because the four functionals $\Phi_{\bar{\beta},\beta,\iota} (q)$ in
Eq. (\ref{expbsk}) for $\beta =c,s$ and $\iota =\pm 1$ all exactly vanish. The corresponding 
one-electron spectral functions behavior is instead $\delta$-function like. For the corresponding $k$ 
momentum intervals one recovers the exact $U=0$ up- and down-spin one-electron spectra \cite{Carmelo_17}. 
Indeed, in the metallic phase the non-interacting spectral functions
are smoothly obtained upon continuously decreasing $u$. 

In contrast, for the Mott-Hubbard 
insulator the exponents $\xi_{\bar{\beta}}^{\sigma} (k)$ of the branch lines whose spectra
give rise to the metallic half-filled non-interacting spectra do not read $-1$ as $u\rightarrow 0$ when $m<1$.
Inspection of Figs. \ref{figure4}-\ref{figure6} and \ref{figure13}-\ref{figure18} reveals that all branch-line exponents are
typically larger than $-1/2$. This is because under the quantum phase transition from the Mott-Hubbard insulator to the metallic half-filling
phase occurring at $U=0$ there is a singular change of the spectral-weight line shapes.

An exception is though the $m\rightarrow 1$ limit. In that case the up-spin $c$ branch-line exponent
$\xi_{c}^{\uparrow} (k)$, Eq. (\ref{expccune1}) for $\beta_c=c$, plotted 
in Fig. \ref{figure16} indeed reads $-1$ for $k\in [0,\pi]$ in the $u\rightarrow 0$ limit.
This implies that the line-shape is $\delta$-function like at that line.
Consistently, one recovers the exact $U=0$ up-spin one-electron spectrum
since in that limit that branch-line spectrum, Eq. (\ref{OkuuRLAcc}) for $\beta_c=c$,
is given by $\epsilon_{c,\uparrow} (k) = 2t\,(1 + \cos k)$ for $k\in [0,\pi]$. This behavior 
results from the proximity of the $m=1$ fully-polarized quantum phase 
associated with the $m\rightarrow 1$ limit.

Studies of the spin dynamical structure factors of the half-filled 1D Hubbard model in
magnetic field has shown that the momentum dependent
exponents that control their line shape at and near the cusp singularities depend very little 
on $u=U/4t$ \cite{Carmelo_21,Carmelo_16}. The main effect of correlations was found to be on the
energy bandwidth of the dynamical structure factors's $(k,\omega)$-plane spectra, which increases
upon decreasing $u$, yet preserving the same shape. Hence, the half-filled 1D Hubbard model for {\it any} 
finite $u=U/4t$ value and a suitable choice of units for the spectra's energy bandwidths 
can describe the same spin dynamical properties. 

In this paper we have investigated whether suitable values of the interaction for chain compounds 
at $m=0$ and for ultra-cold atom systems for $m>0$ described by 1D Mott-Hubbard insulators can be settled by the 
agreement with experimental results on the up- and down-spin one-particle spectral 
functions. Analysis of the spectra plotted in Figs. \ref{figure2},\ref{figure3} and Figs. \ref{figure7}-\ref{figure12}
reveals that again their energy bandwidth increases upon decreasing $u$, yet preserving the same shape.
For $m=0$ (Figs. \ref{figure4}-\ref{figure6}) and low $m$ values (Figs. \ref{figure13}-\ref{figure18}),
the momentum dependent exponents are little $u$ dependent, the same applying to the 
$s$ branch-line exponents (Figs. \ref{figure15},\ref{figure18}) for all spin densities.

In the down-spin case for $m>0$, the exponents that are negative do not depend much on $u$.
The $c$ and $c'$ branch-line exponents though depend much on $m$. 
The $s$ branch-line exponent depends less on $m$ and stays negative.
On the other hand, the up-spin case is different: The exponents are less dependent 
on $m$. Both for up- and down-spin, the $c$ and $c'$ branch-line exponents
(Figs. \ref{figure13},\ref{figure14} and \ref{figure16},\ref{figure17}) are quite
$u$ dependent upon increasing $m$. Indeed, changing $u$ at some fixed $m$ values
leads to corresponding changes of the $k$ intervals for which the exponents are positive 
and negative. This refers to cusp singularities not emerging and emerging in
the spectral functions, respectively. 

Hence we conclude that at $m=0$ changing the interaction values changes little the one-electron
spectral properties of chain compounds described by 1D Mott-Hubbard insulators, since
a suitable choice of units for the spectra's energy bandwidths 
can describe the same spectral properties for different $u$ values.
This implies that physical quantities other than the spin dynamical structure
factors \cite{Carmelo_21,Carmelo_16} and the one-particle spectral functions studied in this paper should be
used to find the interaction values suitable to such compounds. Specifically
and in spite of the lack of superconductivity in 1D, the gapped
$(k,\omega)$-plane distribution of the cusp singularities at and just below the spectra's thresholds
of the spin singlet and triplet pairs spectral functions could provide physically
important information on the issue under consideration both for the
Mott-Hubbard insulator and the corresponding doped insulator.
Such a study could be interesting for instance for the physics of
chain cuprates \cite{Raczkowski_15}. 

An interesting related problem that could be studied elsewhere is whether 
a description in terms of $\eta$-spins of projection $-1/2$ and $+1/2$,
which replace the $u\gg 1$ doublons and holons \cite{Zawadzki_19}, respectively, for
$u>0$, could at $m=0$ and for values of $u$ not necessarily large
to reproduce the characteristic behaviors of the experimental
$\chi^{(3)}$ observed in 1D Mott-Hubbard insulators \cite{Kishida_00}.
This is a problem of technological interest for third-harmonic generation spectroscopy \cite{Shinjo_18,Kishida_00}.

On the other hand, in the case of ultra-cold atom systems on optical lattices 
described by 1D Mott-Hubbard insulators, upon
increasing the spin density $m$ one reaches a regime where changing $u$ at
fixed $m$ changes the one-particle spectral properties. Hence at large enough 
spin densities $m$, suitable values of the interaction for such atomic systems
could indeed be settled by the agreement with experimental results on the up- and down-spin 
one-particle spectral functions. This involves the specific $u$-dependent $k$ intervals for which
the exponents plotted in Figs. \ref{figure4}-\ref{figure6} and \ref{figure13}-\ref{figure18}
as a function of momentum $k$ for different $m$ and $u$ values
are negative and positive, respectively. This predicts a $u$-dependent $(k,\omega)$-plane 
distribution of cusp singularities in the up- and down-spin spectral functions, Eq. (\ref{Bkomega}),
whose comparison with experiments could settle the $u$ value suitable to
the ultra-cold atoms. (Cusp singularities also emerge 
in the $(k,\omega)$-plane regions at and near the $c-s$ boundary lines.)

The interacting spin-$1/2$ fermions described by the model under study here can either be electrons
or spin-$1/2$ atoms. In condensed matter materials at zero magnetic field, ARPES directly 
measures the spectral function of the electrons \cite{Damascelli_03}, which are removed via the photoelectric effect. 
On the other hand, the 1D Hubbard model with one 
fermion per site in a magnetic field can be implemented with ultra-cold atoms \cite{Batchelor_16,Guan_13,Zinner_16,Dao_2007,Stewart_08,Clement_09,Febbri_12}. 
Momentum-resolved radio-frequency (rf) spectroscopy \cite{Dao_2007,Stewart_08} and Bragg spectroscopy 
\cite{Clement_09,Febbri_12} are techniques to measure the spectral functions in ultra-cold atomic systems. 
In particular, momentum-resolved rf is a tool to achieve an analog of ARPES
for ultra-cold atomic systems. The spectroscopy takes advantage of the many spin states of the atoms in these cold systems. 

Can the $(k,\omega)$-plane distribution of spectral peaks at finite magnetic field associated with the
cusp singularities of our theoretical predictions be accessed by ultra-cold spin-$1/2$ atomic experiments?
In a magnetic field, the degeneracy of the atoms's spin states is split by the Zeeman interaction at magnetic field 
strengths around the Feshbach resonance. This Zeeman splitting is much larger than other energy scales in the system. 
Fortunately, all the spin-relaxation mechanisms available to atoms in some of their spin-states are either forbidden 
or strongly suppressed, so a system of atoms in those spin states stays that way without relaxing.

Momentum-resolved rf spectroscopy, takes advantage of the fact that the rf photon has a negligible momentum compared to the 
momentum of the atom. As a result, the spin-flip transition does not change its momentum state. In the language of photoemission 
spectroscopy, this is a vertical transition. The momentum of the spin-flipped atom, and thus the momentum of the atom inside 
the interacting system, can be measured in a time-of-flight experiment. 
Importantly, with this information, one {\it can indeed} reconstruct the one-particle spectral function and thus use the present
results as a theoretical prediction and check their relevance and consequences for actual physical systems \cite{Dao_2007,Stewart_08}. 

Hence our theoretical predictions are of experimental interest for condensed-matter 
chain compounds at $m=0$ where they could be tested by ARPES and for systems of ultra-cold spin-$1/2$ atoms 
on optical lattices for finite spin densities.

\acknowledgements
We thank Karlo Penc for illuminating discussions.
J. M. P. C. would like to thank the Boston University's Condensed Matter Theory Visitors Program for support and
Boston University for hospitality during the initial period of this research. He acknowledges support from
FCT through Grants PTDC/FIS-MAC/29291/2017, SFRH/BSAB/142925/2018, and POCI-01-0145-FEDER-028887.
J. M. P. C. and T. \v{C}. acknowledge support from FCT through Grant UID/FIS/04650/2013.
T. \v{C}. gratefully acknowledges support by the Institute for Basic Science in Korea (IBS-R024-D1).
P. D. S. acknowledges support from FCT through Grants UID/CTM/04540/2013 and UID/CTM/04540/2019.



\appendix

\section{Rotated-fermion related fractionalized particles}
\label{A}

In this appendix the general rotated-fermion representation of the 1D Hubbard model 
for its full Hilbert space is briefly described. It applies both when
its operators $c_{j,\sigma}^{\dagger}$ and $c_{j,\sigma}$ in the expression, Eqs. (\ref{H}) and (\ref{HH}),
refer to electrons and spin-$1/2$ atoms. Rotated fermions are generated from the 
fermions associated with such operators by a unitary transformation. 
The unique definition of the corresponding
fermion - rotated-fermion unitary operator in terms of its $4^{N_a}\times 4^{N_a}$
matrix elements between the model's $4^{N_a}$ energy eigenstates
is given in Ref. \onlinecite{Carmelo_17}, which considers electrons.

As for fermions in the $u\rightarrow\infty$ limit, up- and down-spin single site occupancy,
double site occupancy, and site no occupancy are for the whole $u>0$ range good quantum numbers
for rotated fermions. This means that in terms of such fermions, the local
occupancy configurations whose superposition generates any energy
eigenstate have fixed values for the numbers $L_{s,\pm 1/2}$ of sites singly occupied by rotated 
fermions of spin projection $\pm 1/2$, $L_{\eta,+1/2}$ of sites unoccupied, 
and $L_{\eta,-1/2}$ of sites doubly occupied by them.
Hence $L_{s}=L_{s,+1/2}+L_{s,-1/2}$ gives the total number of sites singly occupied
by rotated fermions and  $L_{\eta}=L_{\eta,+1/2}+L_{\eta,-1/2}$ that 
of sites doubly occupied and unoccupied by rotated fermions.

The rotated-fermion degrees of freedom naturally separate for $u>0$ into occupancy configurations of 
three basic fractionalized particles that generate the exact irreducible representations of the groups associated
with the independent spin and $\eta$-spin $SU(2)$ symmetries and the $c$ lattice $U(1)$ symmetry, respectively, in the
model global $[SU(2)\otimes SU(2)\otimes U(1)]/Z_2^2$ symmetry \cite{Carmelo_17,Carmelo_18A}.
(See operators's relations in Eq. (32) of Ref. \onlinecite{Carmelo_17}.)
For $u>0$ a number $L_s$ of spin $1/2$'s emerge from the rotated fermions that 
singly occupy sites and describe their spin degrees of freedom, 
a number $L_{\eta}$ of $\eta$-spin $1/2$'s emerge from the rotated fermions that doubly occupy sites
and from the unoccupied sites and describe the $\eta$-spin degrees of freedom of such site occupancies, 
and a number $N_c = L_s = N_a - L_{\eta}$ of $c$ particles and $N_c^h = L_{\eta} = N_a - L_{s}$ 
of $c$ holes also emerge and describe the translational degrees of freedom associated with
the relative positions of spin $1/2$'s and $\eta$-spin $1/2$'s, respectively.

The $\eta$-spin $1/2$'s only see the set of $L_{\eta}=N_a -L_s$ sites unoccupied 
and doubly occupied by rotated fermions. The $\eta$-spin $1/2$'s 
thus live in an $\eta$-spin squeezed effective lattice \cite{Karlo_97,Ogata_90,Kruis_04} with $L_{\eta}=N_a -L_s$ sites 
that corresponds to an $\eta$-spin-$1/2$ $XXX$ chain.
The spin $1/2$'s only ``see'' the set of $L_s=N_a-L_{\eta}$ sites singly occupied by rotated fermions. 
They live in a spin squeezed effective lattice with $L_s=N_a-L_{\eta}$ sites that corresponds to a spin-$1/2$
$XXX$ chain. These 1D squeezed lattices are known in the 
$u\rightarrow\infty$ limit in which the rotated fermions become fermions \cite{Karlo_97,Ogata_90,Kruis_04}.
On the other hand, the $c$ particles live on an effective lattice identical to the original lattice. 

The irreducible representations of the $c$ lattice $U(1)$ symmetry group store full information on the relative positions in
the model's original lattice of the spin and $\eta$-spin squeezed effective lattices sites, respectively.
The irreducible representations of the groups associated with the $\eta$-spin and spin $SU(2)$ symmetries
are generated by $\eta$-spin $1/2$'s and spin $1/2$'s occupancy configurations,
respectively, in their squeezed effective lattices. They are similar to those of an $\eta$-spin-$1/2$ and a spin-$1/2$ $XXX$ 
chain on a lattice with $L_{\eta}$ and $L_s $ sites, respectively. 

Let us denote generally by $S_{\alpha}$ and $S_{\alpha}^z$ the $\eta$-spin ($\alpha =\eta$)
and spin ($\alpha =s$) numbers. The spin $1/2$'s and
$\eta$-spin $1/2$'s are then generally named $\alpha$-spin $1/2$'s. The $\alpha$-spin
squeezed effective lattice's local configurations of each energy eigenstate
contain a set of $M_{\alpha}=2S_{\alpha}$ $\alpha$-spin $1/2$'s 
that participate in a $\alpha$-spin multiplet configuration and a complementary set of even number 
$2\Pi_{\alpha} = L_{\alpha}-2S_{\alpha}$ of $\alpha$-spin $1/2$'s that participate in 
$S_{\alpha}^z = S_{\alpha}=0$ $\alpha$-spin singlet configurations. 
The energy eigenstates are superpositions of such configuration terms, 
each being characterized by a different partition of $L_{\alpha}$ $\alpha$-spin $1/2$'s into 
$M_{\alpha}=2S_{\alpha}$ $\alpha$-spin $1/2$'s that participate in a 
$2S_{\alpha}+1$ $\alpha$-spin multiplet and a product of $\alpha$-spin singlets involving the remaining
even number $2\Pi_{\alpha} = L_{\alpha}-2S_{\alpha}$ of $\alpha$-spin $1/2$'s that form a tensor 
product of $\alpha$-spin singlet states. 

For $\alpha =s$ the numbers of {\it unpaired spins $1/2$} and 
{\it paired spins $1/2$} thus are $M_{s} \equiv 2S_{s}$ and $2\Pi_{s}\equiv L_{s}-2S_{s}$, 
respectively. For $\alpha =\eta$ the numbers of {\it unpaired $\eta$-spins $1/2$} and 
{\it paired $\eta$-spins $1/2$} are $M_{\eta} \equiv 2S_{\eta}$ and $2\Pi_{\eta} =
L_{\eta}-2S_{\eta}$, respectively. Hence, $L_{s} = L_{s,+1/2} + L_{s,-1/2} = 2\Pi_{s} + M_{s}$ 
and $L_{\eta} = L_{\eta,+1/2} + L_{\eta,-1/2} = 2\Pi_{\eta} + M_{\eta}$ where
$L_{s,\pm 1/2} = \Pi_{s} + M_{s,\pm 1/2}$ and $L_{\eta,\pm 1/2} = \Pi_{\eta} + M_{\eta,\pm 1/2}$. 
Here $M_{s,\pm 1/2} = S_{s}\mp S_{s}^{z}$ gives the number of unpaired spins $1/2$
of projection $\pm 1/2$ and $M_{\eta,\pm 1/2} = S_{\eta}\mp S_{\eta}^{z}$
that of unpaired $\eta$-spin $1/2$'s of projection $\pm 1/2$, respectively.
$\Pi_{s}\equiv L_{s}/2-S_{s}$ is the number of paired spins of both projections $\pm 1/2$
and $\Pi_{\eta} = L_{\eta}/2-S_{\eta}$ that of paired $\eta$-spins of both projections $\pm 1/2$.

The basic fractionalized particles are directly related to the quantum numbers of the Bethe-ansatz solution whose occupancy
configurations generate the exact energy eigenstates \cite{Carmelo_17,Carmelo_18A}. One has that
$\Pi_{s} = \sum_{n=1}^{\infty}n\,N_{s n}$ and $\Pi_{\eta} = \sum_{n=1}^{\infty}n\,N_{\eta n}$.
Here $N_{s n}$ is for $n=1$ the number called here $N_s$ of $s$ particles  (named $s1$ pseudoparticles 
in Ref. \onlinecite{Carmelo_17}) and of spin $n$-strings for $n>1$ and $N_{\eta n}$ is for $n=1$
the number called here $N_{\eta}$ of $\eta$ particles (named $\eta 1$ pseudoparticles in Ref. \onlinecite{Carmelo_17}) 
and of charge $n$-strings for $n>1$. Such particles and strings correspond to 
Bethe-ansatz quantum numbers, spin and charge $n$-strings being described by non-real complex
rapidities and $c$, $s$, and $\eta$ particles by only real rapidities \cite{Carmelo_17,Carmelo_18A}.

The internal degrees of freedom of one $s$ particle refers to one {\it unbound spin-singlet pair} of
spin $1/2$'s. The internal degrees of freedom of one spin $n$-string refers to 
$n>1$ spin-singlet pairs of spin $1/2$'s bound within it. The internal degrees of freedom of 
one $\eta$ particle refers to one {\it unbound $\eta$-spin-singlet pair} of
$\eta$-spin $1/2$'s. The internal degrees of freedom of one charge $n$-string refers to 
$n>1$ $\eta$-spin-singlet pairs of $\eta$-spin $1/2$'s bound within it \cite{Carmelo_17,Carmelo_18A}. 

The Bethe-ansatz quantum numbers are distributed by 
$c$- and $\alpha n$-bands whose number of occupied values are the above 
numbers $N_c$ and $N_{\alpha n}$, respectively, where $\alpha = \eta,s$ and $n=1,2,...,\infty$.
The translational degrees of freedom of the $M_s=2S_s$ unpaired spin $1/2$'s
(and $M_{\eta} = 2S_{\eta}$ unpaired $\eta$-spin $1/2$'s) are described by the $sn$-band holes 
(and $c$-band and $\eta n$-band holes\cite{Carmelo_18A}.) For the subspace of this paper, only the
$c$- and $s$-bands whose quantum numbers $q_j = {2\pi\over L}I_j^c$
and $q_j = {2\pi\over L}I_j^s$, respectively, are given in Eqs. (\ref{q-j}) and (\ref{Ic-an}) of Appendix \ref{B}
have finite occupancy. 

Finally, the usual holons and spinons refer to the $c$- and $s$-band holes
in excited energy eigenstates of $S_{\eta}=0$ and $S_s=0$ ground states,
respectively. They describe translational degrees of freedom of the unpaired 
$\eta$-spin $1/2$'s and unpaired spin $1/2$'s, respectively.

\section{Some useful and needed quantities}
\label{B}

The 1D Hubbard model with one fermion per site in the subspace of the
quantum problem studied in this paper involves the following Bethe-ansatz equations, 
\begin{eqnarray}
&& q_j = {2\over L} \sum_{j'=1}^{N_a}\, N_{c}(q_{j'})\arctan\left({\Lambda (q_j)-\sin k (q_{j'})\over u}\right)
\nonumber \\
&& \hspace{0.75cm} - {2\over L}\sum_{j'=1}^{N_{\uparrow}}\, N_{s}(q_{j'})\arctan
\left({\Lambda (q_j)-\Lambda (q_{j'})\over 2u}\right) 
\nonumber \\
&& {\rm where} \hspace{0.20cm}  j = 1,...,N_{a_s}\hspace{0.20cm} {\rm and}\hspace{0.20cm}N_{a_s} = N_{\uparrow} 
\nonumber \\
\nonumber \\
&& k (q_j) = q_j 
\nonumber \\
&& \hspace{0.35cm} - {2\over L}
\sum_{j'=1}^{N_{\uparrow}}\,N_{s}(q_{j'})\arctan\left({\sin k (q_j)-\Lambda (q_{j'}) \over u}\right)
\nonumber \\
&& {\rm where} \hspace{0.5cm} j = 1,...,N_{a_c}\hspace{0.20cm} {\rm and}\hspace{0.20cm}N_{a_c} = N = N_a   \, .
\label{TapcoS}
\end{eqnarray}
Here,
\begin{equation}
q_j = {2\pi\over L}\,I^{\beta}_j \hspace{0.20cm}{\rm for}\hspace{0.20cm}
\beta = c,s \, ,
\label{q-j}
\end{equation} 
where the $j=1,...,N_{a_{\beta}}$ quantum numbers $I^{\beta}_j$ are for $\beta = c,s$ either integers or half-odd integers 
according to the following boundary conditions \cite{Takahashi_72},
\begin{eqnarray}
I_j^{c} & = & 0,\pm 1,\pm 2,... \hspace{0.50cm}{\rm for}\hspace{0.15cm}N_{\downarrow}\hspace{0.15cm}{\rm even} 
\nonumber \\
& = & \pm 1/2,\pm 3/2,\pm 5/2,... \hspace{0.50cm}{\rm for}\hspace{0.15cm}N_{\downarrow}\hspace{0.15cm}{\rm odd} 
\nonumber \\
I_j^{s} & = & 0,\pm 1,\pm 2,... \hspace{0.50cm}{\rm for}\hspace{0.15cm}N_{\uparrow}\hspace{0.15cm}{\rm odd} 
\nonumber \\
& = & \pm 1/2,\pm 3/2,\pm 5/2,... \hspace{0.50cm}{\rm for}\hspace{0.15cm}N_{\uparrow}\hspace{0.15cm}{\rm even} \, .
\label{Ic-an}
\end{eqnarray}

The energy dispersions $\varepsilon_c (q)$ and $\varepsilon_s (q')$ that appear in the one-particle spectra 
are defined as ,
\begin{eqnarray}
\varepsilon_c (q) & = & {\bar{\varepsilon}_c} (k (q)) 
\hspace{0.20cm}{\rm and}\hspace{0.20cm} \varepsilon_{s} (q') = {\bar{\varepsilon}}_{s} (\Lambda (q')) 
\hspace{0.20cm}{\rm where}
\nonumber \\
{\bar{\varepsilon}}_c (k) & = & - \Delta_{MH} + \int_{\pi}^{k}dk^{\prime}\,2t\,\eta_c (k^{\prime})\hspace{0.20cm}{\rm and}
\nonumber \\
{\bar{\varepsilon}}_{s} (\Lambda) & = & \int_{B}^{\Lambda}d\Lambda^{\prime}\,2t\,\eta_{s} (\Lambda^{\prime}) \, .
\label{equA4}
\end{eqnarray}
and the corresponding group velocities are given by,
\begin{equation}
v_c (q) = {\partial\varepsilon_c (q)\over\partial q}
\hspace{0.20cm}{\rm and}\hspace{0.20cm} 
v_s (q') = {\partial\varepsilon_s (q')\over\partial q'} \, .
\label{equA4B}
\end{equation}

In Eq. (\ref{equA4}), $\Delta_{MH}$ refers to the Mott-Hubbard gap $2\Delta_{MH}$, Eq. (\ref{2DeltaMHallm}),
and the distributions $2t\,\eta_c (k)$ and $2t\,\eta_{s} (\Lambda)$ 
are solutions of the coupled integral equations,
\begin{eqnarray}
2t\,\eta_c (k) & = & 2t\sin k + \frac{\cos k}{\pi\,u} \int_{-B}^{B}d\Lambda\,{2t\,\eta_{s} (\Lambda)\over 1 
+  \left({\sin k - \Lambda\over u}\right)^2} \, ,
\nonumber \\
2t\,\eta_{s} (\Lambda) & = & {1\over\pi\,u}\int_{-\pi}^{\pi}dk\,{2t\,\eta_c (k)\over 1 +  \left({\Lambda-\sin k\over u}\right)^2} 
\nonumber \\
& - & \frac{1}{2\pi\,u} \int_{-B}^{B}d\Lambda^{\prime}\,{2t\,\eta_{s} (\Lambda^{\prime})\over 1 +  \left({\Lambda -
\Lambda^{\prime}\over 2u}\right)^2} \, .
\label{equA5}
\end{eqnarray}

The rapidity distribution functions $k (q)$ for $q \in [-\pi,\pi]$ and $\Lambda (q')$ 
for $q' \in [-k_{F\uparrow},k_{F\uparrow}]$ in the arguments of the dispersions
${\bar{\varepsilon}_c}$ and $ {\bar{\varepsilon}}_{s}$,  Eq. (\ref{equA4}), are defined in terms of their inverse 
functions,
\begin{eqnarray}
q (k) & = & k + \frac{1}{\pi} \int_{-B}^{B}d\Lambda\,2\pi\sigma (\Lambda)\, \arctan \left({\sin k -
\Lambda\over u}\right) 
\nonumber \\
& & {\rm for}\hspace{0.20cm} k \in [-\pi,\pi] \hspace{0.20cm}{\rm and}
\nonumber \\
q' (\Lambda) & = & {1\over\pi}\int_{-\pi}^{\pi}dk\,2\pi\rho (k)\, \arctan \left({\Lambda-\sin k\over u}\right) 
\nonumber \\
& - & \frac{1}{\pi} \int_{-B}^{B}d\Lambda^{\prime}\,2\pi\sigma (\Lambda^{\prime})\, \arctan \left({\Lambda -
\Lambda^{\prime}\over 2u}\right) 
\nonumber \\
& & {\rm for}\hspace{0.20cm} \Lambda \in [-\infty,\infty] \, ,
\label{equA7}
\end{eqnarray}
respectively. The parameter $B$ in Eqs. (\ref{equA4}), (\ref{equA5}), and (\ref{equA7}) is such that,
\begin{eqnarray}
B & = & \Lambda (k_{F\downarrow}) \hspace{0.20cm}{\rm with}
\nonumber \\
\lim_{m\rightarrow 0} B & = & \infty\hspace{0.20cm}{\rm and}\hspace{0.20cm}
\lim_{m\rightarrow 1} B = 0 \, .
\label{QB-r0rs}
\end{eqnarray}
Furthermore, the distributions,
\begin{equation}
2\pi\rho (k) = {d q (k)\over dk}\hspace{0.20cm}{\rm and}\hspace{0.20cm}
2\pi\sigma (\Lambda) = {d q' (\Lambda)\over d\Lambda} \, ,
\label{equA8}
\end{equation}
 in Eq. (\ref{equA7}) are solutions of the coupled equations,
\begin{eqnarray}
2\pi\rho (k) & = & 1 + \frac{\cos k}{\pi\,u} \int_{-B}^{B}d\Lambda\,{2\pi\sigma (\Lambda)\over 1 
+  \left({\sin k - \Lambda\over u}\right)^2} \, ,
\nonumber \\
2\pi\sigma (\Lambda) & = & {1\over\pi\,u}\int_{-\pi}^{\pi}dk\,{2\pi\rho (k)\over 1 +  \left({\Lambda-\sin k\over u}\right) ^2} 
\nonumber \\
& - & \frac{1}{2\pi\,u} \int_{-B}^{B}d\Lambda^{\prime}\,{2\pi\sigma (\Lambda^{\prime})\over 1 +  \left({\Lambda -
\Lambda^{\prime}\over 2u}\right)^2} \, ,
\label{equA10}
\end{eqnarray}
and obey the sum rules,
\begin{equation}
{1\over\pi}\int_{-\pi}^{\pi}dk\,2\pi\rho (k) = 2
\hspace{0.20cm}{\rm and}\hspace{0.20cm}
\frac{1}{\pi} \int_{-B}^{B}d\Lambda\,2\pi\sigma (\Lambda) = (1-m) \, .
\label{equA10B}
\end{equation}

At $h=0$ and $m=0$ the $c$- and $s$-bands energy dispersions, Eq. (\ref{equA4}), can 
for $u>0$ be written as,
\begin{eqnarray}
\varepsilon_{c} (q) & = & {\bar{\varepsilon}}_{c} (k_c (q)) 
\hspace{0.20cm}{\rm for}\hspace{0.20cm}q \in \left[-\pi,\pi\right] 
\hspace{0.20cm}{\rm where}
\nonumber \\
{\bar{\varepsilon}}_{c} (k) & = & {U\over 2} - 2t\cos k
\nonumber \\
& & - 4t\int_0^{\infty}d\omega\,{\cos (\omega\,\sin k)\over\omega
\left(1 + e^{2\omega\,u}\right)}\,J_1 (\omega)
\nonumber \\
& & {\rm for}\hspace{0.20cm}k \in [-\pi,\pi] \hspace{0.20cm}{\rm and} 
\nonumber \\
\varepsilon_{s} (q) & = & {\bar{\varepsilon}}_{s} (\Lambda_s (q)) 
\hspace{0.20cm}{\rm for}\hspace{0.20cm}q \in \left[-{\pi\over 2},{\pi\over 2}\right] 
\hspace{0.20cm}{\rm where}
\nonumber \\
{\bar{\varepsilon}}_{s} (\Lambda) & = & 
- 2t\int_0^{\infty}d\omega\,{\cos (\omega\,\Lambda)\over\omega\cosh (\omega\,u)}\,J_1 (\omega) 
\nonumber \\
& & {\rm for}\hspace{0.20cm}\Lambda\in [-\infty,\infty] \, ,
\label{varesm0}
\end{eqnarray}
respectively. At $m=0$, the inverse functions of $k_c (q)$ and $\Lambda_s (q)$, Eq. (\ref{equA7}), read,
\begin{eqnarray}
q & = & q_c (k) =  k + 2\int_0^{\infty}d\omega\,{\sin (\omega\,\sin k)\over\omega\left(1 + e^{2\omega\,u}\right)}\,J_0 (\omega)  \, ,
\nonumber \\
q & = & q_s (\Lambda) = \int_0^{\infty}d\omega\,{\sin (\omega\,\Lambda)\over\omega\cosh (\omega\,u)}\,J_0 (\omega) \, .
\label{qLambdam0}
\end{eqnarray}
Here and above $J_0 (\omega)$ and $J_1 (\omega)$ are Bessel functions.

For $m=0$ and $u>0$ the bare $c$- and $s$-band energy dispersions 
defined below in Appendix \ref{C} are given by,
\begin{eqnarray}
{\bar{\varepsilon}}_{c}^0 (k) & = & - {U\over 2} - 2t\cos k
\nonumber \\
& & - 4t\int_0^{\infty}d\omega\,{\cos (\omega\,\sin k)\over\omega
\left(1 + e^{2\omega\,u}\right)}\,J_1 (\omega)
\nonumber \\
& & {\rm for}\hspace{0.20cm}k \in [-\pi,\pi] 
\nonumber \\
{\bar{\varepsilon}}_{s}^0 (\Lambda) & = & 
- 2t\int_0^{\infty}d\omega\,{\cos (\omega\,\Lambda)\over\omega\cosh (\omega\,u)}\,J_1 (\omega) 
\nonumber \\
& & {\rm for}\hspace{0.20cm}\Lambda\in [-\infty,\infty] \, .
\label{varecm02}
\end{eqnarray}

The $\beta = c,s$ phase shifts in Eqs. (\ref{cfunDc}) and (\ref{cfunDs}) read,
\begin{eqnarray}
&& 2\pi\,\Phi_{\beta,\beta'}(q,q') = 2\pi\,\bar{\Phi }_{\beta,\beta'} \left(r,r'\right) 
\hspace{0.20cm}{\rm where}\hspace{0.20cm}\beta,\beta'=c,s
\nonumber \\
&& r = {\sin k (q)\over u}\vert_{\beta = c} 
\hspace{0.20cm}{\rm and}\hspace{0.20cm}
r = {\Lambda (q)\over u}\vert_{\beta = s}
\nonumber \\
&& r' = {\sin k (q')\over u}\vert_{\beta' = c} 
\hspace{0.20cm}{\rm and}\hspace{0.20cm}
r' = {\Lambda (q')\over u}\vert_{\beta' = s} \, ,
\label{Phi-barPhi}
\end{eqnarray}
and the rapidity phase shifts obey the integral equations,
\begin{eqnarray}
\bar{\Phi }_{s,s}\left(r,r'\right) & = & {1\over \pi}\arctan\left({r-r'\over 2}\right) 
\nonumber \\
& + & \int_{-B/u}^{B/u} dr''\,G (r,r'')\,{\bar{\Phi}}_{s,s} (r'',r') \, ,
\nonumber \\
\bar{\Phi }_{s,c}\left(r,r'\right) & = & -{1\over \pi}\arctan(r-r')
\nonumber \\
& + & \int_{-B/u}^{B/u} dr''\,G (r,r'')\,{\bar{\Phi }}_{s,c} (r'',r') \, ,
\nonumber \\
\bar{\Phi }_{c,c}\left(r,r'\right) & = & {1\over{\pi}}\int_{-B/u}^{B/u} dr''{\bar{\Phi}_{s,c} (r'',r') \over {1+(r-r'')^2}} \, ,
\nonumber \\
\bar{\Phi }_{c,s}\left(r,r'\right) & = & -{1\over \pi}\arctan (r-r')
\nonumber \\
& + & {1\over{\pi}}\int_{-B/u}^{B/u} dr''{\bar{\Phi}_{s,s} (r'',r') \over {1+(r-r'')^2}} \, ,
\label{Phicsn-m}
\end{eqnarray}
where kernel $G (r,r')$ is for $u>0$ given by,
\begin{equation}
G(r,r') = - {1\over{2\pi}}\left({1\over{1+((r-r')/2)^2}}\right) \, .
\label{Gne1}
\end{equation} 

The phase shifts appearing in Eqs. (\ref{cfunDc})-(\ref{cfunDs}) read,
\begin{eqnarray}
\Phi_{s,s}\left(\iota k_{F\downarrow},q'\right) & = & \bar{\Phi }_{s,s} \left(\iota {B\over u},{\Lambda (q')\over u}\right) 
\nonumber \\
\Phi_{s,c}\left(\iota k_{F\downarrow},q\right) & = & \bar{\Phi }_{s,c} \left(\iota {B\over u},{\sin k (q)\over u}\right) 
\nonumber \\
\Phi_{c,c}\left(\iota\pi,q\right) & = & \bar{\Phi }_{c,c} \left(0,{\sin k (q)\over u}\right) 
\nonumber \\
\Phi_{c,s}\left(\iota\pi,q'\right) & = & \bar{\Phi }_{c,s} \left(0,{\Lambda (q')\over u}\right) 
\hspace{0.05cm}{\rm for}\hspace{0.05cm}\iota = \pm 1 \, .
\label{Phis-all-qq}
\end{eqnarray}
The phase-shift related spectral parameters are given by,
\begin{eqnarray}
 \xi_{\beta\,\beta'} & = & \delta_{\beta,\beta'} 
+ \sum_{\iota=\pm 1} (\iota)\,\Phi_{\beta,\beta'}\left(q_{F\beta},\iota q_{F\beta'}\right)
\nonumber \\
{\rm where} & &  \beta, \beta' = c, s \, ,
\label{x-aa}
\end{eqnarray}
and except for unimportant $1/L$ contributions,
\begin{equation}
q_{Fs} = k_{F\downarrow}\hspace{0.20cm}{\rm and}\hspace{0.20cm}q_{Fc} = \pi \, .
\label{qq}
\end{equation}

For the present Mott-Hubbard insulator one has that,
\begin{eqnarray}
 \xi_{s\,c} & = & 0\hspace{0.20cm}{\rm for}\hspace{0.20cm}u>0\hspace{0.20cm}{\rm at}\hspace{0.20cm}n_f =1
\nonumber \\
 \xi_{c\,c} & = & 1 \hspace{0.20cm}{\rm for}\hspace{0.20cm}u> 0\hspace{0.20cm}{\rm at}\hspace{0.20cm}n_f =1 \, .
\label{xi1sc}
\end{eqnarray}
The entries of the conformal-field theory dressed-charge matrix $Z$, 
which are the parameters, Eq. (\ref{x-aa}),\cite{Carmelo_93} read,
\begin{eqnarray}
Z & = & \left[\begin{array}{cc}
 \xi_{c\,c} &  \xi_{c\,s}  \\
 \xi_{s\,c}   &  \xi_{s\,s}  
\end{array}\right] =
\left[\begin{array}{cc}
1 &  \xi_{c\,s}  \\
0  &  \xi_{s\,s} 
\end{array}\right] \, . 
\label{ZZ-gen}
\end{eqnarray}
(Here the dressed-charge matrix definition
of Ref. \onlinecite{Woy_89} has been used, which is the transposition of that of Ref. \onlinecite{Frahm_90}.)
For $u>0$, the parameter $ \xi_{c\,s}$ continuously decreases 
upon increasing the spin density $m$ from $ \xi_{c\,s}=1/2$ at $m=0$ to
$ \xi_{c\,s}=0$ for $m=1$. On the other hand, for $u>0$ the parameter $ \xi_{s\,s}$ 
continuously increases from $ \xi_{s\,s}=1/\sqrt{2}$ at $m=0$ to
$ \xi_{s\,s}=1$ for $m=1$.  

For $m=0$ and $u>0$ the matrix, Eq. (\ref{ZZ-gen}), and the $c$ and $s$
particle phase shifts, Eq. (\ref{Phis-all-qq}), simplify to,
\begin{eqnarray}
Z & = & \left[\begin{array}{cc}
1 &  \xi_{c\,s}  \\ 
0  &  \xi_{s\,s} 
\end{array}\right] =
 \left[\begin{array}{cc}
1 & 1/2 \\
0 & 1/\sqrt{2}  
\end{array}\right] \, , 
\label{ZZ-m0}
\end{eqnarray}
and
\begin{eqnarray}
\Phi_{c,c}\left(\iota\pi,q\right) & = & \Psi \left({\sin k_c (q)\over u}\right) 
\nonumber \\
\Phi_{c,s}\left(\iota\pi,q\right) & = & {1\over 2\pi} \arctan\left(\sinh\left({\pi\over 2}{\Lambda_s (q)\over u}\right)\right)
\nonumber \\
\Phi_{s,c}\left(\iota \pi/2,q\right) & = & - {\iota\over 2\sqrt{2}}
\nonumber \\
\Phi_{s,s}\left(\iota \pi/2,q\right) & = & {\iota\over 2\sqrt{2}}
\hspace{0.2cm}{\rm for}\hspace{0.2cm}q\neq \iota \pi/2
\nonumber \\
& = &\iota\left({3\over 2\sqrt{2}} - 1\right)
\hspace{0.2cm}{\rm for}\hspace{0.2cm}q= \iota \pi/2 \, ,
\label{Phicccs}
\end{eqnarray}
respectively, where $u>0$, $\iota = \pm 1$, and the function $\Psi (x)$ is given by,
\begin{eqnarray}
\Psi (x) & = & {1\over\pi}\int_0^{\infty}dz\,{\sin (z\,x)\over z\left(1 + e^{2z}\right)} 
\nonumber \\
& = & {i\over 2\pi}\,\ln\left({\Gamma \left({1\over 2} + i {x\over 4}\right)\Gamma \left(1 - i {x\over 4}\right)
\over\Gamma \left({1\over 2} - i {x\over 4}\right)\Gamma \left(1 + i {x\over 4}\right)}\right) \, .
\label{Psix}
\end{eqnarray}
Here $\Gamma (x)$ is the usual gamma function.

\section{Derivation of the Mott-Hubbard gap for $m\in [0,1]$}
\label{C}

The derivation of the Mott-Hubbard gap involves ground states
and excited states described by only real Bethe-ansatz rapidities.
We start by considering fermionic densities $n_f\in [0,1]$ and spin densities $m\in [0,n_f]$ 
for the 1D Hubbard model in the subspaces spanned by energy 
eigenstates described by only real rapidities. This involves addition to the Hamiltonian,
Eq. (\ref{H}), of the term $\mu\,{\hat{N}}=\mu\,\sum_{\sigma=\uparrow ,\downarrow }\,\hat{N}_{\sigma}$
where $\mu$ is the chemical potential.
For such subspaces, the Bethe-ansatz equations 
have for $n_f\in [0,1]$ the form, Eq. (\ref{TapcoS}) of Appendix \ref{B}, only 
differing from the $n_f=1$ case in the occupancies of the $c$-band momentum 
distribution $N_{c}(q_{j'})$, which are such that $N_c = N$ and $N_c^h = N_a-N$.

The first step of our derivation involves the introduction of $c$- and $s$-band momentum distribution 
deviations for the excited energy eigenstates under consideration,
\begin{eqnarray}
\delta N_{\beta} (q_j)  & = & N_{\beta} (q_j) - N^0_{\beta} (q_j) \hspace{0.20cm}{\rm for}\hspace{0.20cm}
\beta = c,s \hspace{0.20cm}{\rm where}
\nonumber \\
 j & = & 1,...,N_{a_{\beta}}\, ,  \hspace{0.20cm}N_{a_c} = N\, ,  \hspace{0.20cm}N_{a_s} = N_{\uparrow} \, .
\label{DNq}
\end{eqnarray}
Here,
\begin{eqnarray}
N_c^0 (q_j) & = & \theta (q_j - q_{Fc}^{-})\,\theta (q_{Fc}^{+} - q_j)  
\nonumber \\
N_{s}^0 (q_j) & = & \theta (q_j - q_{Fs}^{-})\,\theta (q_{Fs}^{+} - q_j)  \, ,
\label{N0q1DHm}
\end{eqnarray}
are the corresponding ground-state $c$- and $s$-band momentum distributions
for densities $n_f\in [0,1]$ and $m\in [0,n_f]$.
The $c$- and $s$-band Fermi momentum values $q_{F\beta}^{\pm}$ where $\beta =c,s$
appearing here are provided in Eqs. (C.4)-(C.11) of Ref. \onlinecite{Carmelo_04}. Ignoring 
${\cal{O}} (1/L)$ corrections within the present thermodynamic limit simplifies the 
ground-state distributions, Eq. (\ref{N0q1DHm}), to $N_{\beta}^0 (q_j) = \theta (q_{F\beta} - \vert q_j\vert)$
where, 
\begin{equation}
q_{Fc} = 2k_F = \pi\,n_f \hspace{0.20cm}{\rm and}\hspace{0.20cm}q_{Fs} = k_{F\downarrow} = \pi\,n_{f,\downarrow} \, . 
\label{q0Fcs}
\end{equation}

The energy spectrum of the present subspace's eigenstates is within the
Bethe-ansatz solution given by,
\begin{eqnarray}
E & = & \sum_{j=1}^{N_a} N_{c} (q_j)\left(- {U\over 2} - 2t\cos k (q_j)\right)
\nonumber \\
& - & 2\mu_B\,h\left({1\over 2} \sum_{j=1}^{N_a}N_{c} (q_j) -  \sum_{j=1}^{N_{\uparrow}}N_{s} (q_j)\right)
\nonumber \\
& + & \mu \sum_{j=1}^{N_a}N_{c} (q_j) \, ,
\label{E}
\end{eqnarray}
where the rapidity function $k (q_j)$ specific to each state is
defined by the Bethe-ansatz equations.

Next, we derive an energy functional suitable to our goal by using in the Bethe-ansatz equations, 
Eq. (\ref{TapcoS}) of Appendix \ref{B}, and energy spectrum, Eq. (\ref{E}), the $c$- and $s$-band momentum 
distribution functions $N_{\beta} (q_j) = N^{0}_{\beta} (q_j) +  \delta N_{\beta} (q_j)$. 
The combined and consistent solution of those equations and spectra up to second order in the 
deviations $\delta N_{\beta} (q_j)$, Eq. (\ref{DNq}), leads to,
\begin{eqnarray}
E & = & E_0 + \delta E\hspace{0.20cm}{\rm where}
\nonumber \\
\delta E & = & \sum_{\beta=c,s}\sum_{j=1}^{L_{\beta}}\varepsilon_{\beta} (q_j)\delta N_{\beta} (q_j) 
\nonumber \\
& + & {1\over L}\sum_{\beta=c,s}\sum_{\beta'=c,s}\sum_{j=1}^{L_{\beta}}\sum_{j'=1}^{L_{\beta'}}
{1\over 2}\,f_{\beta\,\beta'} (q_j,q_{j'})
\nonumber \\
& \times & \delta N_{\beta} (q_j)\delta N_{\beta'} (q_{j'}) \, .
\label{DE-fermions}
\end{eqnarray}
Here the zeroth-order term $E_0$ is the energy of the ground state corresponding to given values of the densities 
$n_f\in [0,1]$ and $m\in [0,n_f]$. The $f$ functions in the second-order terms involve the $c$ and $s$ 
group velocities and phase shifts defined in Appendix \ref{B}. Their expressions, though, are not needed 
for the present derivation since the corresponding second-order contributions in the deviations 
lead to $1/L$ corrections to the Mott-Hubbard gap that vanish in the thermodynamic limit.

Only the first-order terms in the deviations contribute to the quantities calculated here.
The $c$- and $s$-band energy dispersions $\varepsilon_{\beta} (q_j)$ 
in such first-order terms are found to be given by $\varepsilon_c (q) = {\bar{\varepsilon}_c} (k (q))$
and $\varepsilon_{s} (q') = {\bar{\varepsilon}}_{s} (\Lambda (q'))$,
Eq. (\ref{equA4}) of Appendix \ref{B}, where the rapidity
functions $k (q)$ and $\Lambda (q')$ are defined by the Bethe-ansatz equations.
Importantly, the rapidity-dependent $c$- and $s$-band energy dispersions are for densities 
$n_f\in [0,1]$ and $m\in [0,n_f]$ found to obey the following relations,
\begin{eqnarray}
{\bar{\varepsilon}}_c (k) & = & \mu -\mu_B\,h + {\bar{\varepsilon}}_c^0 (k)
\nonumber \\
{\bar{\varepsilon}}_{s} (\Lambda) & = & 2\mu_B\,h + {\bar{\varepsilon}}_{s}^0 (\Lambda) 
= {\bar{\varepsilon}}_{s}^0 (\Lambda) - {\bar{\varepsilon}}_{s}^0 (B) \, .
\label{equA42}
\end{eqnarray}

The zero-energy level of the bare rapidity-dependent $c$- and $s$-band energy dispersions 
${\bar{\varepsilon}}_c^0 (k)$ and ${\bar{\varepsilon}}_{s}^0 (\Lambda)$ in these relations is 
shifted relative to that of the corresponding energy dispersions ${\bar{\varepsilon}}_c (k)$
and ${\bar{\varepsilon}}_{s} (\Lambda)$, respectively. The
former dispersions are found to be given by,
\begin{eqnarray}
{\bar{\varepsilon}}_c^0 (k) & = & - {U\over 2} - 2t\cos k 
\nonumber \\
& + & \frac{1}{\pi} \int_{-B}^{B}d\Lambda\,2t\,\eta_s (\Lambda)\,\arctan \left({\sin k - \Lambda\over u}\right)
\nonumber \\
{\bar{\varepsilon}}_s^0 (\Lambda) & = & \int_{\infty}^{\Lambda}d\Lambda^{\prime}\,2t\,\eta_{s} (\Lambda^{\prime})
\nonumber \\
& = & \frac{1}{\pi} \int_{-Q}^{Q}dk\,2t\,\eta_c (k)\,\arctan \left({\Lambda - \sin k\over u}\right)
\nonumber \\
& - & \frac{1}{\pi} \int_{-B}^{B}d\Lambda'\,2t\,\eta_s (\Lambda')\,\arctan \left({\Lambda - \Lambda'\over 2u}\right) 
\label{equsc}
\end{eqnarray}
where $Q = k (2k_F)$ and the distributions $2t\,\eta_c (k)$ and $2t\,\eta_{s} (\Lambda)$ are solutions 
of the coupled integral equations, Eq. (\ref{equA5}) of Appendix \ref{B}, with the integrals
$\int_{-\pi}^{\pi}dk$ replaced by $\int_{-Q}^{Q}dk$. 
Related bare $c$- and $s$-band energy dispersions that are a function of the
corresponding $c$- and $s$-band momentum values are given by,
\begin{equation}
\varepsilon_c^0 (q) = {\bar{\varepsilon}}_{c}^0 (k (q)) 
\hspace{0.20cm}{\rm and}\hspace{0.20cm}
\varepsilon_{s}^0 (q') = {\bar{\varepsilon}}_{s}^0 (\Lambda (q')) \, ,
\label{equA112}
\end{equation}
where $k (q)$ and $\Lambda (q')$ are rapidity functions defined by the
Bethe-ansatz equations. 

Finally, we use the important relations given in Eq. (\ref{equA42}) for 
$n_f=1$. The zero-energy level of the present quantum problem
corresponds for that fermionic density to vanishing chemical potential, $\mu =0$, at the middle
of the Mott Hubbard gap. Consistently, from the use of the expression
${\bar{\varepsilon}}_c (k) = - \Delta_{MH} + \int_{\pi}^{k}dk^{\prime}\,2t\,\eta_c (k^{\prime})$,
Eq. (\ref{equA4}) of Appendix \ref{B}, one finds that ${\bar{\varepsilon}}_c (\pi) = - \Delta_{MH}$.
On the other hand, from the use of the expression ${\bar{\varepsilon}}_{s} (\Lambda) = 
\int_{B}^{\Lambda}d\Lambda^{\prime}\,2t\,\eta_{s} (\Lambda^{\prime})$ in that equation,
one finds that the zero-energy level of the $s$-band energy dispersion refers to
${\bar{\varepsilon}}_{s} (B) =0$. In terms of the corresponding
momentum dependent energy dispersions $\varepsilon_c (q)$ and $\varepsilon_{s} (q')$,
this gives $\varepsilon_c (q_{Fc}) = \varepsilon_c (\pi) = - \Delta_{MH}$ and 
$\varepsilon_{s} (q_{Fs})=\varepsilon_{s} (k_{F\downarrow})=0$, respectively.

Taking $\mu =0$ at the middle of the Mott-Hubbard gap, and
using ${\bar{\varepsilon}}_c (\pi) = - \Delta_{MH}$, and ${\bar{\varepsilon}}_{s} (B) =0$, 
the relations in Eq. (\ref{equA42}) give,
\begin{equation}
{\bar{\varepsilon}}_c (\pi) = - \Delta_{MH} = -\mu_B\,h + {\bar{\varepsilon}}_c^0 (\pi) \, ,
\label{equA43}
\end{equation}
and ${\bar{\varepsilon}}_{s} (B) = 0 = 2\mu_B\,h + {\bar{\varepsilon}}_{s}^0 (B)$
for the specific values $k=\pi$ and $\Lambda = B$.
It then follows that the Mott-Hubbard gap is given by,
\begin{equation}
2\Delta_{MH} = -2{\bar{\varepsilon}}_c^0 (\pi) - {\bar{\varepsilon}}_s^0 (B)
= -2\varepsilon_c^0 (\pi) - \varepsilon_s^0 (k_{F\downarrow}) \, ,
\label{MHgap2bands}
\end{equation}
and the spin-density curve reads $h (m) = - {\bar{\varepsilon}}_s^0 (B)/(2\mu_B)$,
which can be rewritten as given in Eq. (\ref{magcurve}).

The use of the bare energy dispersions's expressions, Eq. (\ref{equsc}), in that for 
the Mott-Hubbard gap $2\Delta_{MH}$ given in Eq. (\ref{MHgap2bands}) leads to that gap expression
provided in Eq. (\ref{2DeltaMHallm}). Its derivation for the whole spin density range $m\in [0,1]$ was the
main goal of this Appendix. 


\end{document}